\documentclass[11pt]{article}
\usepackage{jcapmod}
\usepackage{enumerate}
\usepackage{empheq}
\usepackage{feynmp}
\usepackage{booktabs}
\usepackage[english]{babel}
\usepackage{amsmath,amssymb,amsbsy,amstext, amsthm, simplewick}
\usepackage{graphicx}
\usepackage{amsfonts}
\usepackage{amssymb}
\usepackage{caption}
\usepackage{subcaption}
\usepackage{upgreek}
 \usepackage{exscale,relsize}
 \usepackage[makeroom]{cancel}
\usepackage{float}
\usepackage{framed}
\usepackage{array}
\newcolumntype{C}[1]{>{\centering\arraybackslash}p{#1}}

\RequirePackage{color}

\usepackage{colortbl}
\definecolor{rp}{cmyk}{0.2, 1, 0.6, 0}
\definecolor{green2}{cmyk}{0, 1, 0.5, 0}
\definecolor{lightgreen}{cmyk}{0.2, 0, 0.2, 0.2}
\definecolor{lightgray}{cmyk}{0.1,0.2,0,0.1}
\definecolor{lightgray2}{cmyk}{0.4,0.4,0,0.8}
\definecolor{black}{cmyk}{1.0,1.0,1.0,1.0}
\definecolor{paper_blue}{rgb}{0.3,0.2,0.75}
\definecolor{paper_red}{rgb}{0.68,0.02824,0.18824}
\allowdisplaybreaks[1]

\usepackage{colortbl}
\definecolor{lightgreen}{cmyk}{0.2, 0, 0.2, 0.2}
\definecolor{lightgray}{cmyk}{0.1,0.2,0,0.1}
\definecolor{lightgray2}{cmyk}{0.1,0.1,0,0.1}

\makeatletter
\newlength{\apb@width}
\newcommand{\autoparbox}[2][c]{\settowidth{\apb@width}{#2}\parbox[#1]{\apb@width}{#2}}

\makeatother

\numberwithin{equation}{section}

\def\beq{\begin{equation}}
\def\eeq{\end{equation}}

\def\bea{\begin{eqnarray}}
\def\eea{\end{eqnarray}}

\def\d{{\rm d}}

\def\beq{\begin{equation}}
\def\eeq{\end{equation}}
\def\bea{\begin{eqnarray}}
\def\eea{\end{eqnarray}}

\def\d{{\rm d}}

\def\d{{\rm d}}

\def\H{{\cal H}}

\def\0{{\boldsymbol 0}}
\def\k{{\boldsymbol{k}}}
\def\q{{\boldsymbol{q}}}
\def\p{{\boldsymbol{p}}}
\def\v{{\boldsymbol{v}}}
\def\x{{\boldsymbol{x}}}

\def\D{{\boldsymbol{ \nabla}}}

\def\knl{k_{\mathsmaller{\rm NL}}}
\def\fnl{f_{\mathsmaller{\rm NL}}}

\DeclareRobustCommand{\SkipTocEntry}[4]{}

\newcommand{\vev}[1]{\langle #1 \rangle}

\begin{document}

\begin{titlepage}

\setcounter{page}{1} \baselineskip=15.5pt \thispagestyle{empty}

\bigskip\

\vspace{1cm}
\begin{center}

{\fontsize{20}{28}\selectfont  \sffamily \bfseries  Effective Theory of Large-Scale Structure \\[10pt] with Primordial Non-Gaussianity}


\end{center}

\vspace{0.2cm}

\begin{center}
{\fontsize{13}{30}\selectfont Valentin Assassi,$^{\bigstar}$ Daniel Baumann,$^{\bigstar}$ Enrico Pajer,$^{\spadesuit}$\\[3pt] Yvette Welling,$^{\clubsuit}$  and Drian van der Woude$^{\spadesuit}$}
\end{center}

\begin{center}

\vskip 8pt
\textsl{$^\bigstar$ Department of Applied Mathematics and Theoretical Physics, \\ Cambridge University, Cambridge, CB3 0WA, UK}
\vskip 7pt

\textsl{$^\spadesuit$ Institute for Theoretical Physics, Utrecht University, \\
Leuvenlaan 4, 3584 CE Utrecht, The Netherlands}
\vskip 7pt

\textsl{$^\clubsuit$ Leiden Observatory, Universiteit Leiden, \\ Niels Bohrweg 2, 2333 CA Leiden, The Netherlands}

\end{center}

\vspace{1.2cm}
\hrule \vspace{0.3cm}
\noindent {\sffamily \bfseries Abstract} \\[0.1cm]
We develop the effective theory of large-scale structure for non-Gaussian initial conditions.  The effective stress tensor in the dark matter equations of motion contains new operators, which originate from the squeezed limit of the primordial bispectrum.  Parameterizing the squeezed limit by a scaling and an angular dependence, captures large classes of primordial non-Gaussianity.  Within this parameterization, we classify the possible contributions to the effective theory.  We show explicitly how all terms consistent with the symmetries arise from coarse graining the dark matter equations of motion and its initial conditions.  We also demonstrate that the system is closed under renormalization and that the basis of correction terms is therefore complete. The relevant corrections to the matter power spectrum and bispectrum are computed numerically and their relative importance is discussed.

\vskip 10pt
\hrule

\vspace{0.6cm}
 \end{titlepage}

\tableofcontents

\newpage
\section{Introduction}

Large-scale structure (LSS) surveys will play an increasingly important role in probing the initial conditions and subsequent evolution of our universe~\cite{Alvarez:2014vva}.
However, extracting primordial information from the observations will be challenging, and
understanding the many sources of late-time nonlinearities will be essential for realizing the full potential of the future data. On sufficiently large scales, fluctuations in the dark matter density are  small and therefore amenable to a perturbative treatment~\cite{Bernardeau:2001qr, Bernardeau:2013oda}. In contrast, dark matter is strongly clustered on small scales and perturbation theory is insufficient to describe its dynamics.  Moreover, gravitational nonlinearities couple short and long modes, so even the perturbative regime isn't immune to our uncertainties about the evolution of LSS on small scales.

\vskip 4pt
Failing to account for the backreaction of the short-scale nonlinearities on the long-wavelength universe can bias the theoretical interpretation of future observations.
Recently, this problem has been addressed using the methods of effective field theory (EFT)~\cite{Baumann:2010tm, Carrasco:2012cv}.  In this approach, dark matter fluctuations are separated into long and short modes. While the short modes are nonperturbative and can only be modelled through numerical simulations, their effects at long distances can be captured systematically by adding corrections to the evolution equations for the long-wavelength perturbations.
These corrections can be organized in a double expansion in powers of the long-wavelength fluctuations and spatial derivatives.  The allowed terms are constrained by symmetries, and only a finite number of terms is required to describe observations at a finite level of precision.

\vskip 4pt
The EFT description of the long-wavelength fluctuations is complete if (and only if) the set of operators correcting the fluid equations is closed under renormalization.
By this we mean that, at a given order, loops don't generate new operators, but only mix the existing operators.  The renormalization of the power spectrum was treated in~\cite{Carrasco:2012cv, Pajer:2013jj} (at one-loop) and in~\cite{Carrasco:2013sva} (at two-loop), while the one-loop renormalization of the bispectrum was presented in~\cite{Baldauf:2014qfa, Angulo:2014tfa} (for Gaussian initial conditions).  Extensions to halo statistics have appeared in~\cite{McDonald:2006mx, McDonald:2009dh, Schmidt:2012ys, Assassi:2014fva, Senatore:2014eva, Angulo:2015eqa},
redshift space distortions have been included in~\cite{Senatore:2014vja}, and baryonic effects were discussed in~\cite{Lewandowski:2014rca}.  Since probing primordial non-Gaussianity is a key motivation underlying these developments, it is essential to systematically develop the EFT approach for these more general initial conditions. In this paper, we work out the extension of the effective theory of large-scale structure (EFT-of-LSS) to non-Gaussian initial conditions.  We will present explicitly the renormalization of the one-loop dark matter power spectrum and bispectrum for a wide class of primordial non-Gaussianities (PNG).

\begin{figure}[h!]
        \centering
                 \hspace{-0.5cm}  \includegraphics[scale=0.75]{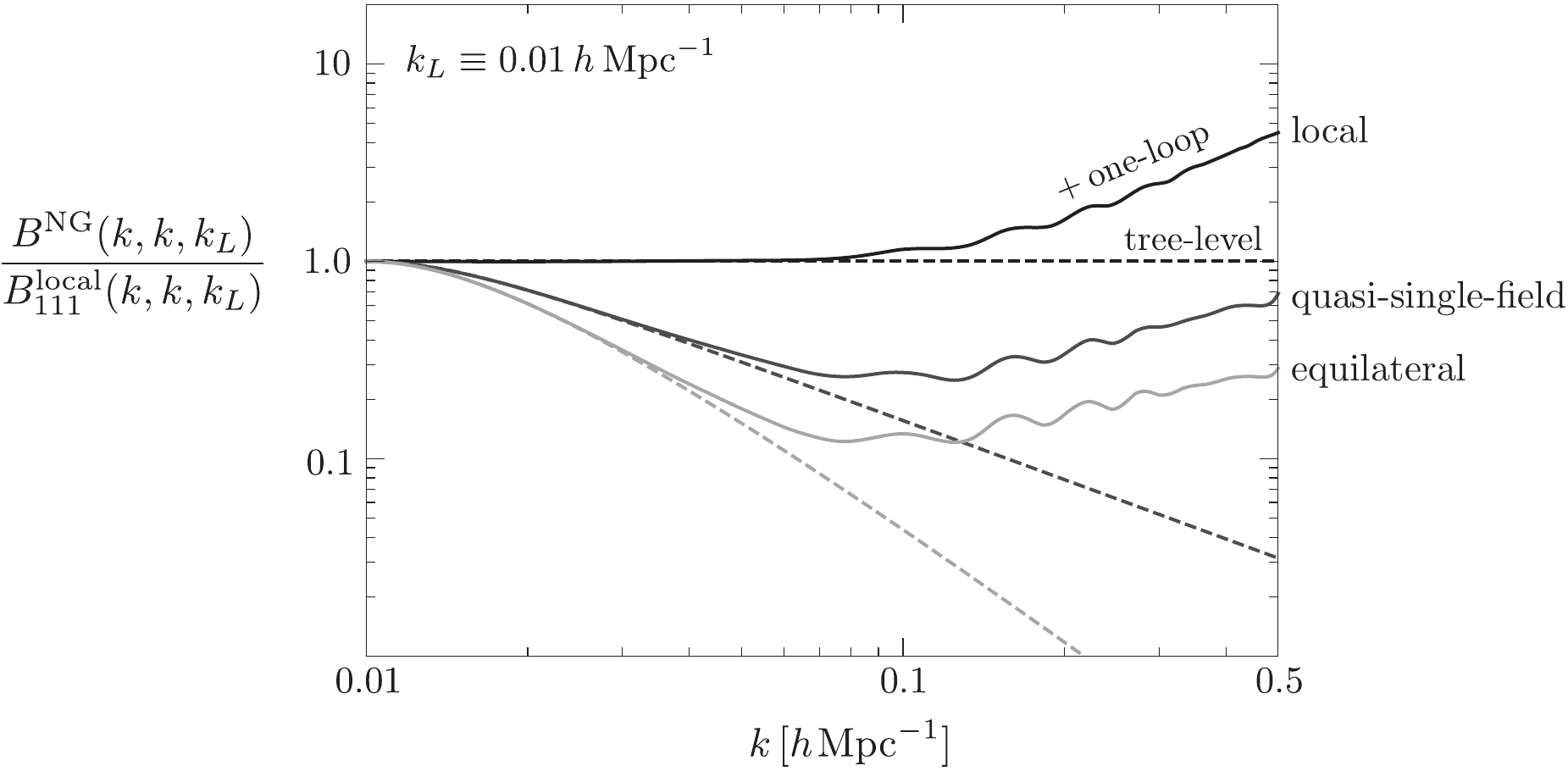}
  \caption{Solid (dashed) lines show the full (tree-level) non-Gaussian SPT contribution at $z=0$ for three representative models of primordial non-Gaussianity (see \S\ref{sec:results} for more details).} 
  \label{fig:intro}
 \end{figure} 

\vskip 4pt
The importance of an accurate treatment of nonlinear corrections to the dark matter bispectrum is illustrated in  fig.~\ref{fig:intro}.  Shown is the bispectrum computed in standard perturbation theory (SPT) for PNG of local, equilateral and quasi-single-field type (see \S\ref{sec:results} for more details).  Dashed lines refer to the tree-level signal, while solid lines contain the leading-order effects of gravitational evolution.  We see that the loop corrections become relevant on relatively large scales and tend to decrease the difference among the three primordial signals.
Even for an idealized observable such as the dark matter bispectrum, linear evolution is therefore insufficient to fully exploit the potential of current and future data sets. In this work, we will show how nonlinear gravitational corrections can be computed in a systematic and self-consistent way.

\vskip 10pt
The outline of the paper is as follows. In Section~\ref{sec:prelim}, we review the shortcomings of standard perturbation theory and explain how they are addressed in the EFT-of-LSS.  We then define the types of non-Gaussian initial conditions studied in this work (see also Appendix~\ref{sec:oddspin}).
In Section~\ref{sec:coarse}, we show how PNG leads to new terms in the stress tensor of the EFT-of-LSS.  We derive these terms both from the `top-down', by coarse graining the equations of motion (cf.~Appendix~\ref{sec:topdown}), and from the `bottom-up', by constructing the most general stress tensor consistent with the symmetries.
In Section~\ref{sec:renorm}, we show that our set of EFT terms is closed under renormalization. We explicitly derive the one-loop counterterms required by the renormalization procedure (with details given in Appendix~\ref{sec:SPT-Appendix}).
In Section~\ref{sec:numerics}, we compute the renormalized one-loop dark matter bispectrum numerically (see also Appendix~\ref{sec:IR}).  
We provide a preliminary analysis of the shapes of the distinct EFT contributions.
A more detailed treatment will appear in~\cite{future}.  Our conclusions are summarized in Section~\ref{sec:conclusions}.


\subsection*{Notation and Conventions}

The most important variables used in this work are collected in Appendix~\ref{sec:notation}.  
Both conformal time $\tau$ and the scale factor $a$ are employed as measures of time evolution.
Three-dimensional vectors will be denoted in boldface ($\x$, $\k$, etc.) or with Latin subscripts ($x_i$, $k_i$, etc.).  The magnitude of vectors is defined as $k \equiv |\k|$ and unit vectors are written as $\hat \k \equiv \k/k$. We sometimes write the sum of $n$ vectors as $\k_{1\ldots n} \equiv \k_1 + \ldots + \k_n$.
We will often use the following shorthand for three-dimensional momentum integrals
$$
\int_\p\ (\ldots) \, \equiv\, \int \frac{\d^3 \p}{(2\pi)^3}\, (\ldots)\ .
$$
We denote the three-dimensional Laplacian by $\bigtriangleup \equiv \delta^{ij}\partial_i \partial_j$. 
A prime on correlation functions, $\vev{\cdots}'$, indicates that an overall momentum-conserving delta function is being dropped.  Our convention for the dimensionless power spectrum is 
$$
\Delta^2(k) \equiv \frac{k^3}{2\pi^2}P(k)\ .
$$
We also define a dimensionless bispectrum as
$$
\mathcal{B}(k_1, k_2, k_3)\equiv \left(\frac{k_1^3}{2\pi^2}\right)^2 B(k_1, k_2, k_3)\ .
$$
We will typically use $\p$ for short-scale fluctuations and reserve $\k$ for long-wavelength modes. We use $X_\ell$ and $X_s$ for the long-wavelength and short-wavelength parts of a quantity $X$.
The Gaussian and non-Gaussian parts of $X$ are $X^{\rm G}$ and $X^{\rm NG}$. Finally, $X^{[L]}$ denotes the spin-$L$ part of $X$.

\vskip 4pt
When we present numerical results, the linear power spectrum is computed with the Boltzmann code {\sf CAMB}~\cite{Lewis:1999bs}, using a flat $\Lambda$CDM cosmology with $\Omega_{m}^0 = 0.27$, $\Omega_{\Lambda}^0 = 0.73$ $h = 0.70$. The initial power spectrum of the gravitational potential is taken to be of a power law form with amplitude $\Delta_\varphi^2 = 8.7 \times 10^{-10}$ and spectral index $n_s = 0.96$, defined at the pivot scale $k_0=0.002 \hskip 2pt h\hskip 1pt {\rm Mpc}^{-1}$.

\section{Preliminaries}
\label{sec:prelim}

We start with a review of some basic background material.
In \S\ref{ssec:SPT}, we summarize the key elements of standard perturbation theory (SPT). This mainly serves to define our conventions and notation. For a much more detailed exposition, we refer the reader to the classic review~\cite{Bernardeau:2001qr}. Further details may also be found in Appendix~\ref{sec:SPT-Appendix}. In \S\ref{ssec:EFT-of-LSS}, we describe the effective theory of large-scale structure (EFT-of-LSS), and explain how it addresses some of the shortcomings of SPT. Finally, in~\S\ref{ssec:NG}, we define our way of parameterizing non-Gaussian initial conditions.

\subsection{Standard Perturbation Theory}
\label{ssec:SPT}

On large scales, dark matter behaves as a pressureless fluid, described by its density contrast $\delta \equiv \rho/\bar \rho-1$ and velocity $\v$. In the Newtonian approximation, the equations governing the evolution of the dark matter perturbations are the continuity equation and the Euler equation:
\begin{align}
\left(\partial_\tau+\v \cdot \D\right)\delta &= -(1+\delta)\D \cdot \v\ ,\\
\left(\partial_\tau+\v \cdot \D\right) \v &= -\H \v-\D\phi\ ,
\end{align}
where $\phi$ is the gravitational potential, which satisfies the Poisson equation
\beq
\bigtriangleup \hskip -2pt\phi = \frac{3}{2}\H^2\Omega_m\delta\ .  \label{equ:poisson}
\eeq
Notice that the matter density parameter $\Omega_m$ is time dependent.
We will sometimes find it convenient to work with the rescaled potential $\Phi \equiv 2 \phi/(3\H^2 \Omega_m)$, so that the Poisson equation reduces to $\bigtriangleup \Phi  = \delta$.
We will also assume that the velocity is irrotational (cf.~\cite{Mercolli:2013bsa}), so that it is fully determined by its divergence $\theta \equiv \D \cdot \v$.

\begin{figure}[h!]
        \centering
        \vskip 6pt
              \includegraphics[scale=1.0]{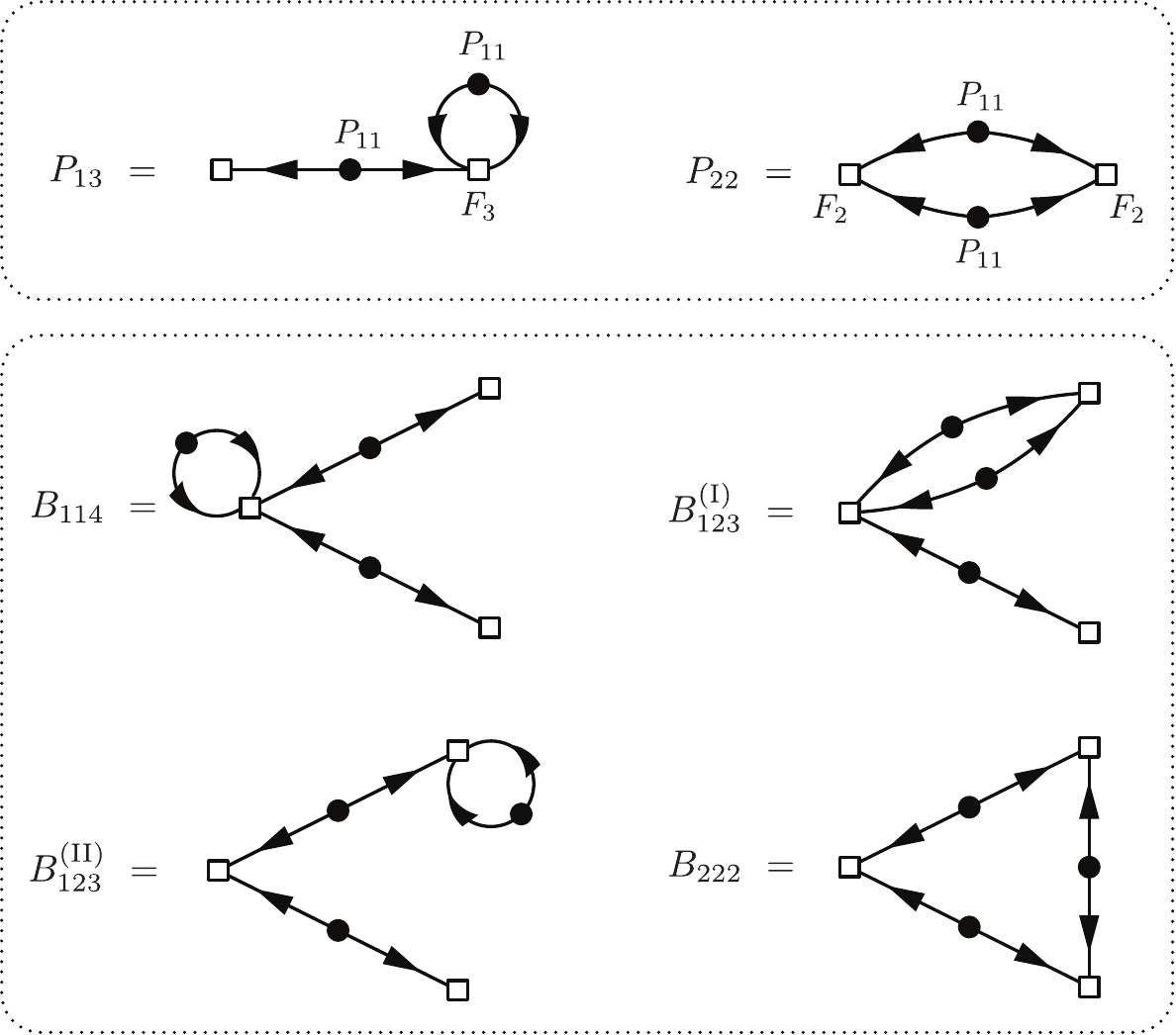}
  \caption{Diagrammatic representation of the one-loop power spectrum and the one-loop bispectrum for Gaussian initial conditions. 
  } \label{fig:LoopB}
 \end{figure}

For small $\delta$ and $\theta$, the equations of motion can be solved perturbatively as an expansion around the solution of the linearized equations, $\delta_{(1)}(\k,\tau) = D_1(\tau) \delta_1(\k)$, where $D_1(\tau)$ is the linear growth function, normalized to $D_1(\tau_{in})\equiv1$ at some initial time $\tau_{in}$ (see \S\ref{ssec:NG}).  The nonlinear solution can be written as a series in powers of $\delta_{1}$:
\beq
\delta(\k,\tau) = \sum_{n=1}^\infty \delta_{(n)}(\k,\tau) \ , \label{dSPT}
\eeq
and similarly for $\theta(\k,\tau)$. The $n$-th order solution can be written as $\delta_{(n)}(\k,\tau)\approx D_1^n(\tau) \delta_n(\k)$, where $\delta_{n}$ is a convolution of $n$ powers of $\delta_1$: 
\beq
\delta_n(\k) = \int_{\k_1}\cdots\int_{\k_n}(2\pi)^3\delta_D\big(\k-\k_{1\ldots n}\big)\,F_{n}(\k_1,\cdots,\k_n)\,\delta_1(\k_1)\cdots \delta_1(\k_n)\ ,\label{eq:deltaSPTn}
\eeq
where $\k_{1\ldots n} \equiv \k_1 + \ldots + \k_n$ and $F_n$ is a kernel function which can be computed iteratively~\cite{Bernardeau:2001qr}.  Diagrammatically, this kernel function can be represented by the following vertex
\vskip 2pt
\beq
\raisebox{-33 pt}{\includegraphics[scale=1.0]{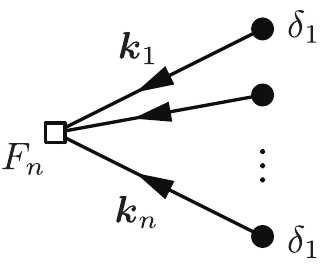}} \hskip 20pt  \equiv \hskip 10pt  F_n(\k_1,\cdots,\k_n)\,(2\pi)^3\delta_D\big(\k-\k_{1\ldots n}\big)\ . \label{equ:delta(n)}
\eeq
\vskip 8pt
\noindent
Correlations of $\delta$ can then be defined in a perturbative expansion. For example, for Gaussian initial conditions, the power spectrum can be written as
\beq
P_\delta(k) \equiv \vev{\delta(\k,\tau) \delta(-\k,\tau)}^\prime = \underbrace{P_{11}(k)}_{\text{tree}} \ +\ \underbrace{P_{13}(k) + P_{22}(k)}_{\text{one-loop}} \ + \ \cdots \ , \label{equ:PSPT}
\eeq
 where $P_{mn}(k) \equiv (\vev{\delta_{(m)}(\k,\tau) \delta_{(n)}(-\k,\tau)}^\prime+{\rm perms})$. Notice that in order to keep later expressions compact we have dropped the explicit time arguments in $P_\delta(k)$ and $P_{mn}(k)$.  Unless stated otherwise, density correlators should always be understood as being evaluated at time $\tau$.
We will also be interested in the late-time matter bispectrum which, for Gaussian initial conditions, has the following loop expansion  
 \beq
B_\delta(k_1,k_2,k_3) \equiv \vev{\delta(\k_1,\tau)\delta(\k_2,\tau)\delta(\k_3,\tau)}^\prime = \underbrace{B_{112}}_{\text{tree}} \ +\ \underbrace{B_{222} + B_{123}^{\rm (I)}+B_{123}^{\rm (II)}+B_{114}}_{\text{one-loop}} \ + \ \cdots \ , \label{equ:BSPT}
\eeq
where $B_{lmn}(k_1,k_2,k_3) \equiv (\vev{\delta_{(l)}(\k_1,\tau)\delta_{(m)}(\k_2,\tau) \delta_{(n)}(\k_3,\tau)}^\prime+{\rm perms})$. 
The diagrammatic representations of the one-loop contributions to the power spectrum and bispectrum are shown in fig.~\ref{fig:LoopB}. (See~\cite{Bernardeau:2001qr} for a more complete description of the Feynman rules of SPT.) 
 
Each loop involves an integral over an unfixed internal momentum. Conceptually, this is problematic since the integrals involve high-momentum (short-wavelength) modes for which $\delta > 1$. Perturbation theory isn't applicable for these modes, so there is no guarantee that the integrals will converge to the right answer.  Indeed, this becomes manifest for power law initial conditions in a pure matter universe. For certain scalings of the power law, perturbation theory would suggest that $P_{13}$ and $P_{22}$ are divergent.  The problem is less apparent in our universe, where the presence of radiation suppresses the contribution from the high-momentum modes. The integrals are then convergent, although a finite error is still being made. If the loops are only performed up to a finite cutoff $\Lambda$, then the answer depends on the arbitrary value of this cutoff scale.
 The EFT-of-LSS fixes these shortcomings of SPT.

\subsection{EFT of Large-Scale Structure}
\label{ssec:EFT-of-LSS}

On small scales, dark matter fluctuations are large and cannot be computed in perturbation theory. The EFT-of-LSS~\cite{Baumann:2010tm, Carrasco:2012cv} provides a framework in which these nonlinear modes are removed consistently, while keeping track of their effects on the large-scale dynamics through additional parameters in the fluid equations. These additional parameters cannot be computed within the EFT, but should be measured in observations or simulations. We briefly sketch the main elements of the EFT-of-LSS, but refer the reader to the literature~\cite{Baumann:2010tm, Carrasco:2012cv, Carrasco:2013sva, Pajer:2013jj, Mercolli:2013bsa} for a more detailed account.

\vskip 4pt
We wish to describe the evolution of fluctuations above a certain cutoff scale $\Lambda^{-1}$. Fluctuations below $\Lambda^{-1}$ are integrated out, but still affect the long-wavelength dynamics through corrections to the equations of motion.
For example, the Euler equation for the long-wavelength velocity receives a contribution from an effective stress tensors $\tau^{ij}$: 
\beq
\left(\partial_\tau +\H +v^j\partial_j \right) v^i  =-\partial^i\phi-\frac{1}{\rho}\partial_j\tau^{ij}\ . \label{equ:Euler}
\eeq
This becomes an effective theory when we expand the stress tensor in terms of the long-wavelength fields ($\phi$ and $v_i$) and their derivatives. This expansion is constrained by symmetries: in particular, the equivalence principle requires the gravitational potential to appear with at least two derivatives acting on it and the velocity with one derivative.  At first order in fluctuations, and to leading order in derivatives, we have~\cite{Baumann:2010tm, Carrasco:2012cv}
\beq
\partial_j\tau^{ij} = \bar\rho\left[c_s^2\,\partial^i\delta-\frac{c^2_{vis}}{\H}\,\partial^i\theta-\frac{\hat c^2_{vis}}{\H}\bigtriangleup\hskip -2pt v^i\right]\ ,
\label{equ:tau}
\eeq
where the tidal tensor $\partial^i \partial^j \phi$ is absent since it is degenerate with $\partial^i\delta$ when we take the divergence of $\tau^{ij}$.
The parameters appearing in the stress tensor ($c_s$, $c_{vis}$ and $\hat c_{vis}$) will depend on the cutoff~$\Lambda$. The part of the long-wavelength solution generated by these terms will therefore also be cutoff dependent.
This is a good thing. The new, cutoff-dependent contributions to the long-wavelength solution act as counterterms, in the sense that their cutoff dependence cancels against the one arising from loops of the long-wavelength fields. In particular, the terms in (\ref{equ:tau}) are precisely the ones required to cancel the cutoff-dependent part of the one-loop contribution $P_{13}(k)$ in (\ref{equ:PSPT}). (The cutoff dependence in $P_{22}$ is removed by a noise term; see below).  In fact, at every order in the loop expansion, the complete solution is cutoff independent if the stress tensor $\tau^{ij}$ is defined up to sufficiently high order in the derivative expansion.  This was demonstrated explicitly for the one-loop power spectrum in~\cite{Carrasco:2012cv, Pajer:2013jj}, for the two-loop power spectrum in~\cite{Carrasco:2013sva}  and for the one-loop bispectrum in~\cite{Baldauf:2014qfa, Angulo:2014tfa}.
So far, the discussion of renormalization in the EFT-of-LSS has been restricted to Gaussian initial conditions.  In this paper, we extend the renormalization of the EFT-of-LSS to allow for non-Gaussian initial conditions.

\subsection{Non-Gaussian Initial Conditions}
\label{ssec:NG}

We choose to define our ``initial" conditions at some time $\tau_{in}$ after matter-radiation equality, but early enough that nonlinearities in the prior evolution can still be ignored.  The linearly-evolved 
potential $\phi_{(1)}(\k,\tau)$ and the linearly-evolved dark matter density contrast $ \delta_{(1)}(\k,\tau)$ can then be written in terms of the primordial potential $\varphi(\k)$: 
\begin{align}
\phi_{(1)}(\k,\tau) &= T(k,\tau)\varphi(\k)\ , \label{equ:T}\\
 \delta_{(1)}(\k,\tau) &= - \frac{2}{3}\frac{k^2}{\H^2 \Omega_m}  \phi_{(1)}(\k,\tau) \equiv M(k,\tau) \varphi(\k)\ , \label{equ:M}
\end{align}
where the transfer function $T(k,\tau)$ captures the evolution of the gravitational potential in the radiation era.  To simplify the notation, we will typically drop the time arguments, but the functions $T(k)$ and $M(k)$ should always be evaluated at time $\tau$. 
We define the power spectrum of the primordial potential as
\beq
P_\varphi(k)  \equiv  \vev{\varphi(\k)\varphi(-\k)}' \equiv \frac{2\pi^2}{k^3}\Delta_\varphi^2 \left(\frac{k}{k_0}\right)^{n_s-1} \ , \label{eq:Pvar}
\eeq
where we have assumed a nearly scale-invariant power law ansatz motivated by inflation.
The
power spectrum of the linearly-evolved density contrast $\delta_{(1)}(\k,\tau)$ then is  
\beq
P_{11}(k) = [M(k)]^2P_\varphi(k)\ . \label{eq:Pin}
\eeq
Recall that both $M(k)$ and $P_{11}(k)$ are to be evaluated at time $\tau$, while $P_\varphi(k)$ is time independent.

 \vskip 4pt
Primordial non-Gaussianity (PNG) leads to higher-order correlations beyond the power spectrum.
The leading diagnostic for PNG is the bispectrum 
\begin{align}
B_\varphi(k_1,k_2,k_3) &\equiv \vev{\varphi(\k_1)\varphi(\k_2)\varphi(\k_3)}'\ , \label{eq:Bvar} \\[2pt]
B_{111}(k_1,k_2,k_3) &= [M(k_1)M(k_2)M(k_3)] B_\varphi(k_1,k_2,k_3)\ .\label{eq:Bin}
\end{align}
For perturbative PNG, the potential $\varphi$ can be expanded around a Gaussian field~$\varphi_g$. For instance, at lowest order, we may write~\cite{Schmidt:2010gw}
\beq
\varphi(\k)= \varphi_g(\k)+\fnl\int_\p{K}_{\mathsmaller{\rm NL}}(\p,\k-\p)\big[\varphi_g(\p)\varphi_g(\k-\p)-P_g(p)\,(2\pi)^3\delta_D(\k)\big] + \cdots\ ,
\label{eq:phiNG}
\eeq
where we have subtracted the power spectrum of the Gaussian field, $P_g(p) \equiv \vev{\varphi_g(\p)\varphi_g(-\p)}^\prime$, to ensure that $\vev{\varphi}=0$.
Eq.~(\ref{eq:phiNG}) is the most general quadratic expansion satisfying the requirements of statistical homogeneity and isotropy.
The momentum-dependent kernel function $K_{\mathsmaller{\rm NL}}(\k_1,\k_2)$ parametrizes the shape of the non-Gaussianity.
Substituting (\ref{eq:phiNG}) into (\ref{eq:Bvar}), and keeping only the leading term in an expansion in $\fnl\varphi_g$, we get 
\begin{align}
B_\varphi(k_1,k_2,k_3) &= 2 \fnl\hskip 1pt K_{\mathsmaller{\rm NL}}(\k_1,\k_2)\hskip 1pt P_\varphi(k_1) P_\varphi(k_2)+ \text{2 perms}\ .\label{equ:Bprimordial}
\end{align}
The permutations in (\ref{equ:Bprimordial}) imply that
the bispectrum alone does not uniquely determine the kernel function $K_{\mathsmaller{\rm NL}}(\k_1,\k_2)$ (i.e.~different kernel functions can give rise to the same bispectum, but different trispectrum).
However, the ambiguity in the choice of kernel disappears in the squeezed limit, $q \equiv k_1/k_2 \ll 1$, where we get 
\beq
K_{\mathsmaller{\rm NL}}(\k_1,\k_2) \ \xrightarrow{\, q \to 0\, } \ \frac{B_\varphi(k_1,k_2,k_3)}{4\fnl P_\varphi(k_1)P_\varphi(k_2)} \Big(1+ {\cal O}(q) \Big)\ .
\eeq
In this limit, the kernel and its leading scaling with $q$ are uniquely defined in terms of the bispectrum.
As we will see, the squeezed limit will be particularly relevant for our investigation.

\vskip 4pt
Statistical homogeneity requires that $K_{\mathsmaller{\rm NL}}(\k_1,\k_2)$ is only a function of $k_1$, $k_2$ and $\hat \k_1 \cdot \hat \k_2$.
It will be convenient to express the angular dependence as an expansion in terms of Legendre polynomials $P_L(\hat \k_1 \cdot \hat \k_2)$, and write the squeezed limit as 
\beq
\boxed{K_{\mathsmaller{\rm NL}}(\k_1,\k_2) \  \xrightarrow{\, q \to 0\, } \   \, \sum_{L,i} a_{L,i} \left(\frac{k_1}{k_2}\right)^{\Delta_i} P_L(\hat \k_1 \cdot \hat \k_2)}\ , \label{eq:angSL}
\eeq
where, by symmetry, $L$ has to be an even integer (see Appendix~\ref{sec:oddspin} for more details). 
The ansatz (\ref{eq:angSL}) captures many physically relevant cases. (Non-Gaussianity in feature models~\cite{Chluba:2015bqa} may fall outside of this paramererization.)  We can organize the different contributions by the order $L$ (the ``spin") of the Legendre polynomial and treat each scaling $\Delta_i$ separately: 

\begin{itemize}
\item {\it Scalar contributions}

First, we consider the scalar contributions; i.e.~we set $a_{L \ge 2}\equiv 0$.  The different squeezed limits are then distinguished by their scaling dimensions $\Delta$:
\begin{itemize}\renewcommand{\labelitemii}{$\circ$}
\item For $\Delta = 0$, the kernel is a momentum-independent constant, $K_{\mathsmaller{\rm NL}}(\k_1,\k_2) = 1$, and the ansatz (\ref{eq:phiNG}) corresponds to a {\it local} expression for $\varphi$ in real space~\cite{Komatsu:2001rj}, i.e.~$\varphi(\x) = \varphi_g(\x) + f_{\mathsmaller{\rm NL}}(\varphi_g^2(\x) - \vev{\varphi_g^2})$. This is the case of {\it local non-Gaussianity}.

\item For $\Delta = 2$, the squeezed limit is suppressed by two additional powers of the low-momentum mode $k_1$.
This is characteristic of higher-derivative interactions in single-field inflation, which produce {\it equilateral non-Gaussianity}~\cite{Alishahiha:2004eh, Cheung:2007st}.

\item A squeezed limit with intermediate momentum scaling, $\Delta \in [0,1.5]$, can arise if the inflaton interacts with massive scalar particles during inflation. In these models of {\it quasi-single-field inflation}~\cite{Chen:2009zp},
nonlinear interactions of the additional scalars can be mediated to the inflaton sector, creating observable non-Gaussianity with a  characteristic signature in the squeezed limit: the scaling dimensions $\Delta$ are functions of the masses of the extra particles.
Coupling the inflaton to operators in a conformal field theory~\cite{Green:2013rd} allows to extend the intermediate momentum scaling to the regime~$\Delta \in [0,2]$.
\end{itemize}

\item {\it Higher-spin contributions}

Various physical mechanisms can lead to an angular dependence in the squeezed limit~\cite{Shiraishi:2013vja}:

\begin{itemize}\renewcommand{\labelitemii}{$\circ$}
\item The inflaton may couple to massive higher-spin particles.  In this case, the angular dependence is given by the Legendre polynomial of order the spin of the particle~\cite{Arkani-Hamed:2015bza}.
At tree level, only particles with even spin contribute to the bispectrum in the squeezed limit.

\item At loop level, the angular dependence induced by the interaction with higher-spin particles can be different from just a simple Legendre polynomial. For example, coupling the inflaton to a $U(1)$ gauge field via the interaction $I(\phi)F^2$~\cite{Barnaby:2012tk} leads to a shape of the form (\ref{eq:angSL}) with $a_2 =  a_0/2$.

\item Curvature perturbations sourced by large-scale primordial magnetic fields~\cite{Shiraishi:2012sn, Shiraishi:2012rm} also can lead to non-zero $a_0$, $a_1$ and $a_2$.

\item The bispectrum produced in solid inflation~\cite{Endlich:2012pz} corresponds to   $a_2 \gg a_0$.

\end{itemize}

\end{itemize}

\section{Coarse Graining}
\label{sec:coarse}

In the EFT-of-LSS the short-scale fluctuations have to be integrated out both in the equations of motion and in the initial conditions. The presence of non-Gaussianity in the initial conditions yields some non-trivial features which have to be taken into account in a consistent renormalization of the loop expansion.
In contrast to a usual fluid (such as air or water), for dark matter there isn't a hierarchy of time scales between the evolution of short-wavelength and long-wavelength fluctuations. This implies that the short scales keep memory of their initial conditions. For Gaussian initial conditions, this isn't very important since all scales are statistically independent. As a result, there are no initial correlations between the long-wavelength fluctuations of the EFT and the short scales which are being integrated out. However, in the presence of primordial non-Gaussianity, the initial statistics of the short scales depends on the long-wavelength fluctuations. This initial dependence will affect the dynamics of the short modes, which in turn will then backreact on the evolution of the long-wavelength fluctuations.  The objective of this section is to understand how this memory effect can be incorporated in the EFT-of-LSS.

\vskip 4pt
In \S\ref{sec:SEE}, we coarse grain the dark matter equations of motion to derive the stress tensor of the effective theory.  As we explained in \S\ref{ssec:EFT-of-LSS}, this stress tensor can be expanded in terms of the long-wavelength fields and their derivatives. In the presence of PNG, the coefficients of this expansion will be spatially modulated. In \S\ref{ssec:IC}, we derive this dependence by coarse-graining the initial condition (\ref{eq:phiNG}). This introduces new, non-dynamical fields in the EFT. We collect the leading-order non-Gaussian corrections in \S\ref{sec:tau}.

\subsection{Smoothing the Equations of Motion}
\label{sec:SEE}

We will follow the notation of \cite{Baumann:2010tm}, and write the long-wavelength fluctuations of any field~$X$ as
\beq
X_{\ell}(\x)\equiv [X]_\Lambda(\x)\equiv\int\d^3\x'\ W_\Lambda(|\x-\x'|)\,X(\x')\ , \label{equ:W}
\eeq
where $W_\Lambda(x)$ is a window function which quickly vanishes for $x>\Lambda^{-1}$.
In momentum space, this convolution becomes a simple product
\beq
X_{\ell}(\k) = F_\Lambda(k) X(\k)\ ,
\eeq
where $F_\Lambda(k)$ is the Fourier transform of $W_\Lambda(x)$. Naturally, the short-scale fluctuations of $X$ are defined as the complement $X_s\equiv X-X_\ell$.
\vskip 4pt
While the continuity equation remains unchanged,\footnote{This is only true if we smooth the momentum density field $\boldsymbol{\pi}\equiv\rho\v$ instead of the velocity field $\v$~\cite{Mercolli:2013bsa}.}
the smoothing of the Euler equation~(or, more generally, the Vlasov equation) generates terms which contribute to an effective stress tensor~\cite{Carrasco:2012cv,Baumann:2010tm,Pietroni:2011iz}
\beq
\tau^{ij} \,=\, \frac{1}{8\pi G a^2}\big[2\partial^i\phi_{s}\partial^{j}\phi_{s}-\delta^{ij}(\partial_k\phi_s)^2\big]_\Lambda \,+\, \big[\rho\hskip 1pt v^i_sv^j_s\big]_\Lambda \,+\, \big[\rho\hskip 1pt\sigma_s^{ij}\big]_\Lambda \ ,\label{eq:taueffX}
\eeq
where $\sigma_s^{ij}$ is the velocity dispersion of the short scales.  
To determine the effect of this stress tensor on the long-wavelength universe,
we need to understand how $\tau^{ij}$ correlates with long-wavelength modes.
When the short modes are perturbative, this backreaction on the long modes can be computed explicitly in perturbation theory (see Appendix~\ref{sec:topdown}). However,  in general, the short scales are not perturbative and therefore cannot be computed analytically. When this is the case, the stress tensor should be written as the most general function of the long-wavelength fluctuations consistent with the symmetries of the problem. To do so, let us first notice that the dynamics of the short-scale fluctuations depends both on their initial conditions and on the 
long-wavelength background in which they evolve.  Assuming that the short scales depend locally on the long-wavelength fluctuations, the most general effective stress tensor consistent with the equivalence principle then is
\beq
\tau^{ij}(\x,\tau) = {\cal F}[\,\phi_s(\q,\tau_{in})\,;\,\partial^i\partial^j\phi_\ell(\x_{fl}(\tau'),\tau')\,,\,\partial^iv^j_\ell(\x_{fl}(\tau'),\tau')\,,\,\cdots]\ ,\label{eq:stresstensor}
\eeq
where $\x_{fl}(\tau')$ is the position of the fluid element which at time $\tau > \tau'$ 
 is located at $\x$. It will be important that the initial short-scale fluctuations are evaluated at the Lagrangian position~$\q\equiv\x_{fl}(\tau_{in})$. The ellipses refer to higher derivatives in the long-wavelength fluctuations. Since there is no hierarchy in the time scales of the evolution of the short and long modes, the stress tensor depends on the evolution (along the fluid trajectory) of the long-wavelength fluctuations.  
However, as explained in~\cite{Baldauf:2014qfa}, when perturbation theory is valid, this expansion can be reorganised in an expansion which is local in time, but in general non-local in space (unless convective time derivatives are added to the expansion, see \cite{Mirbabayi:2014zca}). Working at second order in perturbation theory, one can show that $\delta(\x_{fl}(\tau'),\tau')$ is locally related to the late-time overdensity $\delta(\x,\tau)$ and the tidal tensor $\partial_i\partial_j\phi(\x,\tau)$.\footnote{ Note that this relation relies only on the equations of motion and therefore applies regardless of whether the initial conditions are Gaussian or not.} Hence, in this work, we may assume that the dependence of the stress tensor on the long-wavelength fluctuations is effectively local in time.

\vskip 4pt
At first order in the long-wavelength fluctuations and to leading order in derivatives, we can write \eqref{eq:stresstensor} as
\begin{align}
\tau^{ij}(\x,\tau)\ &=\ c^{ij}(\x,\tau) 
\ +\  c^{ij}{}_{kl}(\x,\tau)\,\partial^k\partial^l\phi_\ell(\x,\tau)\ +\ \hat  c^{ij}{}_{kl}(\x,\tau)\,\partial^kv^l_\ell (\x,\tau)\ +\ \cdots\ .\label{eq:effstresstensor}
\end{align}
We will sometimes suppress the indices on the coefficients in (\ref{eq:effstresstensor}) and collectively refer to them as $c(\x,\tau)$.
Notice that the coefficients $c(\x,\tau)$
 depend on position through their dependence on the initial short-scale fluctuations\hskip 1pt\footnote{In~(\ref{eq:coeff1}), we have assumed that the initial short-scale fluctuations are fully determined by the short-scale potential $\phi_s(\q,\tau_{in})$. We could also have added a dependence on the short-scale velocity $v^i_s(\q,\tau_{in})$ and its dispersion.
 However, at early times, every scale is perturbative so that one can use the linear equations of motion to express the velocity in terms of $\phi_s$. In particular, assuming that these scales are in their growing mode, one can show that the velocity is determined by the gradient of the potential, $v^i_s \propto \partial^i\phi_s$.}
\beq
c(\x,\tau) \equiv c[\phi_s(\q,\tau_{in}),\tau]\ .\label{eq:coeff1}
\eeq
Since the short scales have been removed by coarse graining, the relation in~(\ref{eq:coeff1}) is non-local. More precisely, the coefficients $c(\x,\tau)$ will depend on the value of $\phi_s$ over a patch of size $\Lambda^{-1}$ centred around the Lagrangian coordinate $\q(\x,\tau)$. We will come back to this point in the next section.
\vskip 4pt
When the small-scale fluctuations are replaced by their statistical ensemble averages, the stress tensor becomes
\beq
\tau^{ij} =  \big\langle \tau^{ij} \big\rangle_s + \Delta\tau^{ij}\ ,
\label{eq:tau2}
\eeq
where $\vev{\cdots}_s$ denotes an average over many realizations of the short modes in a fixed long-wavelength background and $\Delta \tau^{ij}$ is a stochastic term which accounts for the statistical deviation from the average. We will sometimes refer to $\vev{\tau^{ij}}_s$ as the ``viscosity'' part of the stress tensor and to $\Delta\tau^{ij}$ as the ``noise'' part.
We get the viscosity contribution to the stress tensor by replacing the coefficients $c(\x,\tau)$ in~(\ref{eq:effstresstensor}) by their statistical averages $\vev{c(\x,\tau)}_s$.
For Gaussian initial conditions, the average over short-scale fluctuations is independent of the long-wavelength fluctuations. As a result, the averaged coefficients $\vev{c(\x,\tau)}_s$ become simple cutoff-dependent parameters of the EFT. However, primordial non-Gaussianity gives rise to correlations between different scales and the averaged coefficients will depend (non-locally) on the initial long-wavelength fluctuations. This dependence can be determined by integrating out the short-scale fluctuations in the initial conditions, which we shall do next.

\subsection{Smoothing the Initial Conditions}
\label{ssec:IC}

We apply the filtering procedure \eqref{equ:W} to the primordial potential (\ref{eq:phiNG}). Isolating the terms which contribute to the coupling between the long and short modes, we get
\begin{align}
\varphi_\ell(\k) &\,\supset\, \fnl\int_{\tilde\p} K_{\mathsmaller{\rm NL}}(\k-\tilde\p,\tilde\p)\big[\varphi_g^{s}(\k-\tilde\p)\varphi_g^{s}(\tilde\p)-P_g(\tilde p)\,(2\pi)^3\delta_D(\k)\big] \ \equiv \ \fnl\psi_J(\k)\ , \label{eq:philong}\\
\varphi_{s}(\p) &\, \supset\, 2\fnl\int_{\tilde \p} K_{\mathsmaller{\rm NL}}(\p-\tilde \p, \tilde\p)\, \varphi_g^{s}(\p- \tilde \p)\,\varphi_g^{\ell}(\tilde \p)\ .\label{eq:phishort}
\end{align}
The field $\psi_J$ in~(\ref{eq:philong}) does not correlate with the long-wavelength fluctuations,  but it has non-trivial correlations with the noise terms of the stress tensor, $\Delta \tau^{ij}$. We will explain this in more detail in the next section. The right-hand side of~(\ref{eq:phishort}) encodes the dependence of the short modes on the long-wavelength fluctuations. Writing $\phi_g^s(\p,\tau_{in}) \equiv T(p,\tau_{in})\varphi_g^s(\p)$ and going to position space, we find 
\beq
\phi_s(\x,\tau_{in}) \,\simeq\, \phi_g^{s}(\x,\tau_{in})+ 2\fnl\int_\k\int_\p K_{\mathsmaller{\rm NL}}(\k,\p)\,\phi_g^{s}(\p,\tau_{in})\hskip 1pt\varphi_g^{\ell}(\k)\,e^{i(\p+\k)\cdot\x}\ .\label{eq:1ptshort}
\eeq
Since $k \ll p$, the result in (\ref{eq:1ptshort}) only depends on the squeezed limit of the kernel function $K_{\mathsmaller{\rm NL}}$ (and the corresponding bispectrum).
In \S\ref{ssec:NG}, we introduced the following ansatz 
\beq
K_{\mathsmaller{\rm NL}}(\k,\p)\ \xrightarrow{\ k\ll p\ }\   \sum_{L,i} a_{L,i} \left(\frac{k}{p}\right)^{\Delta_i} P_L(\hat\k\cdot\hat\p)\Big[1+{\cal O}\left({k^2/p^2}\right)\Big]\ .\label{eq:sep}
\eeq
For each scaling $\Delta$ in this sum, we can treat the different orders (``spins") in the Legendre expansion separately:

\begin{itemize}
\item {\it Spin-0}

For $L=0$, the initial short-scale fluctuations can be written as
\beq
\phi_s(\x,\tau_{in}) = \phi_g^{s}(\x,\tau_{in}) + \fnl\hskip 1pt\alpha^{(s)}(\x)\,\psi(\x)\ ,\label{eq:shortin0}
\eeq
where 
\begin{align}
\boxed{\psi(\k)\equiv   \left(\frac{k}{\mu}\right)^{\Delta} \varphi_g^{\ell}(\k)} \ , \label{equ:psi}
\end{align}
and $ \alpha^{(s)}(\p)\equiv 2a_0\left({\mu/p}\right)^{\Delta} \phi_g^{s}(\p,\tau_{in})$. The scale $\mu$ is an arbitrary momentum scale introduced to make $\psi(\x)$ dimensionless.  
As we shall see, the field $\psi$ will play an important role in the rest of this paper. It describes how the statistics of the small scales is modulated by the presence of the long modes.
 For local non-Gaussianity, we have $\Delta=0$ and therefore $\psi(\x) = \varphi_g(\x)$. 
 Eq.~(\ref{eq:shortin0}) then reduces to a  more familiar expression  
 \beq
\phi_{s}(\x,\tau_{in}) =\Big(1+2\fnl^{\rm local}\varphi_g^{\ell}(\x)\Big) \, \phi_g^s(\x,\tau_{in})\ .
\eeq
We see that local non-Gaussianity simply modulates the amplitude of the small-scale fluctuations.

\vskip 4pt
To compute the dependence of the coefficients $\vev{c(\x,\tau)}_s$ on $\psi$, we substitute~(\ref{eq:shortin0}) into~(\ref{eq:coeff1}).
At first order in an expansion in $\psi$, we find
\beq
\vev{c(\x,\tau)}_s \,=\, \vev{c[\phi_g^{s}(\q,\tau_{in}),\tau]}_s + \fnl\int_{\tilde \q} \left\langle\frac{\delta c(\x,\tau)}{\delta \phi_s(\tilde\q,\tau_{in})}\,\alpha^{(s)}(\tilde{\q},\tau_{in})\right\rangle_s\psi(\tilde\q)
\ ,\label{eq:coeff}
\eeq
where the derivative is evaluated for the Gaussian field configuration $\phi_g^s$. Furthermore, since the integral only has support for $|\tilde\q-\q(\x,\tau)|<\Lambda^{-1}$, we can pull the long-wavelength field $\psi(\tilde\q) \approx \psi(\q)$ out of the integral (up to corrections that are higher order in $\nabla^2/\Lambda^2$).  Translation invariance then guarantees that the remaining integral is independent of the position~$\x$, and eq.~(\ref{eq:coeff}) can be written as
\begin{align}
\vev{c(\x,\tau)}_s
&\,\equiv\, \mathfrak{c}(\Lambda,\tau) +\fnl\hskip 1pt\mathfrak{c}_{\psi}(\Lambda, \tau)\hskip 1pt \Psi(\x,\tau)\ ,\label{eq:coeff2}
\end{align}
where we have defined
\beq
\Psi(\x,\tau) \equiv\psi(\q(\x,\tau)) \ . \label{PSI}
\eeq
In Section~\ref{sec:renorm}, we will see why it is important that field $\psi$ is evaluated at the Lagrangian position $\q$. 
Restoring the indices, but dropping the arguments of the coefficients and the fields,
we get
\begin{align}
\vev{c^{ij}}_s^{[0]} &\,=\, \big[ \mathfrak{c}^{[0]}_{\phantom\psi} + \fnl\hskip 1pt\mathfrak{c}_{\psi}^{[0]} \Psi \big]\delta^{ij}\ ,\label{eq:coeff4}\\[10pt]
\vev{c^{ijkl}}_s^{[0]} &\,=\,  \big[\mathfrak{c}_0^{[0]}+\fnl\hskip 1pt\mathfrak{c}_{\psi,0}^{[0]} \Psi \big]\delta^{ij}\delta^{kl}  + \big[{\mathfrak{c}}_1^{[0]} + \fnl\hskip 1pt{\mathfrak{c}}_{\psi,1}^{[0]} \Psi\big]\big[\delta^{ik}\delta^{jl}+\delta^{il}\delta^{jk} \big]\ , \label{eq:coeff5}
\end{align}
and similarly for the coefficient 
$\vev{\hat c^{ijkl}}_s^{[0]}$. The tensor structure of (\ref{eq:coeff4}) and (\ref{eq:coeff5}) is fixed by statistical isotropy.

\item{\it Spin-2}

For $L=2$, equation~(\ref{eq:1ptshort}) becomes
\beq
\phi_s(\x,\tau_{in}) = \phi_g^s(\x,\tau_{in}) + \fnl\hskip 1pt\alpha^{(s)}_{ij}(\x)\,\psi^{ij}(\x)\ ,\label{eq:shortin2}
\eeq
where 
\begin{align}
\boxed{\psi^{ij}(\k)\equiv \frac{3}{2}{\cal P}^{ij}(\hat \k)\left(\frac{k}{\mu}\right)^{\Delta} \varphi_g^{\ell}(\k)}\ , \label{psiij}
\end{align}
and  $\alpha^{(s)}_{ij}(\p)\equiv a_2\left({\mu/p}\right)^{\Delta}(\hat p_i\hat p_j)\phi_g^{s}(\p,\tau_{in})$. In (\ref{psiij}), we have defined the projection operator ${\cal P}^{ij}(\hat \k)\equiv \hat k^i\hat k^j- \frac{1}{3}\delta^{ij}$.  Note that the tensor $\psi^{ij}$ is traceless and hence contains no scalar contribution. 
Substituting~(\ref{eq:shortin2}) into~(\ref{eq:coeff1}), we find
\begin{align}
\vev{c^{ij}}_s^{[2]} &\,=\,   \mathfrak{c}^{[2]}_{\phantom\psi}\hskip 1pt\delta^{ij} +   \fnl\hskip 1pt\mathfrak{c}_{\psi}^{[2]} \hskip 1pt \Psi^{ij}\ , \label{cij} \\[10pt]
\vev{c^{ijkl}}_s^{[2]} &\,=\, \mathfrak{c}_0^{[2]}\hskip 1pt\delta^{ij}\delta^{kl}+\mathfrak{c}_1^{[2]}\hskip 1pt \big[\delta^{ik}\delta^{jl} + \delta^{il}\delta^{jk} \big]  \label{cijkl} \\[4pt]
&\hspace{0.5cm} +\, \fnl\hskip 1pt\left[\mathfrak{c}_{\psi,0}^{[2]}\Psi^{ij}\delta^{kl} \,+\,\mathfrak{c}_{\psi,1}^{[2]}\Psi^{kl}\delta^{ij} \,+\, \mathfrak{c}_{\psi,2}^{[2]}\big[\Psi^{ik}\delta^{jl}+\Psi^{jk}\delta^{il} + \Psi^{il}\delta^{jk}+\Psi^{jl} \delta^{ik}\big]\right]\ , \nonumber
\end{align}
where we have defined $\Psi^{ij}(\x,\tau) \equiv \psi^{ij}(\q(\x,\tau))$.
Although we have dropped the arguments, the coefficients in (\ref{cij}) and (\ref{cijkl}) of course still depend on the cutoff $\Lambda$ and time $\tau$.

\item{\it Spin-4}

For $L=4$, equation~(\ref{eq:1ptshort}) becomes
\beq
\phi_s(\x,\tau_{in}) = \phi_g^s(\x,\tau_{in}) + \fnl\hskip 1pt\alpha^{(s)}_{ijkl}(\x)\psi^{ijkl}(\x)\ ,\label{eq:shortin4}
\eeq
where 
\begin{align}
\boxed{\psi^{ijkl}(\k)\equiv \frac{35}{8} {\cal P}^{ijkl}(\hat \k)\left(\frac{k}{\mu}\right)^{\Delta} \varphi_g^{\ell}(\k)} \ , \label{eq:psiijkl}
\end{align}
and $\alpha^{(s)}_{ijkl}(\p)\equiv a_4\left({\mu/p}\right)^{\Delta}(\hat p_i\hat p_j\hat p_k\hat p_l)\phi_g^{s}(\p,\tau_{in})$. The projection tensor ${\cal P}^{ijkl}(\hat\k)$ ensures that $\psi^{ijkl}$ is symmetric and traceless in any two of its indices 
\begin{align}
{\cal P}^{ijkl}(\hat \k)&\equiv \hat k^i\hat k^j\hat k^k\hat k^l-\frac{1}{7}\left(\delta^{ij}\hat k^k\hat k^l+{ 5\ \rm perms}\right) +  \frac{1}{35}\left(\delta^{ij}\delta^{kl}+{\rm 2\ perms}\right)\ .\label{eq:spin4}
\end{align}
Let us note that, at leading order in derivatives, the coefficients with only two indices, $c^{ij}$, cannot depend on $\psi^{ijkl}$ since we would need to contract two of the indices. As a result, the field~$\psi^{ijkl}$ only contributes to the coefficients with four indices
\beq
\vev{c^{ijkl}}_s^{[4]} =  \mathfrak{c}_0^{[4]}\delta^{ij}+{\mathfrak{c}}_1^{[4]}[\delta^{ik}\delta^{jl} + \delta^{il}\delta^{jk}] + \fnl\hskip 1pt\mathfrak{c}_{\psi}^{[4]}\Psi^{ijkl}\ ,
\eeq
where, as before, we have defined $\Psi^{ijkl}(\x,\tau)\equiv\psi^{ijkl}(\q(\x,\tau))$.

\item {\it Higher-spin}

It should be clear that higher-spin contributions ($L>4$) will be parametrized by higher-order symmetric traceless tensors. However, at the order we are working at, these higher-order tensors will need to be contracted with one or several Kronecker deltas. 
Hence, they do not contribute to the stress tensor and will therefore not be considered in this work.

\end{itemize}

\subsection{Effective Stress Tensor}
\label{sec:tau}

Putting the results of the previous two sections together, the stress tensor becomes a functional of the long-wavelength fields\footnote{To avoid clutter in the expressions, we will drop the subscripts on the long-wavelength fields from now on: i.e.~we set $\{ v^i_\ell,\Phi_\ell\} \to \{v^i,\Phi\}$.} $\{v^i,\Phi\equiv 2\phi/(3\H^2\Omega_m)\}$, the non-dynamical fields $\{\Psi,\Psi^{ij},\Psi^{ijkl}\}$, and their derivatives. 
In this section, we present the most general expression for the viscosity and noise parts of the stress tensor.

\subsubsection*{Viscosity part}

We write the viscosity part of the stress tensor as
 \beq
\big\langle \tau^{ij} \big\rangle_s = {\cal F}[\partial^i\partial^j\Phi, \partial^iv^j, \Psi,\Psi^{ij},\Psi^{ijkl},\cdots]\ ,\label{eq:taueff}
\eeq
where the ellipses refer to terms with higher derivatives. 
Since the long-wavelength fluctuations are perturbative, we can expand the functional~${\cal F}$ in~(\ref{eq:taueff}) in powers of the fluctuations and their derivatives. We will focus on terms which involve single powers of the new long-wavelength fields~$\{\Psi,\Psi^{ij},\Psi^{ijkl}\}$, since the Gaussian terms have been discussed in~\cite{Carrasco:2012cv, Carrasco:2013mua, Baldauf:2014qfa, Angulo:2014tfa}. The stress tensor can then be written as the sum of three terms organized by spin:
\beq
\big\langle \tau^{ij} \big\rangle_s^{\rm NG}=\fnl\left[\big\langle \tau^{ij} \big\rangle_s^{[0]}+\big\langle \tau^{ij} \big\rangle_s^{[2]}+\big\langle \tau^{ij} \big\rangle_s^{[4]}\right]\ .\label{eq:viscspin}
\eeq
We look at each of these contributions in turn.

\begin{itemize}
\item {\it Scalar contributions}

The leading spin-0 contributions to the stress tensor are 
\begin{align}
\frac{1}{\bar\rho} \big\langle \tau^{ij} \big\rangle_s^{ [0]} &\,=\, \Big[g\hskip 1pt\Psi +c_1\hskip 1pt\Psi\delta +\frac{c_2}{\H} \hskip 1pt \Psi \theta \Big]\delta^{ij} + c_3\hskip 1pt\Psi \partial^{i}\partial^j\Phi + \frac{c_4}{2\H} \hskip 1pt\Psi \big(\partial^{i}v^{j}+\partial^jv^i\big) \ ,\label{eq:tauNG}
\end{align}
where the coefficients $g(\tau)$ and $c_i(\tau)$ are dimensionless and time dependent. At second order in perturbation theory, the expression in~(\ref{eq:tauNG}) simplifies, since some terms are related by the equations of motion. In particular, using $v^i_{(1)} = -[\H f]\partial^i\Phi_{(1)}$, with $f \equiv d \ln D_1/d \ln a$, we get
\beq
\boxed{\frac{1}{\bar\rho} \big\langle \tau^{ij} \big\rangle_s^{ [0]}\,=\, \big[g\Psi + g_1\Psi \delta \big]\delta^{ij} + g_2\Psi s^{ij}}\ ,\label{eq:tauNG0V2}
\eeq
where we have defined $g_1\equiv c_1+\frac{1}{3}c_3-f\,c_2$, $ g_2\equiv c_3-f\,c_4$ and introduced the tidal tensor $s_{ij}\equiv\partial_i\partial_j\Phi-\tfrac{1}{3}\delta_{ij}\hskip -1pt\bigtriangleup \hskip -1pt\Phi$.

\item {\it Higher-spin contributions}

 Similarly, the leading spin-2 and spin-4 contributions to the stress tensor are
 \begin{align}
\frac{1}{\bar\rho} \big\langle \tau^{ij} \big\rangle_s^{ [2]} &\,=\, \tilde g\hskip 1pt\Psi^{ij} +\tilde g_1\hskip 1pt\Psi^{ij}\,\delta +\tilde g_2(\Psi^{ik} s_k{}^j+\Psi^{jk} s_k{}^i)+\tilde g_3\hskip 1pt \Psi^{kl}s_{kl}\hskip 1pt\delta^{ij} \ ,\label{eq:tauNG2}\\
\frac{1}{\bar\rho} \big\langle \tau^{ij} \big\rangle_s^{ [4]} &\,=\, \hat g\hskip 1pt\Psi^{ijkl} s_{kl}\ .\label{eq:tauNG4}
\end{align}

\end{itemize}
 We see that a bispectrum with a squeezed limit of the form~(\ref{eq:sep}), generally yields eight additional parameters in the stress tensor (for each value of $\Delta_i$).  When the squeezed limit is isotropic,  the number of additional parameters reduces to three.

\subsubsection*{Noise part}

The noise term in~(\ref{eq:tau2}) arises because, for any specific realization, the short-scale modes fluctuate away from their ensemble averages. This stochastic contribution to the effective stress tensor can also be expanded in powers of the long-wavelength fields $\{\Phi, \Psi,\Psi^{ij},\Psi^{ijkl} \} $. Just like the viscosity term in~(\ref{eq:viscspin}), the non-Gaussian contributions to the noise term can be organized in terms of spin:
\beq
(\Delta\tau^{ij})^{\rm NG} = \fnl\left[(\Delta\tau^{ij})^{[0]}+(\Delta\tau^{ij})^{[2]}+(\Delta\tau^{ij})^{[4]}\right]\ .
\eeq
We look at each of these contributions in turn.

\begin{itemize}
\item {\it Scalar contributions}

The scalar contribution is given by
\beq
\boxed{\frac{1}{\bar\rho}(\Delta\tau^{ij})^{[0]} \,=\,  J^{ij}_\psi\,\Psi } \ ,\label{eq:noise0}
\eeq
where $J^{ij}_\psi$ is a random variable representing  stochastic noise, i.e.~contributions which are uncorrelated with the long-wavelength fluctuations. This noise term also correlates with the leading-order Gaussian noise term, $(\Delta\tau^{ij})^{\rm G}\supset J^{ij}_0$. As we shall see in Section~\ref{sec:renorm}, the correlation of these two noise terms will be important in the renormalization of non-Gaussian loop diagrams. Furthermore, noise terms are uncorrelated on large scales, i.e.~their correlation  functions in position space are proportional to delta functions.  
This means that, in Fourier space, the correlator of $J^{ij}_\psi$ with $J^{ij}_0$ is
\beq
\vev{J^{ij}_0(\k,\tau)J^{kl}_\psi(\k',\tau)} = {\cal J}^{ijkl}(\k,\tau)(2\pi)^3\delta_D(\k+\k')\ ,
\eeq
where  ${\cal J}^{ijkl}(\k,\tau)$ is an analytic function, which around $\k=\0$ can be expanded as
\beq
{\cal J}^{ijkl}(\k,\tau) = {\cal J}(\Lambda,\tau)\delta^{ij}\delta^{kl}+\hat{\cal J}(\Lambda,\tau)\left[\delta^{ik}\delta^{jl}+\delta^{il}\delta^{jk}\right]+ {\cal O}(k^2)\ .
\eeq
 Furthermore, as advertised earlier, the noise terms of the stress tensor, $J_{0,\psi}^{ij}$, also correlate with the noise term in the initial conditions, $\psi_{J}$: 
 \begin{align}
\vev{\psi_{J}(\k) J_a^{ij}(\k',\tau)}' &= \int_{\p}K_{\mathsmaller{\rm NL}}(\k-\p,\p)\vev{\varphi_g^{s}(\k-\p)\varphi_g^{s}(\p) J_a^{ij}(\k',\tau)}'\nonumber\\[4pt]
&=\tilde{\cal J}_a(\Lambda,\tau)\delta^{ij} + {\cal O}(k^2)\ ,\label{eq:tildeJ}
\end{align}
where $a\in\{0,\psi\}$ and  we have assumed that the integral is analytic in $\k$. 

\item {\it Higher-spin contributions}

The leading spin-2 and spin-4 contributions to the noise part of the stress tensor are
\begin{align}
\frac{1}{\bar\rho}(\Delta\tau^{ij})^{[2]}&= J^{ijkl}_\psi\Psi_{kl} \ ,\\
\frac{1}{\bar\rho}(\Delta\tau^{ij})^{[4]}&= J^{ijklmn}_\psi\Psi_{klmn}\ ,
\end{align}
where the tensors $J^{ijkl}_\psi$ and $J^{ijklmn}_\psi$ are uncorrelated with long-wavelength fluctuations, but can be correlated with $\psi_{J}$ and $J^{ij}_0$. 
\end{itemize}
Both the viscosity and noise parts of the stress tensor are crucial in a consistent renormalization of matter correlation functions. As we will see in the next section, they generate new solutions which have precisely the correct momentum dependence to absorb the divergences arising from the loop diagrams of standard perturbation theory.

\section{Renormalization}
\label{sec:renorm}

In this section, we describe the one-loop renormalization of the power spectrum and the bispectrum in the presence of primordial non-Gaussianity.
We will focus on loop diagrams arising from non-Gaussian initial conditions, since loops coming from Gaussian initial conditions have already been studied in~\cite{Carrasco:2013sva,Baldauf:2014qfa, Angulo:2014tfa}.  Moreover, we will restrict the presentation to the case of isotropic (spin-0) initial conditions. The generalization to PNG with non-trivial angular dependence is straightforward, and left as an exercise to the reader.

\vskip 4pt
In \S\ref{sec:SPT2}, we classify the divergences arising in the non-Gaussian loops of standard perturbation theory. 
We then show, in \S\ref{sec:counter}, that the solutions generated by the stress tensor~(\ref{eq:tauNG0V2}) act as the appropriate counterterms. 
When we compute correlators of the field $\Psi$ it will be convenient to expand the right-hand side~of (\ref{PSI}) around the Eulerian position $\x$\hskip 1pt: 
\begin{align}
\Psi(\x,\tau) 	&\ =\ \psi(\q(\x,\tau)) \nonumber \\[4pt]
			&\ =\ \psi(\x)\ +\ \D\psi(\x)\cdot\D\Phi(\x,\tau) \ +\ \cdots\ ,\label{eq:Psiexp}
\end{align}
where, in the last line, we have used linear perturbation theory to write the velocity $v^i$ in terms of the gradient of the gravitational potential $\partial^i\Phi$. The correlation between $\delta$ and $\psi$ is  
\beq
P_{1\psi}(k)\equiv \vev{\delta_{(1)}(\k,\tau)\psi(-\k)}' = \frac{(k/\mu)^\Delta}{M(k)}P_{11}(k)\ ,\label{eq:psidelta}
\eeq
where $\Delta$ and $M(k)$ were defined in \eqref{equ:psi} and \eqref{equ:M}, respectively.
Finally, in \S\ref{sec:renormalized}, we discuss the time dependence of the renormalized EFT parameters. 

\subsection{Loops in Standard Perturbation Theory}
\label{sec:SPT2}

There is only one non-Gaussian contribution to the one-loop power spectrum
\begin{align}
P_{12}(k) &\ =\ \ \ \raisebox{-8 pt}{\includegraphics[scale=0.8]{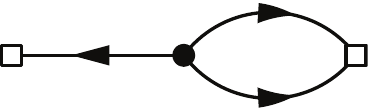}} \ \ \ =\ \int_\p F_{2}(\p,\k-\p)B_{111}(k,p,|\k-\p|) \,+\, \text{perm}\ . \label{equ:P12}
\end{align}
Extracting the UV behavior of this integral, we find the following divergence 
\begin{align}
P_{12}(k)\biggr|_{p\to\infty} &=\, - {\color{paper_red} \frac{1}{21}}\fnl\left[2\hskip 1pt\sigma^2(\Lambda) \, k^2\, P_{1\psi}(k) + \sigma^2_\psi(\Lambda)\, k^2M(k)\right] \,+\, \text{perm}\ ,\label{eq:P12div}
\end{align}
where we have defined 
\begin{align}
\sigma^2(\Lambda)&\equiv \int^\Lambda\frac{\d p}{2\pi^2} \frac{ a_0}{(p/\mu)^\Delta}\,P_{11}(p)\ ,\label{eq:Div1}\\[2pt]
\sigma^2_\psi(\Lambda)&\equiv   \int^\Lambda\frac{\d p}{2\pi^2}\frac{K_{\mathsmaller{\rm NL}}(\p,-\p)}{[M(p)]^2}\,P_{11}^2(p)\ .\label{eq:Div2}
\end{align}
Recall that $M(p)$ and $P_{11}(p)$ are evaluated at time $\tau$ (or scale factor $a$). The coefficients
$\sigma^2(\Lambda)$ and $\sigma^2_\psi(\Lambda)$ are therefore also time dependent.
In (\ref{eq:Div2}), we assumed that $K_{\mathsmaller{\rm NL}}(\p,-\p)$ is independent of the direction of the loop momentum~$\p$.
We see that the first term in~(\ref{eq:P12div}) is proportional to $P_{1\psi}(k)$, and will therefore be renormalized by a counterterm proportional to $\psi$. On the other hand, the second divergence in~(\ref{eq:P12div}) is proportional to $M(k)$ and is  renormalized by the noise contribution to the initial conditions $\psi_{J}$.

\begin{figure}[h!]
        \centering
        \vskip 6pt
              \includegraphics[scale=1.0]{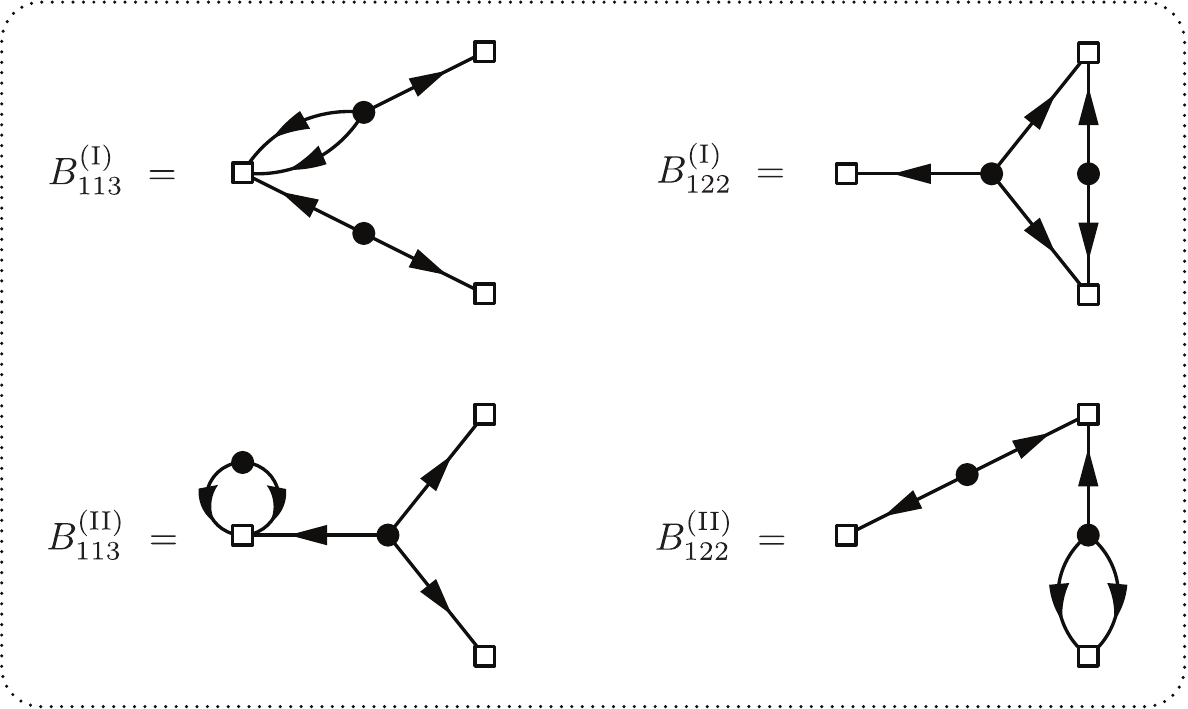}
   \caption{ Diagrammatic representation of the non-Gaussian contributions to the one-loop bispectrum. The diagrams of type II are renormalized by the same counterterms that renormalize the one-loop power spectrum. }
\label{fig:NGLoop}
 \end{figure}

\vskip 4pt
There are four non-Gaussian contributions to the bispectrum (see fig.~\ref{fig:NGLoop}):
\begin{align}
B^{{(\rm I)}}_{113} &= 3P_{11}(k_2)\int_\p F_{3}(\k_1+\p,-\p,\k_2)B_{111}(k_1,p,|\k_1+\p|) + \text{5 perms}\ , \label{equ:B113-v1}\\
B^{{(\rm II)}}_{113} &= 3B_{111}(k_1,k_2,k_3)\int_\p F_{3}(\k_1,\p,-\p)P_{11}(p) + \text{2 perms}\ , \label{equ:B113-v2}\\
B^{{(\rm I)}}_{122} &= 4\int_\p F_2(\k_3+\p,-\p)F_{2}(\p,\k_2-\p)B_{111}(k_1,|\k_3+\p|,|\k_2-\p|)P_{11}(p) + \text{2 perms}\ , \label{equ:B122-v1}\\
B^{{(\rm II)}}_{122} &= F_{2}(\k_1,\k_2)P_{11}(k_2)P_{12}(k_1) + \text{5 perms}\ . \label{equ:B122-v2}
\end{align}

\vspace{-0.3cm}

\noindent
 We note that the divergences appearing in $B_{122}^{\rm (II)}$ already arise in $P_{12}$ and are therefore renormalized by the same counterterms. Hence, we turn our attention to the diagrams $B_{113}^{\rm (I)}$, $B_{113}^{\rm (II)}$ and $B_{122}^{\rm (I)}$\hskip 1pt:

\begin{itemize}
\item The UV limit of $B_{113}^{\rm (I)}$ is
\begin{align}
B_{113}^{\rm (I)}\biggr|_{p\to\infty} &=\, -\bigg[\frac{4}{105}k_3^2 +{\color{paper_red}\frac{1}{21}}\frac{\k_1\cdot\k_2}{k_2^2}\,k_3^2+\frac{7}{45}\frac{ (\k_3\cdot\k_2)^2}{k_2^2}\bigg] \nonumber\\[4pt]
&\ \ \ \, \times \fnl\bigg[ 2\hskip 1pt\sigma^2(\Lambda) P_{1\psi}(k_1)+ \sigma^2_\psi(\Lambda)M(k_1)\bigg]P_{11}(k_2)\ +\ 5\ {\rm perms} \ , \label{eq:B113Idiv}
\end{align}
where $\sigma^2(\Lambda)$ and $\sigma^2_\psi(\Lambda)$ were defined in~(\ref{eq:Div1}) and (\ref{eq:Div2}), respectively.
Let us focus on the term proportional to $\k_1\cdot\k_2$. 
It is easy to see that this contribution is renormalized by the counterterm $\D\psi\cdot\D\Phi$. However, in order for this term to transform correctly under boosts, it has to appear in the combination
\beq
\psi(\x)+\D\psi(\x)\cdot\D\Phi(\x,\tau)\ ,
\eeq
i.e.~the field $\psi$ has to be evaluated at the Lagrangian position $\q(\x,\tau)$; cf.~eq.~\eqref{eq:Psiexp}.  For this to happen, the divergences in (\ref{eq:P12div}) and (\ref{eq:B113Idiv}) have to have the same numerical coefficient (highlighted in red).
It is reassuring that our computation reproduces this fact.

\item The UV limit of $B_{113}^{\rm (II)}$ is
\begin{align}
B_{113}^{\rm (II)}\biggr|_{p\to\infty} &=\,-\frac{61}{630}\sigma^2_{v}(\Lambda)\left(k_1^2+k_2^2+k_3^2\right)B_{111}(k_1,k_2,k_3)\ , \label{equ:B113}
\end{align}
where we have defined  
\beq
\sigma^2_{v}(\Lambda) \equiv \int^\Lambda\frac{\d p}{2\pi^2} \hskip 1pt P_{11}(p)\ . \label{equ:sigmav}
\eeq
Since the divergence in (\ref{equ:B113}) is the same as that appearing in $P_{13}$ and $B_{123}^{\rm (II)}$ (cf.~fig.~\ref{fig:LoopB}), it will be renormalized by the same (Gaussian) counterterm. 

\item The UV limit of $B_{122}^{\rm (I)}$ is
\begin{align}
B_{122}^{\rm (I)}\biggr|_{p\to\infty} &=  - \frac{26}{147}\fnl\left[k_2^2k_3^2 - \frac{40}{13}(\k_2\cdot\k_3)^2\right] \nonumber\\
&\hskip 46pt\times\bigg[2\hskip 1pt\hat\sigma^2(\Lambda)P_{1\psi}(k_1)+\hat\sigma^2_\psi(\Lambda)\, M(k_1)\bigg]+2\ \text{perms}\ ,\label{eq:b122div}
\end{align}
where we have defined 
\begin{align}
\hat \sigma^2(\Lambda)&\equiv\int^\Lambda\frac{\d p}{2\pi^2} \hskip 1pt \frac{a_0}{p^2(p/\mu)^\Delta} \hskip 1pt P_{11}^2(p)\ , \label{equ:Div3}\\[2pt]
\hat \sigma^2_\psi(\Lambda)&\equiv\int^\Lambda\frac{\d p}{2\pi^2} \hskip 1pt\frac{K_{\mathsmaller{\rm NL}}(\p,-\p)}{p^2[M(p)]^2} \hskip 1pt P_{11}^3(p)\ . \label{equ:Div4}
\end{align}
We notice that all terms in~(\ref{eq:b122div}) are {\it analytic} in two of the external momenta. For instance, the first permutation is analytic in $k_2$ and $k_3$. In position space, these divergences therefore are proportional to derivatives of delta functions. As we will explain below, these terms are renormalized by the noise term in the effective stress tensor~(\ref{eq:noise0}).

\end{itemize}

\subsection{Renormalization in the EFT-of-LSS}
\label{sec:counter}

Next, we demonstrate that the stress tensor~(\ref{eq:tauNG0V2}) generates new solutions which precisely cancel the cutoff dependence arising from the loop diagrams in SPT. Many of the computational details will be relegated to Appendix~\ref{sec:SPT-Appendix}.

\vskip 4pt
Including the stress tensor, the equations of motion for $\delta$ and $\theta$ are\begin{align}
{\cal D}_\delta \delta&\equiv\H^2\left[-a^2\partial_a^2+\left(\frac{3}{2}\Omega_m-3\right)a\partial_a+\frac{3}{2}\Omega_m\right]\delta={\cal S}_\beta-\H\partial_a(a{\cal S}_\alpha)+\tau_\theta\ ,\label{eq:Ddelta}\\
{\cal D}_\theta \theta&\equiv\H^2\left[+a^2\partial_a^2+\left(4-\frac{3}{2}\Omega_m\right)a\partial_a+(2-3\Omega_m)\right]\theta=\partial_a(a{\cal S}_\beta)-\frac{3}{2}\Omega_m\H{\cal S}_\alpha+\partial_a(a\tau_\theta)\ ,\label{eq:Dtheta}
\end{align}
where ${\cal S}_\alpha$ and ${\cal S}_\beta$ are nonlinear source terms defined in Appendix~\ref{sec:SPT-Appendix} and 
\beq
\tau_\theta\equiv-\partial_i\left[\frac{1}{\rho}\partial_j\tau^{ij}\right]\ . \label{tauT}
\eeq
Substituting~(\ref{eq:tauNG0V2}) and (\ref{eq:noise0}), we find that $\tau_\theta$ is the sum of a contribution from $\vev{\tau^{ij}}_s$ and one from the stochastic term $\Delta\tau^{ij}$:
\beq
\tau_\theta=\tau_v+\tau_n\ , \label{eq:viscnoise}
\eeq
where, to second order in the fluctuations and including the Gaussian contributions (see~\cite{Baldauf:2014qfa,Angulo:2014tfa}), we have 
\begin{align}
\tau_v&= -d^2\hskip -1pt\bigtriangleup\hskip -2pt\delta-e_1\hskip -1pt\bigtriangleup\hskip -2pt(\delta^2) -e_2\hskip -1pt\bigtriangleup\hskip -2pt(s^2) -e_3\hskip 1pt\partial_i(s^{ij}\partial_j\delta)\nonumber\\[2pt]
&\hskip 12pt-\fnl\big[g\big(\hskip -1pt\bigtriangleup \hskip -2pt\Psi-\partial_i(\delta\partial^i\Psi)\big)+g_1 \hskip -1pt\bigtriangleup \hskip -2pt(\Psi\delta)+g_2\partial_{i}\partial_j(\Psi s^{ij})\big]\ ,\label{eq:tauv}\\[8pt]
\tau_n&= -\partial_{i}\big[\partial_j J^{ij}_0-(\delta\partial_{j}J^{ij}_0)\big] -\partial_{i}\partial_{j}(J^{ij}_1\delta) -\partial_{i}\partial_{j}(J^{ij}_2{}_{kl} s^{kl})-\fnl\partial_{i}\partial_{j}(J^{ij}_\psi\Psi)\ . \label{eq:taun} 
\end{align}
The full solution  $\{\delta,\theta\}$ can then be written as a sum of three terms
\beq
\delta = \delta^{{\rm SPT}}+\delta^c+\delta^J\quad{\rm and}\quad \theta=\theta^{{\rm SPT}}+\theta^c+\theta^J\ ,
\eeq
where $\{\delta^{{\rm SPT}},\theta^{{\rm SPT}}\}$ is the SPT solution (see Appendix~\ref{sec:SPT-Appendix}), while $\{\delta^c,\theta^c\}$  and $\{\delta^J,\theta^J\}$  are the solutions generated by $\tau_v$ and $\tau_n$, respectively. 
In (\ref{dSPT}), we expanded the SPT solution in powers of the first-order initial condition~$\delta_1$. A similar expansion can be defined for  the counterterms 
\begin{align}
\delta^{c,J}(a)=\sum_{n=1}^\infty \delta^{c,J}_{(n)}(a)\ ,\label{eq:expct}
\end{align}
where $\delta^{c}_{(n)}\propto (\delta_{1})^n$ and $\delta^J_{(n)}\propto (\delta_{1})^{n-1}$, and equivalently for $\theta$.  
In particular, the viscosity counterterm $\delta_{(n)}^c$ can be expressed as the sum of a Gaussian and a non-Gaussian contribution
\begin{align}
\delta^c_{(n)}(\k,a) &= \int_{\k_1}\ldots\int_{\k_n}(2\pi)^3\delta_D\big(\k- \k_{1\ldots n}\big)\,F_n^{c}(\k_1,\ldots,\k_n|a)\,\delta_{(1)}(\k_1,a)\ldots\delta_{(1)}(\k_n,a)
\nonumber\\
 			&\ \ \ + \fnl\int_{\k_1}\ldots\int_{\k_n}(2\pi)^3\delta_D\big(\k- \k_{1\ldots n}\big)\,H_n^{c}(\k_1,\ldots,\k_n|a)\,\psi(\k_1)\ldots\delta_{(1)}(\k_n,a)  \ , \label{equ:deltac(n)}
\end{align}
where the kernel functions  $F_{n}^{c}$ and $H_{n}^{c}$ have the following diagrammatical representations:
\begin{align}
\raisebox{-33 pt}{\includegraphics[scale=1.0]{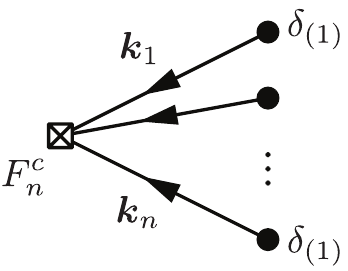}} \hskip 20pt
&\equiv \hskip 10pt F_n^{c}(\k_1,\ldots,\k_n|a)\,(2\pi)^3\delta_D\big(\k-\k_{1\ldots n}\big)\ , \\[10pt]
\raisebox{-33 pt}{\includegraphics[scale=1.0]{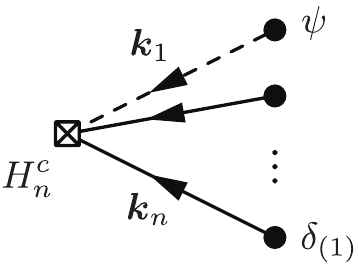}} \hskip 20pt
 & \equiv \hskip 10pt\fnl\hskip 1ptH_n^{c}(\k_1,\ldots,\k_n|a)\,(2\pi)^3\delta_D\big(\k-\k_{1\ldots n}\big)\ .
\end{align}

\vskip 6pt
\noindent 
Our goal in this section is to compute $\delta^{c}$ and $\delta^{J}$ up to second order. 
Naturally, both $\delta^c$ and $\delta^J$ have contributions from the Gaussian and non-Gaussian terms of the stress tensor. However, since the Gaussian contributions have already been computed in \cite{Baldauf:2014qfa,Angulo:2014tfa,Carrasco:2013mua}, we will mainly focus our attention on the non-Gaussian contributions. Furthermore, since we are only interested in showing that the counterterms cancel the cutoff dependence arising from the SPT loop diagrams, we will, for simplicity, restrict the analysis to the Einstein-de Sitter cosmology. The extension to $\Lambda$CDM will be explained in more detail in the next section. 

\subsubsection*{Viscosity counterterms}

$\bullet$\ \ The first-order counterterm $\delta^c_{(1)}$ is the solution to 
\beq
{\cal D}_\delta\, \delta^c_{(1)}= -d^2(a)\bigtriangleup\delta_{(1)}-g(a)\fnl\bigtriangleup\psi\ . \label{equ:d1E}
\eeq 
The Green's functions associated with the operator ${\cal D}_\delta$ is defined in Appendix~\ref{sec:SPT-Appendix}. It allows us to write the solution of (\ref{equ:d1E}) as
\beq
\delta^{c}_{(1)}(\k,a) =  - \xi(a)k^2\delta_{(1)}(\k,a)-\gamma(a)\fnl k^2\psi(\k)\ ,\label{eq:delta1c}
\eeq
where we have defined\hskip 1pt\footnote{Note that the parameter $\gamma$ of \cite{Baldauf:2014qfa} corresponds to the parameter $\xi$ in this paper.}
\begin{align}
\xi(a)&\equiv -\frac{1}{a}\int_{a_{in}}^a\d a'\, G_\delta(a, a')\hskip 2pt a' d^2( a')\ , \label{equ:xi}\\
\gamma(a)&\equiv-\int_{a_{in}}^a\d a'\, G_\delta(a, a') \hskip 2pt g( a')\ . \label{equ:gamma}
\end{align}
The non-Gaussian contribution to $\delta_{(1)}^c$ cancels the loop divergence in $P_{12}$\,:
\begin{align}
P_{12}+P_{1c}&\,\supset\ \ \ \raisebox{-9 pt}{\includegraphics[scale=0.8]{Figures/P12Eq}}\ \ \ +\ \ \ \raisebox{0 pt}{\includegraphics[scale=0.8]{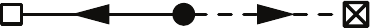}}\nonumber\\[10pt]
&\,= -\fnl 
\left(\frac{4}{21}\sigma^2(\Lambda)+2\hskip 1pt\gamma(a)\right)k^2P_{1\psi}(k)\ =\ {\rm finite}\ ,\label{eq:P12renorm}
\end{align}
where, for the moment, we have only focused on the divergence proportional to $P_{1\psi}(k)$. (The divergence proportional to $M(k)$ in \eqref{eq:P12div} will be cancelled by a  noise term.)
Hence, the cutoff dependence of the one-loop power spectrum is removed if
\beq
\gamma(\Lambda,a)=-\frac{2}{21}\sigma^2(\Lambda,a)+{\rm finite}\quad \Leftrightarrow\quad g(\Lambda,a) = -\frac{1}{3\hskip 1pt a}\sigma^2(\Lambda,a) + {\rm finite}\ ,\label{eq:gctr}
\eeq
where we have added the explicit arguments to highlight that both $\gamma$ and $\sigma$ are functions of the cutoff $\Lambda$ and the time $a$.
The finite piece of (\ref{eq:gctr}) will be discussed in \S\ref{sec:renormalized}.

\vskip 4pt
\newpage
The Gaussian contribution to $\delta_{(1)}^c$ cancels the loop divergence of $B_{113}^{\rm (II)}$:
\begin{align}
B_{113}^{\rm (II)}+B_{11c} &\ \supset \ \ \ \ \raisebox{-28 pt}{\includegraphics[scale=0.8]{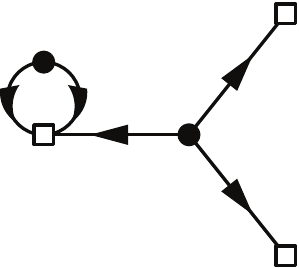}}\hskip 20pt+\hskip 20pt\raisebox{-28 pt}{\includegraphics[scale=0.8]{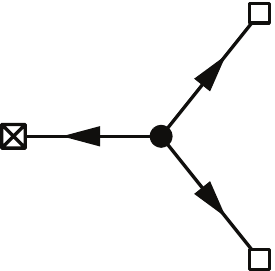}}\nonumber\\[15pt]
&\ =\, -\left[\frac{61}{630}\sigma^2_v(\Lambda)+\xi(a)\right](k_1^2+k_2^2+k_3^2)B_{111}(k_1,k_2,k_3)\ .
\end{align}
Hence, renormalization requires
\beq
\xi(\Lambda,a) = -\frac{61}{630}\sigma^2_v(\Lambda,a)+{\rm finite}\quad\Leftrightarrow\quad d^2(\Lambda,a) = -\frac{61}{180}\frac{1}{a}\sigma^2_v(\Lambda,a)+{\rm finite}\ .\label{eq:xidiv}
\eeq
Note that the parameter $\xi$ is fixed by the renormalization of the Gaussian one-loop diagram~$P_{13}$~\cite{Carrasco:2012cv}. 

\vskip 10pt
\noindent
$\bullet$\ \  The second-order counterterm $\delta_{(2)}^c$ satisfies
\beq
{\cal D}_\delta\,\delta_{(2)}^c = {\cal S}_\beta^{(2)}-\H\partial_a(a{\cal S}_\alpha^{(2)})+\tau_v^{(2)}\ ,
\eeq
where ${\cal S}_{\alpha,\beta}^{(2)}$ are obtained by replacing one of the $\delta$ (or $\theta$) in the nonlinear source terms by their first-order SPT solution and the other one by the first-order counterterm $\delta^c_{(1)}$ (or $\theta^c_{(1)} = -\dot\delta^{c}_{(1)}$). 
Focusing on the solution generated by the non-Gaussian terms, we find
\begin{align}
\delta_{(2)}^{c,{\rm NG}}(\k,a)	&=\int_{a_{in}}^a\d a'\ G_\delta(a,a')\left[{\cal S}_\beta^{(2)}(a')-\H\partial_{a'}(a'{\cal S}_\alpha^{(2)}(a'))+\tau_v^{(2)}(a')\right]\nonumber\\[2pt]
	&=\fnl\int_\p H_{2}^c(\p,\k-\p|a)\,\psi(\p)\delta_{(1)}(\k-\p,a)\ .
\end{align}
The kernel $H_{2}^c$ receives contributions from ${\cal S}_{\alpha,\beta}$ and $\tau_v$,
and can be written as
\beq
H_{2}^c(\k_1,\k_2|a) \equiv -\gamma(a)\big[G_\Psi(\k_1,\k_2)+G_{\alpha\beta}(\k_1,\k_2)\big] -\gamma_1(a)G_{1}(\k_1,\k_2)-\gamma_2(a)G_{2}(\k_1,\k_2)\ ,
\eeq
where we have defined
\beq
\gamma_{i}(a) \equiv -\frac{1}{a}\int_{a_{in}}^a\d a'\ G_\delta(a,a')\,a'g_{i}(a')\ . \label{equ:gammai}
\eeq
Explicit expressions for the kernels $G_{\{\Psi,\alpha\beta,i\}}$ are given in Appendix~\ref{sec:SPT-Appendix}.
Let us note that the value of $\gamma(a)$ has already been fixed by the renormalization of the power spectrum and therefore can no longer be adjusted. Substituting~(\ref{eq:gctr}) for $\gamma(a)$, the sum of the loop diagram and the counterterm becomes
\begin{align}
B_{113}^{\rm (I)}+B_{11c} &\ \supset \ \ \ \raisebox{-28 pt}{\includegraphics[scale=0.8]{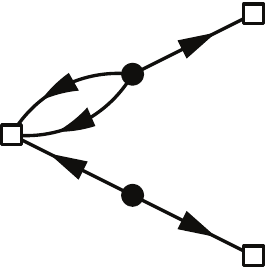}}\hskip 20pt+\hskip 20pt\raisebox{-28 pt}{\includegraphics[scale=0.8]{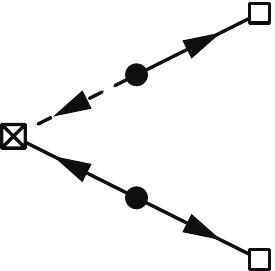}}\nonumber\\[15pt]
&\ =\,-\fnl\left[\left(\frac{74}{567}\sigma^2(\Lambda)+\gamma_1(a) \right)k_{3}^2
+\left(\frac{52}{135}\sigma^2(\Lambda)+\gamma_2(a)\right)\left(\frac{(\k_{3}\cdot\k_2)^2}{k_2^2}-\frac{1}{3}k_{3}^2\right) \right]\nonumber\\
&\ \ \ \ \ \ \ \ \ \ \ \times 
P_{1\psi}(k_1)P_{11}(k_2)\ +\ \text{5 perms}\ .
\end{align}
The cutoff dependence is therefore cancelled if
\begin{align}
\gamma_1(\Lambda,a)&=-\frac{74}{567}\sigma^2(\Lambda,a)+{\rm finite}\quad \Leftrightarrow\quad g_1(\Lambda,a)=-\frac{74}{63}\frac{1}{a}\sigma^2(\Lambda,a) +{\rm finite} \ ,\\[2pt]
\gamma_2(\Lambda,a)&=-\frac{52}{135}\sigma^2(\Lambda,a)+{\rm finite}  \quad \Leftrightarrow\quad g_2(\Lambda,a)=-\frac{52}{15}\frac{1}{a}\sigma^2(\Lambda,a) +{\rm finite} \ .
\end{align}

\vskip 10pt
We have shown that every counterterm generated by the viscosity contribution to the stress tensor, $\tau_v$, yields the correct momentum dependence to remove some of the loop divergences in the power spectrum and the bispectrum. Next, we show that the remaining divergences are cancelled by the noise counterterms.  This part is for aficionados and can be skipped on a first reading.

\subsubsection*{Noise counterterms}

Noise terms appear both  in the initial conditions (\ref{eq:philong}) and in the stress tensor~(\ref{eq:noise0}).  They are essential in the renormalization procedure as they remove some divergences in $P_{12}$ and $B_{113}^{\rm (I)}$, as well as all the divergences in $B_{122}^{\rm (I)}$ .

\vskip 4pt
As indicated in~(\ref{eq:expct}), the noise term can be expanded in powers of the long-wavelength fluctuations. The $n$-th order solution $\delta^J_{(n)}$ receives contributions from both Gaussian and non-Gaussian initial conditions:
\beq
\delta^J_{(n)}(\k,a) = \delta^{J,\hskip 1pt\rm G}_{(n)}(\k,a) +\delta^{J,\hskip 1pt\rm NG}_{(n)}(\k,a)\ .\label{eq:deltaJ}
\eeq
Specifically, at one-loop order, we will require the first-order Gaussian contribution 
\begin{align}
\delta^{J,\hskip 1pt \rm G}_{(1)}(\k,a) &= k_ik_jN_0^{ij}(\k,a)\ ,
\end{align}
and the first and second-order non-Gaussian contributions
\begin{align}
\delta^{J,\hskip 1pt \rm NG}_{(1)}(\k,a) &= \fnl\hskip 1ptM(k)\hskip 1pt\psi_J(\k)\ , \label{dJ1}\\
\delta^{J,\hskip 1pt \rm NG}_{(2)}(\k,a) &= \fnl \int_{\p}\left[k_ik_jN^{ij}_\psi(\k-\p,a)\psi(\p)+2F_{2}(\k-\p,\p)\delta_{(1)}(\k-\p,a)M(p)\psi_{J}(\p)\right] \, , \label{dJ2}
\end{align}
where 
\beq
N^{ij}_{0,\psi}(\k,a)\equiv\int_{a_{in}}^a\d a'\ G_{\delta}(a,a') \hskip 1pt J^{ij}_{0,\psi}(\k,a')\ .\label{eq:Nij}
\eeq
We will now show that these counterterms cancel the divergences proportional to $M(k)$ in both $P_{12}$ and $B_{113}^{\rm (I)}$ and all the divergences in $B_{122}^{\rm (I)}$\hskip 1pt:
\begin{itemize}
\item The divergence proportional to $M(k)$ in $P_{12}$ is renormalized by the counterterm~$\delta^{J}_{(1)}$. More precisely, we have  
\begin{align}
P_{12} + P_{JJ}&\,\supset\, -\frac{2}{21}\fnl \sigma_\psi^2(\Lambda)\, k^2M(k)+2\hskip 2pt \vev{ \delta^{J,\hskip 1pt {\rm NG}}_{(1)}(\k,a)\delta_{(1)}^{J,\hskip 1pt{\rm G}}(-\k,a)}'\nonumber\\[3pt]
&\,=\, \fnl\left[-\frac{2}{21}\sigma_\psi^2(\Lambda)+ 2 \int_{a_{in}}^a\d a'\ G_{\delta}(a,a')\tilde{\cal J}_0(a') \right]k^2M(k)\ =\ {\rm finite}\ ,
\end{align}
where $\tilde{\cal J}_0(a')$ was defined in~(\ref{eq:tildeJ}).
Hence, the cutoff dependence is eliminated provided that
\beq
\tilde {\cal J}_0(\Lambda,a)=-\frac{1}{6 \hskip 1pt a}\sigma_\psi^2(\Lambda,a)  + {\rm finite}\ .
\eeq
Similarly, the divergence proportional to $M(k)$ in $B_{122}^{\rm(II)}$ (which is the same as the divergence in $P_{12}$), is renormalized by the second term in~(\ref{dJ2}). 

\item Let us consider the following correlation between the first-order SPT solution and the noise terms
\begin{align}
B_{1JJ} &\,\supset\, \vev{ \delta_{(1)}^{\rm SPT}(\k_1,a)\delta^{J,\hskip 1pt {\rm NG}}_{(1)}(\k_2,a)\delta^{J,\hskip 1pt {\rm G}}_{(2)}(\k_3,a)}' \ +\ \vev{ \delta_{(1)}^{\rm SPT}(\k_1,a)\delta^{J,\hskip 1pt {\rm G}}_{(1)}(\k_2,a)\delta^{J,\hskip 1pt {\rm NG}}_{(2)}(\k_3,a)}'  \nonumber  
\\[4pt]
&\,\equiv\,  \hspace{2.8cm}B_{1JJ}^{(A)} \hspace{2.6cm}+\hspace{2.5cm} B_{1JJ}^{(B)}
\end{align}
The first term, $B_{1JJ}^{(A)}$, cancels the divergence proportional to $M(k)$ in $B_{113}^{\rm (I)}$, while the second term, $B_{1JJ}^{(B)}$, cancels the first divergence in $B_{122}^{\rm (I)}$. We will demonstrate the second fact explicitly and leave the first as an exercise to the reader.

Using (\ref{dJ1}) and (\ref{dJ2}), we can write
\begin{align}
B_{1JJ}^{(B)} &=(k_{2})_i(k_2)_j(k_3)_k(k_3)_l\vev{N^{ij}_0(\k_2,a)N^{kl}_\psi(-\k_2,a)}'\, P_{1\psi}(k_1) +{5\ \rm perms}\ ,
\end{align}
where $N^{ij}_{0,\psi}$ was defined in (\ref{eq:Nij}).   We will assume that the cutoff-dependent part of $J^{ij}_{0,\psi}$ satisfies $
J^{ij}_{0,\psi}\propto a$, so that $N^{ij}_{0,\psi} = -\tfrac{2}{7}aJ^{ij}_{0,\psi}$ and
\beq
\vev{N^{ij}_0(\k,a)N^{kl}_\psi(-\k,a)}' =\frac{4}{49}a^2\left[{\cal J}(a)\delta^{ij}\delta^{kl}+\hat {\cal J}(a)\left(\delta^{ik}\delta^{jl}+\delta^{il}\delta^{jk}\right)\right]\ .
\eeq
Hence, we find that 
\begin{align}
B_{122}^{\rm (I)}+B_{1JJ}^{(B)}  &\, \supset\, \fnl\bigg[\left(\frac{4}{49}{\cal J}(a)a^2-\frac{52}{147}\hat \sigma^2(\Lambda)\right)k_2^2k_3^2\nonumber\\
						&\qquad \ \ + \left(\frac{8}{49}\hat{\cal J}(a)a^2+\frac{160}{147}\hat\sigma^2(\Lambda)\right)(\k_2\cdot\k_3)^2\bigg] P_{1\psi}(k_1) \, + \, {\rm perms}\nonumber\\[10pt]
						&\,=\ {\rm finite}\ .\label{eq:b122renorm}
\end{align}
We see that $B_{1JJ}^{(B)}$ has the right momentum dependence to absorb the first divergence in~$B_{122}^{\rm (I)}$. In particular, the cutoff dependence is cancelled if
\beq
{\cal J}(\Lambda,a) = \frac{13}{3a^2}\tilde \sigma^2(\Lambda,a)+{\rm finite}\quad\ {\rm and}\quad\ \hat{\cal J}(\Lambda,a)=-\frac{20}{3} \frac{1}{a^2}\hat\sigma^2(\Lambda,a)+{\rm finite}\ .
\eeq
\item Finally, we consider 
\begin{align}
B_{JJJ}&\supset \vev{\delta^{J,\hskip 1pt\rm NG}_{(1)}(\k_1,a)\delta^{J,\hskip 1pt\rm G}_{(1)}(\k_2,a)\delta^{J,\hskip 1pt\rm G}_{(1)}(\k_3,a)}'+{\rm perms}\nonumber\\
&= \fnl\hskip 1ptM(k_1) (k_2)_i(k_2)_j(k_3)_k(k_3)_l\vev{\psi_J(\k_1)J^{ij}_0(\k_2,a)J^{kl}_0(\k_3,a)}'\ .\label{eq:bJJJ}
\end{align}
It is difficult to determine the precise momentum dependence of this correlation function, since the probability distributions of the noise terms are unknown. 
We will assume that it can be expanded around $\k_i=\0$, so that we can write
\beq
\vev{\psi_J(\k_1)J^{ij}_0(\k_2,a)J^{kl}_0(\k_3,a)}' = {\cal N}(a)\delta^{ij}\delta^{kl}+\hat{\cal N}(a)\big(\delta^{ik}\delta^{jl}+\delta^{il}\delta^{jk}\big)+{\cal O}(k_i^2)\ .
\eeq
In that case, we get
\begin{align}
B_{122}^{\rm (I)}+B_{JJJ}&\,\supset\, \fnl\bigg[ \left(-\frac{26}{147}\hat\sigma^2_\psi(\Lambda)+{\cal N}(a)\right)k_2^2k_3^2\nonumber\\
&\hskip 30pt+\left(\frac{80}{147}\hat\sigma^2_\psi(\Lambda)+2\hskip 2pt\hat{\cal N}(a)\right)(\k_2\cdot\k_3)^2 \bigg]M(k_1)+{\rm perms}\nonumber\\
&\,=\,{\rm finite}\ .
\end{align}
We see that $B_{JJJ}$ has the correct momentum dependence to absorb the cutoff dependence of the second divergence in~(\ref{eq:b122div}). More precisely, we have 
\beq
{\cal N}(\Lambda,a) = \frac{26}{147}\hat \sigma_\psi^2(\Lambda,a)+{\rm finite}\quad \ {\rm and}\quad \ \hat{\cal N}(\Lambda,a)=-\frac{40}{147}\hat\sigma^2_\psi(\Lambda,a)+{\rm finite}\ .
\eeq
\end{itemize}
We have shown that the solutions generated by the stress tensor derived in Section~\ref{sec:coarse} have the correct momentum dependence to cancel all divergences coming from SPT loop diagrams. So far, we have focused on the cutoff-dependent parts of the EFT parameters. In the next section, we study their finite (or renormalized) parts.

\subsection{Renormalized EFT Parameters}
\label{sec:renormalized}

Any EFT parameter can be written as the sum of a cutoff-dependent part (the counterterm) and a finite (or ``renormalized'') part:
\beq
g(\Lambda,a) = g^{(0)}(\Lambda,a) + g^{(R)}(a)\ .
\eeq
Ultimately, every long-wavelength observable must be independent of the cutoff. Indeed, once the large-scale correlation functions have been properly renormalized, one can send the cutoff to infinity.
In $\Lambda$CDM, all loop integrals are convergent and the cutoff can be taken to infinity even {\it before} renormalization. This may seem to go against the EFT philosophy since we are including modes which are outside the regime of validity of the EFT. However, the mistakes one makes in doing so, can be absorbed into a shift of the renormalized EFT parameters. 

\vskip 4pt
The time dependence of the renormalized parameters may be different from that of the counterterms. Moreover, it cannot be computed within the EFT framework, but must be determined from simulations or observations. 
An exception is a ``scaling universe" --- i.e.~a matter-dominated Einstein-de Sitter (EdS) cosmology with scale-free initial conditions $P_{11}(k, a_{in}) \propto k^n$.
In this special case, an additional symmetry constrains the time dependence of the parameters in the EFT. 
To see this, we first note that the equations of motion in EdS are invariant under a {\it Lifshitz scaling}
\beq
\x\mapsto\lambda_x\x\quad\ {\rm and}\quad \ 
a\mapsto\lambda_aa \ .\label{eq:Lifshitz}
\eeq
The dimensionless power spectrum of the initial conditions, $\Delta_{11}^2(k,a_{in}) \propto a_{in}^2 k^{n+3}$, is also invariant under the Lifshitz scaling iff
\beq
\lambda_x\equiv \lambda_a^{2/(n+3)}\ .\label{eq:powerlaw}
\eeq
When~(\ref{eq:powerlaw}) holds, then the Lifshitz scaling maps one EdS solution to another with a different realization of the same statistical initial condition. 
The linearly-evolved power spectrum can be written in a manifestly self-similar form, $\Delta_{11}^2(k,a) = (k/\knl)^{n+3}$, where the nonlinear scale satisfies $\knl \propto a^{-2/(n+3)}$.  
Nonlinear corrections to $\Delta_{11}^2$ appear as higher powers of $k/\knl$~\cite{Scoccimarro:1995if, Pajer:2013jj}. 
 Assuming that the short-scale fluctuations satisfy the Lifshitz scaling, the terms in the effective stress tensor must also have the right transformation properties. 
 In particular, the symmetry fixes the combination of powers of $k$ and 
$a$ which can appear in the equations of motion and therefore constrains the time dependence of the renormalized parameters. For instance, using such an argument,\footnote{Let us remark that this argument does not determine the time dependence of the counterterms $g^{(0)}(\Lambda, a)$, since the cutoff introduces a new scale and therefore breaks the Lifshitz scaling symmetry~\cite{Pajer:2013jj}.} one finds that the speed of sound in a scaling universe must satisfy $c_s^2(a)\propto a^{(1-n)/(n+3)}$~\cite{Pajer:2013jj}. 

\vskip 4pt
For non-Gaussian initial conditions, we also need to impose that the higher-point correlation functions are invariant under the Lifshitz scaling~\eqref{eq:Lifshitz} with $\lambda_x$ and 
$\lambda_a$ related by~(\ref{eq:powerlaw}). More specifically, let us assume that the dimensionless bispectrum satisfies 
\beq
{\cal B}_{111}\left(\frac{k_1}{\lambda_x},\frac{k_2}{\lambda_x},\frac{k_3}{\lambda_x},\lambda_a\hskip 1pta_{in}\right)=\lambda_x^m\lambda_a^3\hskip 2pt{\cal B}_{111}\left(k_1,k_2,k_3,a_{in}\right)\ , \label{eq:Bscaling}
\eeq
where we have used that ${\cal B}_{111}(k_i,a_{in})\propto a_{in}^3$.
Given~(\ref{eq:powerlaw}), the initial bispectrum is invariant under~(\ref{eq:Lifshitz}) iff $m =-\frac{3}{2}(n+3)$. 
 Furthermore, if the primordial Gaussian potential $\varphi_g$ is scale-invariant, 
then the field $\psi$ must transform as $\psi(\x)\mapsto  \lambda_x^{\Delta}\,\psi(\lambda_x\x)$ under the Lifshitz scaling;  cf.~eq.~(\ref{equ:psi}).   
 Using that the Lagrangian coordinate transforms as 
 $\q(\x,a)\mapsto \lambda_x^{-1}\q(\lambda_x\x,\lambda_aa)$,
  we find that the field $\Psi(\x,a) = \psi(\q(\x,a))$ transforms as 
\beq
\Psi(\x,a) \,\mapsto \, \lambda_x^{\Delta}\,\Psi(\lambda_x\x,\lambda_aa)\ .
\eeq
This implies that the terms in the stress tensor~(\ref{eq:tauNG0V2}) preserve the Lifshitz scaling iff their time dependence is
\begin{align}
\big\{g^{(R)}(a)\ ,\ g_{1,2}^{(R)}(a)\big\}&\propto a^{{(1-n+2\Delta)/(n+3)}} \propto [D_1(a)]^{{(1-n+2\Delta)/(n+3)}} \ .
\end{align}

\begin{figure}[h!]
        \centering
                      \hspace{-1cm}  \includegraphics[scale=0.8]{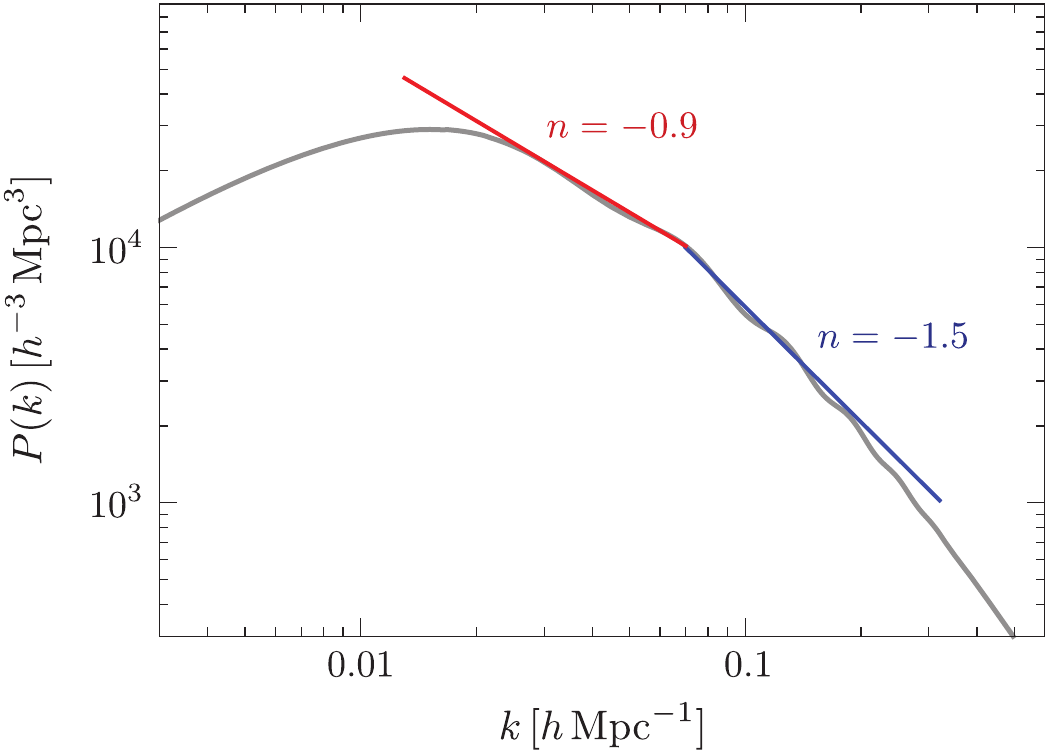}
  \caption{Scaling of the initial matter power spectrum. We see that in the momentum regions $[0.02,0.07]\,h\hskip 1pt{\rm Mpc}^{-1}$ and $[0.07,0.25]\,h\hskip 1pt{\rm Mpc}^{-1}$ the power spectrum is well approximated by a power law with $n\approx -0.9$ and $\knl\approx 0.16\,h\hskip 1pt{\rm Mpc}^{-1}$ (red line) and $n\approx -1.5$ and $\knl\approx 0.23\,h\hskip 1pt{\rm Mpc}^{-1}$ (blue line), respectively.}\label{fig:scaling}
 \end{figure}

In $\Lambda$CDM and for general initial conditions, the Lifshitz scaling isn't a symmetry and the initial conditions aren't scale-free. However, in fig.~\ref{fig:scaling} we show that the initial matter power spectrum has momentum regimes in which a power law ansatz is a good approximation.
This explains why, for Gaussian initial conditions, the ansatz deduced from the Lifshitz scaling symmetry has worked remarkably well \cite{Baldauf:2014qfa,Angulo:2014tfa,Carrasco:2013mua}.  We will extend these results and assume that the EFT parameters have the following time dependence\hskip 1pt\footnote{Notice that we have included a factor $(\H f)^2$ which is not present in \cite{Baldauf:2014qfa}. We found that this ansatz improves the accuracy of~(\ref{eq:approx1}) by a few percents.} 
\beq
g^{(R)}(a) = [\H(a)f(a)]^2 [D_1(a)]^{m_g+1} \, \bar g^{(R)}\ ,\label{eq:gtime}
\eeq
where $m_g$ is a real parameter and $\bar g^{(R)}$ is a constant. 
At one loop, we don't need to know the precise time dependence of $g_{1,2}^{(R)}$ to compute the counterterms.  The ansatz (\ref{eq:gtime}) is therefore only required for the parameter $g(a)$.
In general, the value of $m_g$ is unknown. In what follows we will assume that the time dependence matches the ansatz of the scaling universe, i.e.~we will use $m_g=(1-n+2\Delta)/(n+3)$. 
From fig.~\ref{fig:scaling} we deduce that $n\in [-1.5,-0.9]$ for $k\in[0.02,0.25]\,h\hskip 1pt {\rm Mpc}^{-1}$. For these values of $n$, we have $m_g\in[0.9+ \Delta,1.7+1.3\hskip 1pt\Delta]$.  For our numerical computations, we will choose $m_g = 1.7+1.3\hskip 1pt\Delta$. However, we have checked that the counterterms only change by at most a few percents as the value of $m_g$ runs over this interval, so our numerical results will not depend sensitively on this choice. 

\section{Numerical Analysis}
\label{sec:numerics}

In this section, we present a numerical computation of the renormalized power spectrum and bispectrum for non-Gaussian initial conditions.
We start, in \S\ref{sec:summary}, with a summary of the different contributions (SPT and EFT) to the dark matter correlation functions.
In \S\ref{sec:scaling}, we estimate the relative sizes of the terms in a scaling universe.
Finally, in \S\ref{sec:results}, we make some preliminary observations on the shapes of the various contributions. A more detailed analysis will appear in~\cite{future}.

\subsection{Renormalized Correlation Functions}
\label{sec:summary}

In \S\ref{sec:tau}, we wrote the effective stress tensor of the dark matter as a sum of terms made from the long-wavelength fields and their derivatives
\beq
\tau_v = \sum_n a_n \hskip 1pt {\cal O}_n\ ,
\eeq
where the coefficients $a_n$ and the corresponding operators ${\cal O}_n$ relevant for this paper are listed in Table~\ref{tab:parameters}. 

\begin{table}[h!]
\begin{center}
\begin{tabular}{c |  c  c c c c c c }
${\cal O}_n$ & \ $\bigtriangleup \delta$\ & $\bigtriangleup(\delta^2)$ & $\bigtriangleup(s^2)$ & $\partial^i(s_{ij}\partial^j\delta)$ & $\bigtriangleup \hskip -1pt\Psi-\partial_i(\delta\partial^i\Psi)$ & $\bigtriangleup(\Psi \delta)$ & $\partial^i \partial^j(\Psi s_{ij})$ \\[2pt]
\hline \\[-10pt]
$a_n$ & \ $d^2$\ & $e_1$ & $e_2$ & $e_3$ & $g$ & $g_1$ & $g_2$ \\[2pt]
$\alpha_n$ & \ $\xi$\ & $\epsilon_1$ & $\epsilon_2$ & $\epsilon_3$  & $\gamma$ &  $\gamma_1$ &  $\gamma_2$
\end{tabular}
	\caption{Operators and parameters in the EFT-of-LSS with primordial non-Gaussianities.\label{tab:parameters}  }
	\end{center}
\end{table}

\noindent
In \S\ref{sec:counter}, we showed that these operators gives rise to counterterms which renormalize the one-loop power spectrum and bispectrum of standard perturbation theory. Schematically, the solution can be written as (see \S\ref{sec:counter} and \S\ref{sec:perturb} for more details)
\beq
\delta^{c}(a) \simeq \int_{a_{in}}^{a} \d a' \,G_{\delta}(a,a') \hskip 1pt \tau_v(a') \, \simeq\, \sum_n \alpha_{n}(a) \hskip 1pt {\cal O}_n(a) \ ,
\eeq
where the parameters $\alpha_n$ can be expressed as integrals over time of the corresponding parameters~$a_n$.
The SPT solution $\delta^{\rm SPT}(a)$ and the EFT counterterms~$\delta^c(a)$ generate correlation functions for the renormalized dark matter density contrast $\delta(a)$ which depend both on the cosmological parameters of the $\Lambda$CDM model ${\boldsymbol \theta} \equiv \{\Omega_m^{0}, \Omega_\Lambda^{0}, h, \Delta_\varphi, n_s \}$ 
and on the EFT parameters~$\boldsymbol{\alpha}\equiv \{\xi,\epsilon_{1,2,3}, \gamma, \gamma_{1,2}\}$.

\begin{itemize}
\item  The dark matter power spectrum can be written as $P = P^{\rm G} + \fnl P^{\rm NG}$,
where $P^{\rm G}$ arises from the nonlinear evolution of Gaussian initial conditions~\cite{Carrasco:2012cv} and $P^{\rm NG}$ is the contribution from non-Gaussian initial conditions. Both $P^{\rm G}$ and $P^{\rm NG}$ can be split into an SPT part and an EFT part. At one loop, we have
\begin{align}
P^{\rm G_{\phantom 1}}_{\rm SPT} &= P_{11} + P_{13}+P_{22}\ , \qquad P^{\rm G_{\phantom 1}}_{\rm EFT} = -2 \xi\hskip 1pt k^2 P_{11} \ , \label{equ:PGEFT} \\[4pt]
P^{\rm NG_{\phantom 1}}_{\rm SPT} &= P_{12}\ , \hspace{2.03cm}\qquad P^{\rm NG_{\phantom 1}}_{\rm EFT} = - 2\gamma\hskip 1pt k^2 P_{1 \psi}\ , \label{equ:PNGEFT}
\end{align}
where $P_{1\psi}(k)$ was defined in~(\ref{eq:psidelta}).

\item The total dark matter bispectrum is $B = B^{\rm G} + \fnl B^{\rm NG}$,
where $B^{\rm G}$ is the part arising from the nonlinear evolution of Gaussian initial conditions~\cite{Baldauf:2014qfa, Angulo:2014tfa}, and $B^{\rm NG}$ is the part associated with non-Gaussian initial conditions.
Both $B^{\rm G}$ and $B^{\rm NG}$ can be split into an SPT part and an EFT part.
At one loop, we have
\begin{align}
B^{\rm G_{\phantom 1}}_{\rm SPT}&=B_{112} + \Big[B_{114}+B_{123}^{\rm (I)}+B_{123}^{\rm (II)}+B_{222}\Big]\ , \label{eq:BGloop}\\[10pt]
B^{\rm NG_{\phantom 1}}_{\rm SPT}&=B_{111} + \Big[B_{113}^{\rm (I)}+B_{113}^{\rm (II)}+B_{122}^{\rm (I)}+B_{122}^{\rm (II)}\Big]\ ,\label{eq:BNGloop} \\[8pt]
B^{\rm G_{\phantom 1}}_{\rm EFT} &=  \xi B_\xi^{\rm G_{\phantom 1}} + \sum_{i=1}^3 \epsilon_i B_{\epsilon_i}\ , \label{BG} \\
B^{\rm NG_{\phantom 1}}_{\rm EFT} &= \xi B_\xi^{\rm NG_{\phantom 1}} + \gamma B_\gamma + \sum_{i=1}^2 \gamma_i  B_{\gamma_i} \ . \label{BNG}
\end{align}
The individual bispectra in (\ref{BG}) and (\ref{BNG}) can be written as
\begin{align}
B_\xi^{\rm G_{\phantom 1}}&\equiv-2\big[E_{\alpha\beta}(\k_1,\k_2)+E_{\delta}(\k_1,\k_2)\big]P_{11}(k_1)P_{11}(k_2) + \text{2 perms}\ ,\\[4pt]
B_\xi^{\rm NG_{\phantom 1}} &\equiv-(k_1^2+k_2^2+k_3^2)B_{111}(k_1,k_2,k_3)\ \label{equ:BxiNG},\\[4pt]
B_{\epsilon_i}&\equiv-2\hskip 1ptE_i(\k_1,\k_2)\hskip 1ptP_{11}(k_1)P_{11}(k_2) + \text{2 perms}\ ,\\[4pt]
B_{\gamma}&\equiv -\big[G_{\alpha\beta}(\k_1,\k_2)+G_{\Psi}(\k_1,\k_2)\big]P_{11}(k_1)P_{1\psi}(k_2)+\text{5 perms}\ ,\\[4pt]
B_{\gamma_i}&\equiv -G_i(\k_1,\k_2)P_{11}(k_1)P_{1\psi}(k_2)+\text{5 perms}\ ,
\end{align}
where the kernel functions $E_{\cdot \cdot}(\k_1,\k_2)$ and $G_{\cdot \cdot }(\k_1,\k_2)$ are defined explicitly in Appendix~\ref{sec:SPT-Appendix}.

For simplicity, we will take the cosmological parameters to be fixed and allow only the EFT parameters to vary (for a more complete treatment see~\cite{future}). Furthermore, we will assume that the sound speed parameter has been measured in the power spectrum and its value is fixed $\xi\equiv 1.5 \,h^{-2}\,{\rm Mpc}^{2}$~\cite{Baldauf:2014qfa}. 
The following parts of both the Gaussian and the non-Gaussian bispectra are therefore predicted
\beq
B_0^{\rm I} \equiv B^{\rm I}_{\rm SPT}+\xi B^{\rm I}_{\xi}\ , \quad {\rm I}={\rm G}, {\rm NG}\ .  \label{equ:B0}
\eeq
The total bispectrum can then be written as
\beq
\boxed{B = B_0^{\rm G} + B_c^{\rm G} +\fnl \left( B_0^{\rm NG} + B_c^{\rm NG} \right) } \ ,
\eeq
where we have defined the sum of the additional counterterm contributions as
\begin{align}
B_c^{\rm G} &\equiv \sum_{i=1}^3\epsilon_iB_{\epsilon_i}\ , \label{equ:BcG}\\
B_c^{\rm NG} &\equiv \gamma B_\gamma +\sum_{i=1}^2 \gamma_{i}B_{\gamma_i} \ . \label{equ:BcNG}
\end{align}

In general, the coefficients in (\ref{equ:BcG}) and (\ref{equ:BcNG}) cannot be predicted, but need to be measured in N-body simulations or in observations. 
For the numerical results of \S\ref{sec:results}, we will estimate 
the sizes of these coefficients by looking at the one-loop divergences that they cancel. 
 In particular, since in our universe loop integrals are convergent, they are in practice evaluated with the cutoff $\Lambda$ taken to infinity. In that case, the SPT loop diagrams contain integrals of the linear power spectrum extrapolated to scales beyond the linear regime (i.e.~$k>\knl$). 
This is not a problem, since, as we explained in \S\ref{sec:renormalized}, this finite error is removed by adjusting the counterterms.  
More precisely, from~(\ref{eq:gctr}) we infer that the term in $\gamma$ which cancels this known error is
\beq
|\gamma|\, \sim\, \frac{2}{21}\int_{\knl}^\infty\frac{\d p}{2\pi^2}\frac{a_0}{(p/\mu)^\Delta} P_{11}(p) + \cdots\ ,\label{eq:gest}
\eeq
where $\knl$ is the nonlinear scale. We will choose $\knl=0.2\,h\hskip 1pt{\rm Mpc}^{-1}$, which corresponds to the nonlinear scale found in \cite{Baldauf:2014qfa,Angulo:2014tfa}. For local non-Gaussianity (i.e.~$\Delta=0$), we then find $|\gamma|\sim 1.6\,h^{-2}\hskip 1pt{\rm Mpc}^{2}$. 
Note that, for $\Delta\neq0$, the precise value of $\mu$ is unimportant as it cancels in the product $\gamma B_\gamma$.

Of course, there is also an unknown, finite contribution to $\gamma$ that accounts for the effects of the nonperturbative short scales on the long-wavelength dynamics. We will assume that there is no fine-tuned cancellation between this finite part and the term shown in~(\ref{eq:gest}). The result in~(\ref{eq:gest}) then gives an approximate lower bound on the size of the counterterm.  Applying the same estimate to $\gamma_1$ and $\gamma_2$, we find 
\beq
\gamma_1\approx \frac{37}{27}\gamma\quad{\rm and}\quad \gamma_2\approx\frac{182}{45}\gamma\ . \label{equ:gest2}
\eeq
  Let us stress that the estimates in (\ref{eq:gest}) and (\ref{equ:gest2}) only provide approximated lower bounds and should not be considered precise evaluations. 
\end{itemize}

\noindent
In \S\ref{sec:results}, we will present numerical results for three different primordial bispectrum shapes:
\begin{itemize}
\item Local non-Gaussianity is obtained by setting $K_{\mathsmaller{\rm NL}} = 1$ in~(\ref{equ:Bprimordial}). The primordial bispectrum then is 
\beq
B_\varphi^{\rm local}(k_1,k_2,k_3) = \ 2\fnl^{\rm local}A^2\bigg[\frac{1}{k_1^3k_2^3}+\frac{1}{k_1^3k_3^3}+\frac{1}{k_2^3k_3^3}\bigg]\ ,
\eeq
where $A\equiv 2\pi^2\Delta_\varphi^2$.  For simplicity, we have written the bispectrum for scale-invariant initial conditions, $n_s=1$.
The latest CMB constraint on the amplitude of the local bispectrum is $\fnl^{\rm local}= 1.8 \pm 5.6$~\cite{PlanckNG}.
\item  Higher-derivative corrections to slow-roll inflation produce equilateral non-Gaussianity~\cite{Alishahiha:2004eh, Cheung:2007st}.
In the effective field theory of inflation~\cite{Cheung:2007st}, this is captured by two cubic operators, $\dot\pi^3$ and $\dot\pi(\partial_i\pi)^2$, for the Goldstone boson of broken time translations, $\pi$.
Both operators produce very similar equilateral bispectra.
For purposes of illustration, we focus on the PNG produced by the operator~$\dot\pi^3$, which has a bispectrum of the form
\beq
B^{\rm equil}_{\varphi}(k_1,k_2,k_2) = 162\, \fnl^{\rm equil}\cdot\frac{A^2}{k_1k_2k_3}\frac{1}{K^3}\ ,\label{equ:Bequi}
\eeq
where $K \equiv k_1+k_2+k_3$. The latest CMB constraint on the amplitude of the equilateral bispectrum is $\fnl^{\rm equil}= -9.2 \pm 69$~\cite{PlanckNG}.

\item The shape of the bispectrum in quasi-single-field inflation cannot be computed analytically. However, it is well approximated by the following ansatz~\cite{Chen:2009zp,Sefusatti:2012ye}
\beq
B_{\varphi}^{\rm QSFI}(k_1,k_2,k_3) = 18\sqrt{3}\, \fnl^{\rm QSFI}\cdot\frac{A^2}{k_1k_2k_3} \frac{1}{K^{3}} \cdot \frac{1}{\sqrt{\kappa}}\frac{N_\nu[8 \kappa]}{N_\nu[8/27] }\ ,\label{equ:BQSFI}
\eeq
where $\kappa \equiv k_1k_2k_3/K^3$ and $N_\nu$ is the Neumann function of order $\nu$. The parameter $\nu$ depends on the mass of the hidden sector field to which the inflaton field couples during inflation. It determines the scaling of the bispectrum in the squeezed limit via $\Delta\equiv\frac{3}{2}-\nu$ in~(\ref{eq:angSL}).
For purposes of illustration, we choose $\nu =\frac{1}{2}$ corresponding to $\Delta=1$ which is precisely intermediate between the scaling of local and equilateral non-Gaussianity. 
\end{itemize}

 \vskip 4pt
In our numerical computations, we consider a flat $\Lambda$CDM cosmology with the following parameters: $\Omega_m^{0} = 0.27$, $\Omega_\Lambda^{0} = 0.73$, $h=0.70$. The amplitude of the primordial potential is $\Delta_\varphi^2(k_0)\simeq 8.7\times 10^{-10} $, defined at the pivot scale $k_0=0.002\,h\,{\rm Mpc}^{-1}$. The one-loop correlators are computed numerically with {\sf Mathematica}. For validation purposes, we wrote two independent codes. Both codes yielded results which are in very good agreement.\footnote{As a technical aside, let us note that the expression of the one-loop bispectrum in~(\ref{eq:BNGloop}) is ill-suited for numerical evaluation as the different contributions each contain IR divergences. The equivalence principle guarantees that these divergences cancel when all diagrams are summed~\cite{Scoccimarro:1995if}. However, to avoid cancelling large numbers against each other, it is preferable to rewrite the integrand in a form which makes this cancellation manifest. In Appendix~\ref{sec:IR}, we present the ``IR-safe" integrands for the non-Gaussian loops.} 

\begin{figure}[h!]
        \centering
                        \includegraphics[scale=0.71]{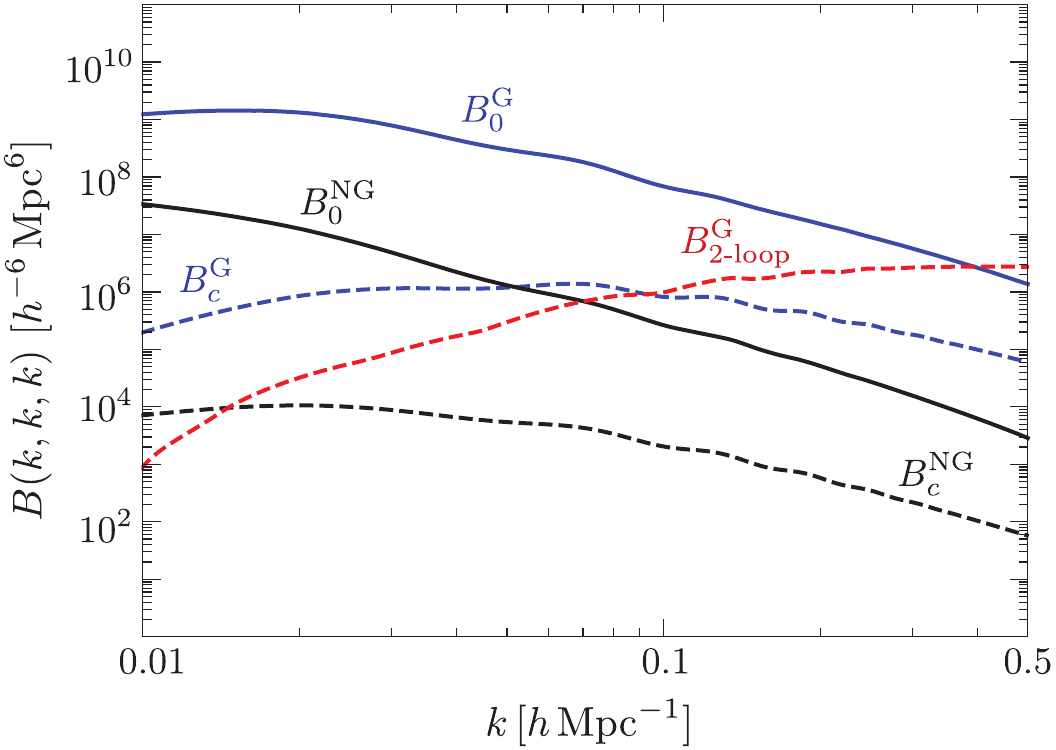} \hspace{0.5cm}  \includegraphics[scale=0.71]{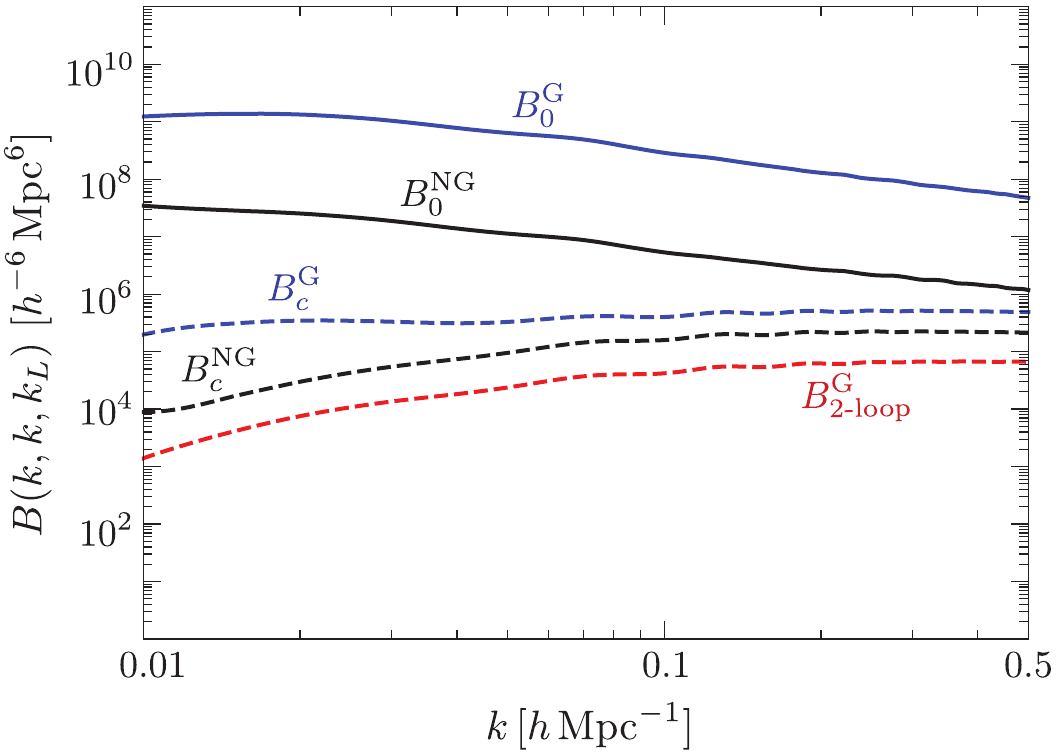}
  \caption{Contributions to the bispectrum for local non-Gaussianity with $\fnl^{\rm local}=10$,
 evaluated in the equilateral configuration ({\it left}) and for fixed $k_L \equiv 0.01\, h \hskip 1.5pt{\rm Mpc}^{-1}$ ({\it right}).  The definitions of $B_0^{\rm G, NG}$ and $B_c^{\rm G, NG}$ can be found in (\ref{equ:B0}), (\ref{equ:BcG}) and  (\ref{equ:BcNG}).  }\label{fig:Local}
 \end{figure}

\begin{figure}[h!]
        \centering
                        \includegraphics[scale=0.71]{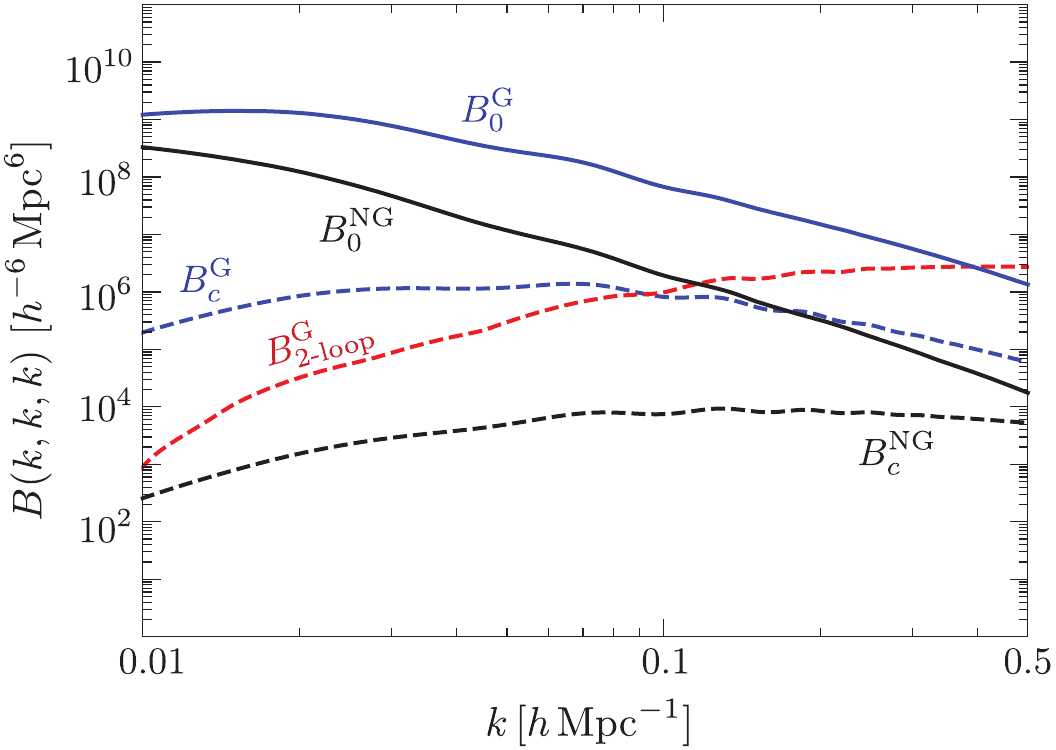} \hspace{0.5cm}  \includegraphics[scale=0.71]{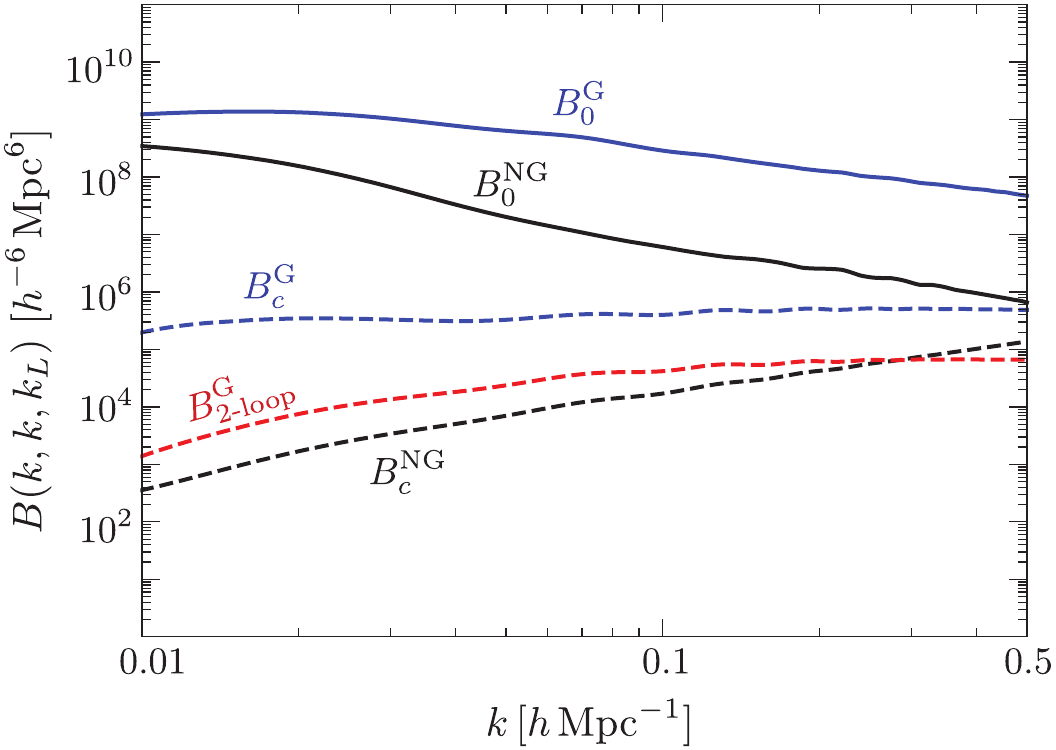}
  \caption{Same as fig.~\ref{fig:Local} but for equilateral non-Gaussianity  with $\fnl^{\rm equil} = 100$. }\label{fig:equil}
 \end{figure}

\begin{figure}[h!]
        \centering
                        \includegraphics[scale=0.71]{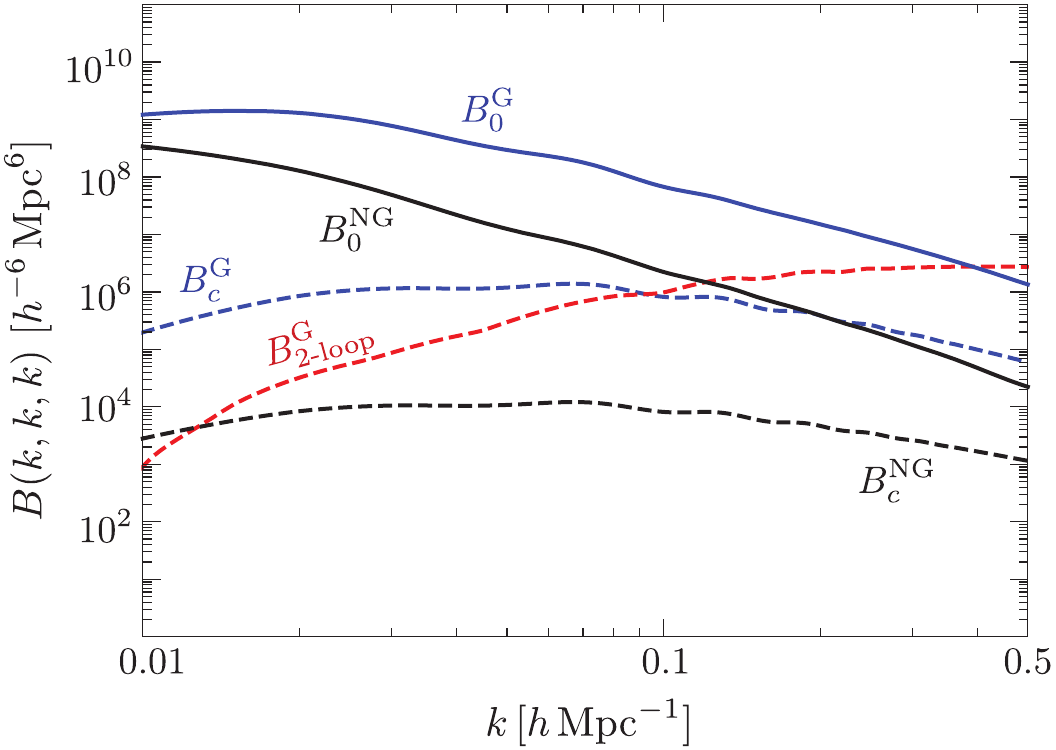} \hspace{0.5cm}  \includegraphics[scale=0.71]{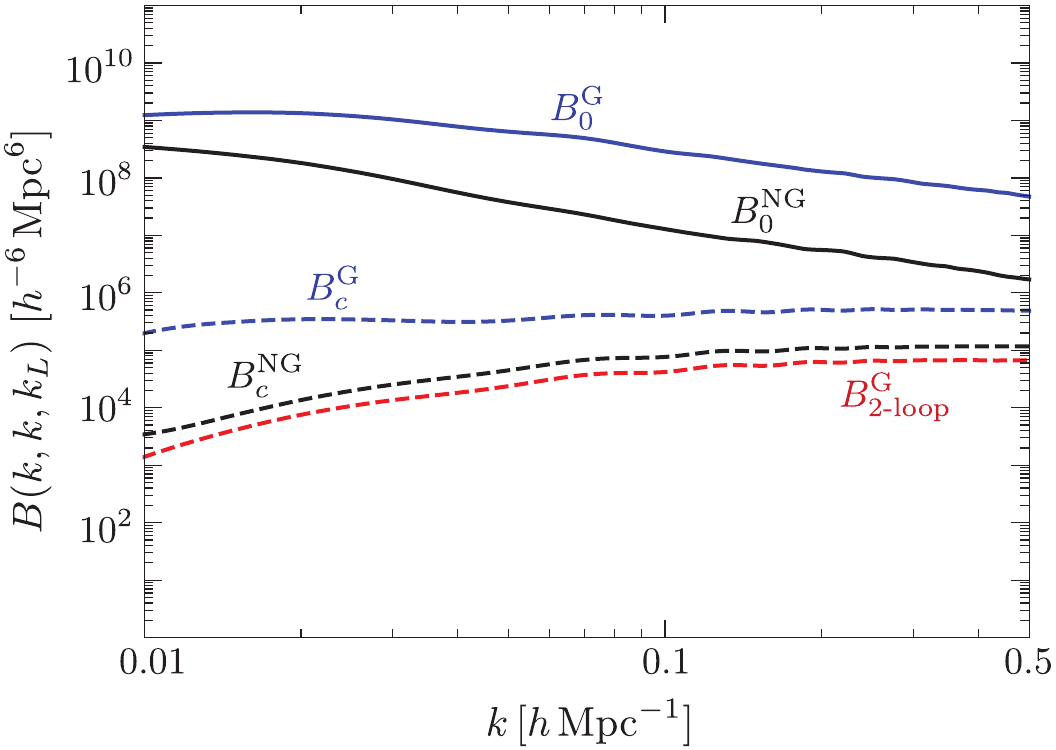}
  \caption{Same as fig.~\ref{fig:Local} but for quasi-single-field inflation with $\Delta = 1$ and  $\fnl^{\rm QSFI} = 100$.}\label{fig:QSFI}
 \end{figure}

\vskip 4pt
Our results are summarized in figs.~\ref{fig:Local}, \ref{fig:equil} and \ref{fig:QSFI} for local, equilateral and quasi-single-field PNG, respectively.  The plots show the SPT and counterterm contributions to the bispectrum separated into Gaussian terms (which are the same in every plot) and non-Gaussian terms. Here, and in the following, the results are evaluated at redshift $z=0$. Notice that while PNG has a larger effect on larger scales, we have chosen to show the ``mildly nonlinear'' range of scales, where the EFT counterterms can play an important role. For each type of PNG, we represent the corresponding matter bispectrum in the equilateral configuration (left) and for fixed $k_L \equiv 0.01\, h \hskip 1.5pt{\rm Mpc}^{-1}$ (right). We see that, in the equilateral configuration, the two-loop Gaussian contribution becomes comparable to the non-Gaussian contribution on rather large scales. In~\S\ref{sec:scaling}, we provide an analytical estimate of the critical scale $k_c$ at which the two-loop Gaussian contribution becomes relevant. On the other hand, we see that the two-loop contribution is much more subdominant in squeezed configurations. This is also the momentum configuration in which the shape of PNG (and more precisely, the scaling $\Delta$ of the squeezed limit) can leave an appreciable imprint on the matter bispectrum. In~\S\ref{sec:results}, we discuss the possibility of extracting information about the shape of PNG from the matter bispectrum.

\vspace{0.3cm}
 
\subsection{Estimates in a Scaling Universe}
\label{sec:scaling}

In this section, we determine the critical scale $k_c$ at which the two-loop Gaussian contribution becomes relevant. For purpose of illustration, we will consider local PNG, although our result won't depend sensitively on the precise shape of the non-Gaussianity. 
To estimate the relative sizes of the various contributions to the power spectrum and bispectrum we consider a scaling universe (i.e.~EdS with power-law spectrum $P_{\delta}(k,\tau_{in}) \propto k^n$).  This extends the results of~\cite{Pajer:2013jj,Baldauf:2014qfa} to non-Gaussian initial conditions.

\vskip 4pt
As explained in \S\ref{sec:renormalized}, in a scaling universe, Lifshitz transformations of the form~(\ref{eq:powerlaw}) leave the initial dimensionless power spectrum unchanged, which can therefore be written as a polynomial in~$k/\knl$. Because the evolution is self-similar, the same applies to the nonlinear power spectrum
\beq
\Delta^2_{\delta}(k,\tau)\equiv\frac{k^3}{2\pi^2}P_{\delta}(k,\tau) = \Delta^2_{\delta}(k/\knl)\ . \label{equ:Deltad}
\eeq
Furthermore, the transformation~(\ref{eq:Bscaling}) implies that the initial bispectrum is also self-similar, and that
the nonlinear dimensionless bispectrum is therefore a polynomial in $k_i/\knl$: 
\beq
\mathcal{B}_{\delta}(k_1, k_2, k_3,\tau)\equiv \left(\frac{k_1^3}{2\pi^2}\right)^2 B_\delta(k_1, k_2, k_3,\tau) =\mathcal{B}_{\delta}\big(k_1/ \knl, k_2/ \knl, k_3/ \knl)\ . \label{equ:calBd}
\eeq
The different contributions to the dimensionless power spectrum and bispectrum will scale as different powers of $k/\knl$.

\subsubsection*{Power Spectrum}

Let us first collect the terms that contribute to the power spectrum $\Delta^2_{\delta}(k/\knl)$. The momentum scalings of the Gaussian contributions were derived in~\cite{Pajer:2013jj}: 
\begin{align}
\Delta_{\delta, \rm G}^2(k) \ \subset\ {s}(n)\left(\frac{k}{\knl}\right)^p \ \leftarrow \ \left\{\begin{array}{lll} p = n+3 &\approx 1.5 & \hspace{1cm} \text{tree}  \\[4pt]
p = 2(n+3) &\approx 3.0& \hspace{1cm} \text{1-loop}\\[4pt]
p = n+ 5  &\approx 3.5 & \hspace{1cm} \text{LO vis.}\\[4pt]
p = 3(n+3)&\approx 4.5 & \hspace{1cm} \text{2-loop}\\[4pt]
p = 2 n+ 8  &\approx 5.0 & \hspace{1cm} \text{NLO vis.}\\[4pt]
p =  n+ 7  &\approx 5.5 & \hspace{1cm} \text{NLO h.d.}\\[4pt]
p = 4(n+3)&\approx 6.0 & \hspace{1cm} \text{3-loop}\\[4pt]
p = 7 &\approx 7.0 & \hspace{1cm} \text{noise}
\end{array}\right. \nonumber
\end{align}
where `tree' refers to the linear power spectrum $\Delta^2_{11}$, `loop' stands for the SPT loop contributions, `LO vis.' is the leading-order viscosity counterterm defined in (\ref{equ:PGEFT}),
`NLO vis.' are the next-to-leading-order viscosity counterterms~\cite{Pajer:2013jj}, `NLO h.d.' are next-to-leading-order higher-derivative terms, and `noise' refers to $\Delta_{JJ}^2$.  
The overall coefficient $s(n)$ is a (computable) order-one number, which is different for the different contributions (see Appendix~A of \cite{Pajer:2013jj}).
For the numerical estimates of the scaling index~$p$ we have used $n \approx - 1.5$, which corresponds to the scaling of the linear power spectrum in the regime $k\in[0.07,0.25] \,h\hskip 1pt{\rm Mpc}^{-1}$ (see fig.~\ref{fig:scaling}). 
The scaling for $k\in[0.02,0.07] \,h\hskip 1pt{\rm Mpc}^{-1}$, which is $n \approx -0.9$, will also be relevant, but hasn't been shown explicitly.

\vskip 4pt
For the non-Gaussian contributions to the power spectrum, we find 
\begin{align}
\Delta_{\delta,\rm NG}^2(k)\ \subset\  s(n)\, \fnl \Delta_\varphi\left(\frac{k}{\knl}\right)^p \ \leftarrow \ \left\{ \begin{array}{lll} p = \frac{3}{2}(n+3) &\approx 2.25 & \hspace{1cm} \text{1-loop}\\[4pt]
p = \frac{1}{2}(n+7) + \Delta  &\approx 2.75 + \Delta & \hspace{1cm} \text{LO vis.} \\[4pt]
p = \frac{5}{2}(n+3) &\approx 3.75  & \hspace{1cm} \text{2-loop}\\[4pt]
p = \frac{7}{2}(n+3) &\approx 5.25 & \hspace{1cm} \text{3-loop}\\[4pt]
p = \frac{1}{2}(n+13) &\approx 5.75 & \hspace{1cm} \text{noise}
\end{array} \right. \nonumber
\end{align}
where `LO vis.' refers to the leading order non-Gaussian viscosity counterterm, defined in (\ref{equ:PNGEFT}), `loop' stands for the SPT loop contributions and `noise'  stands for~$\Delta_{JJ}^2$. The scalings of the viscosity counterterm and the noise term
are fixed by the $k$-dependent parts of the UV-limit of $\Delta_{12}^2$~(see \S\ref{sec:SPT2}). 
We note that the amplitude of the non-Gaussian contributions is suppressed by a factor of~$\fnl\Delta_\varphi\sim 3 \times 10^{-5}\fnl$.
Higher-loop corrections are suppressed by additional factors of $\Delta_{11}^2(k)= (k/\knl)^{n+3}$.
As before, the overall coefficient $s(n)$ is a computable order-one number which is different for the different contributions.
 
\vskip 4pt
\newpage
We see that the non-Gaussian contributions to the power spectrum are highly suppressed.   We should therefore ask when the Gaussian two-loop terms (which we haven't included in our analysis) are of the same order. 
The two-loop Gaussian contribution is estimated through a single representative term, namely~$P_{33}^{(\rm I)}$~\cite{Carrasco:2013sva}:

\beq
P_{\text{2-loop}}^{\rm G} \ \subset\ P_{33}^{\rm (I)} \ \ =\ \ \raisebox{-10pt}{\includegraphics[scale=0.8]{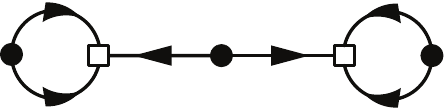}} 
\eeq

\begin{figure}[h!]
        \centering
                        \includegraphics[scale=0.71]{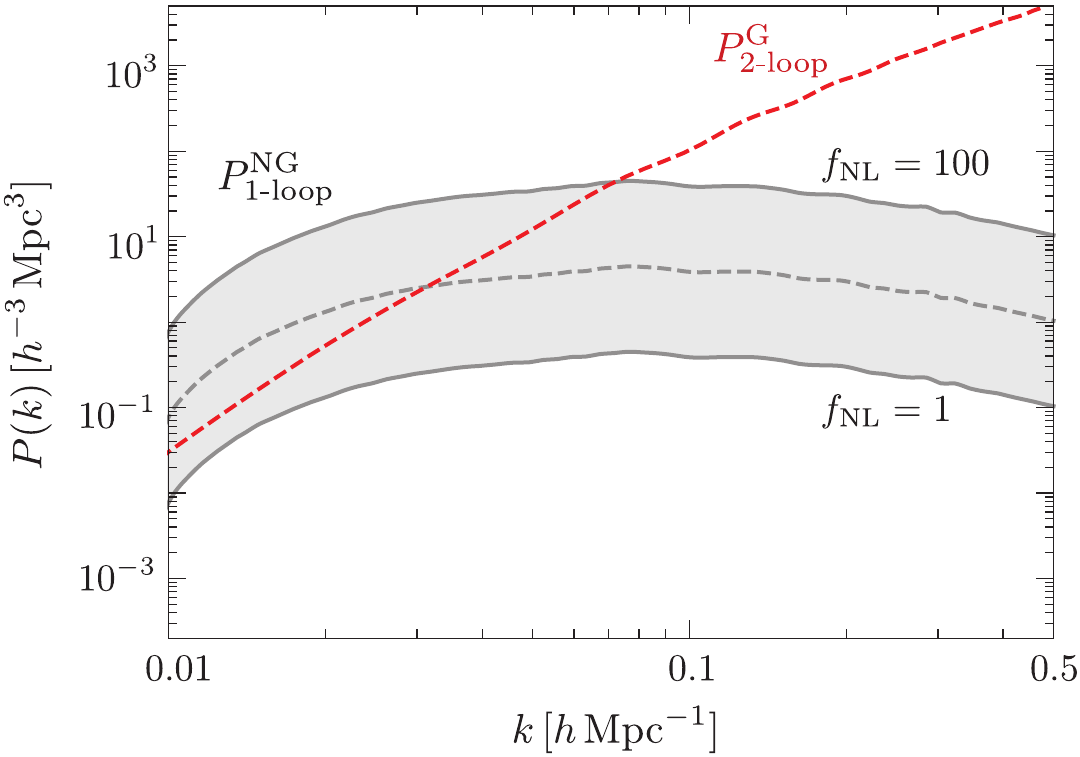} \hspace{0.5cm}  \includegraphics[scale=0.71]{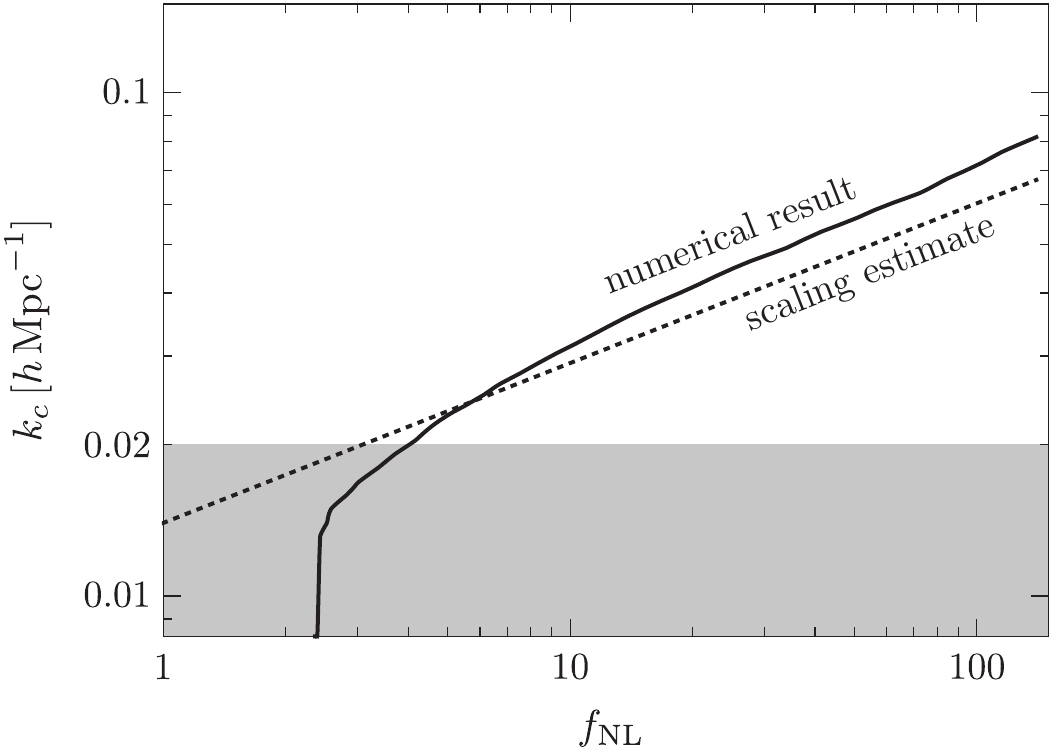}
  \caption{Comparison between the 1-loop contribution to the power spectrum for local PNG with  $\fnl \in [1,100]$ (grey band) and an estimate of the 2-loop Gaussian contribution (red dashed line). The right plot shows the critical momentum value $k_c$ at which the two-loop Gaussian contribution becomes larger than the one-loop non-Gaussian contribution as a function of $\fnl$, evaluated numerically (solid line) and using the estimate~(\ref{eq:kcPS}) (dashed line). The numerical result is for a $\Lambda$CDM cosmology, while the estimate is for a scaling universe. The shaded region represents the momentum range where the scaling $n\approx-0.9$ is no longer valid in the real universe.}  \label{fig:Scaling-PS}
 \end{figure}

\vskip 6pt
\noindent
A rough estimate of the scale $k_c$ at which the two-loop Gaussian contribution equals the one-loop non-Gaussian contribution is 
\beq
\frac{P_{\text{2-loop}}^{\rm G}}{P_{\text{1-loop}}^{\rm NG}} \approx \frac{1}{\fnl\Delta_\varphi}\frac{s^{\rm (I)}_{33}}{s_{12}}\left(\frac{k_c}{\knl}\right)^{\frac{3}{2}(n+3)} = 1 \quad \longrightarrow \quad k_c\approx 0.18\hskip 2pt\knl \left(\frac{\fnl}{10}\right)^{1/3}\ ,\label{eq:kcPS} 
\eeq
where $s_{33}^{\rm (I)}$ and $s_{12}$ are the coefficients of $P_{33}^{\rm (I)}$ and $P_{12}$, respectively.
For the numerical estimate in~(\ref{eq:kcPS}) we have used\footnote{These values are applicable to the momentum range $k\in[0.02,0.07] \,h\hskip 1pt{\rm Mpc}^{-1}$ (see fig.~\ref{fig:scaling}), which is roughly where the two-loop Gaussian contribution is expected to be larger than the one-loop non-Gaussian contribution (see fig.~\ref{fig:Scaling-PS}).} $n\approx -0.9$ , $\knl\approx 0.16\,h\hskip 1pt{\rm Mpc}^{-1}$ and $s^{\rm (I)}_{33}/s_{12}\simeq 0.06$.    In fig.~\ref{fig:Scaling-PS}, we compare this estimate for $k_c(\fnl)$ with the value obtained from a numerical computation in a $\Lambda$CDM universe.  We see that already on relatively large scales the two-loop Gaussian term is of the same order as the one-loop non-Gaussian term.

 \newpage
\subsubsection*{Bispectrum}

For the bispectrum, we focus our attention on the equilateral configuration,~$\mathcal{B}_{\delta}(k,k,k)$, for which the two-loop corrections are largest. 
For Gaussian initial conditions, the scalings of the various contributions were derived in~\cite{Baldauf:2014qfa}: 
\begin{align}
{\cal B}_{\delta}^{ \rm G}(k) \ \subset\ s(n)\left(\frac{k}{\knl}\right)^p \ \leftarrow \ \left\{\begin{array}{lll} p = 2(n+3) &\approx 3.0 & \hspace{1cm} \text{tree}  \\[4pt]
p = 3(n+3) &\approx 4.5& \hspace{1cm} \text{1-loop}\\[4pt]
p = 2(n+4) &\approx 5.0 & \hspace{1cm} \text{NLO vis.} \\[4pt]
p = 4(n+3) &\approx 6.0& \hspace{1cm} \text{2-loop}\\[4pt]
p = 3 n+ 11  &\approx 6.5 & \hspace{1cm} \text{NNLO vis.}\\[4pt]
p =  2n+ 10  &\approx 7.0 & \hspace{1cm} \text{NNLO h.d.}\\[4pt]
p = 5(n+3) &\approx 7.5 & \hspace{1cm} \text{3-loop}\\[4pt]
p = n+10 &\approx 8.5 & \hspace{1cm} \text{noise}\\[4pt]
\end{array}\right. \nonumber
\end{align}
where `NLO vis.' and `NNLO vis.' refer to the next-to-leading-order and next-to-next-to-leading-order contributions to the Gaussian counterterm, respectively. More precisely, the `NLO vis.' term corresponds to the term ${\cal B}_{\rm EFT}^{\rm G_{\phantom 1}}$ defined in~(\ref{BG}), while the `NNLO vis.'~term~corresponds to the one-loop diagrams formed with the leading Gaussian counterterms.  The term `NNLO h.d.' represents higher-derivative contributions and the term `noise' corresponds to ${\cal B}_{1JJ}$. 

For the non-Gaussian contributions to the bispectrum, we find 
\begin{align}
{\cal B}_{\delta}^{\rm NG}(k)\ \subset\ s(n)\,\fnl \Delta_\varphi \left(\frac{k}{\knl}\right)^p \ \leftarrow \ \left\{ \begin{array}{lll} p = \frac{3}{2}(n+3) &\approx 2.25 & \hspace{1cm} \text{tree}  \\[4pt]
p = \frac{5}{2}(n+3) &\approx 3.75 & \hspace{1cm} \text{1-loop}\\[4pt]
p =  \frac{1}{2}(3n+13)  &\approx 4.25   & \hspace{1cm} \text{LO vis.\hskip 2pt(1)}  \\[4pt]
p =  \frac{1}{2}(3n+13) + \Delta  &\approx  4.25 + \Delta  & \hspace{1cm} \text{LO vis.\hskip 2pt(2)}  \\[4pt]
p =  \frac{7}{2}(n+3)  &\approx  5.25 & \hspace{1cm} \text{2-loop} \\[4pt]
p = \frac{1}{2}(5n+19) + \Delta &\approx  5.75 + \Delta & \hspace{1cm} \text{NLO vis.} \\[4pt]
p =  \frac{1}{2}(3n+17) + \Delta &\approx 6.25 + \Delta  & \hspace{1cm} \text{NLO h.d.} \\[4pt]
p = \frac{1}{2}(3n+19) &\approx 7.25 & \hspace{1cm} \text{noise\hskip 2pt(A)} \\[4pt]
p =  \frac{1}{2}(n+17) + \Delta &\approx 7.75 + \Delta & \hspace{1cm} \text{noise\hskip 2pt(B)}\\[4pt]
p =  \frac{1}{2}(n+23)  &\approx 10.75 & \hspace{1cm} \text{noise\hskip 2pt(C)}
\end{array} \right. \nonumber
\end{align}
 where `LO vis.' and `NLO vis.'~refer to the leading-order and next-to-leading-order contributions to the non-Gaussian counterterm ${\cal B}_{c}^{\rm NG}$, respectively. More precisely,  the term `LO vis.\hskip 2pt(1)' refers to ${\cal B}_\xi^{\rm NG}$ (\ref{equ:BxiNG}), the term `LO vis.\hskip 2pt(2)' refers to ${\cal B}_c^{\rm NG}$ (\ref{equ:BcNG}), and the term `NLO vis.' refers to the finite part of the one-loop diagrams formed with one counterterm. 
 The term `NLO h.d.'~represents higher-derivative contributions and the terms `noise\hskip 2pt(A)', `noise\hskip 2pt(B)' and `noise\hskip 2pt(C)' refer to the noise terms ${\cal B}_{1JJ}^{(A)}$, ${\cal B}_{1JJ}^{(B)}$ and ${\cal B}_{JJJ}$ (cf.~\S\ref{sec:counter}). As before, the scalings of the viscosity and noise contributions to the bispectrum are determined by the $k$-dependence of the UV-limit of the loop diagrams they cancel (cf.~\S\ref{sec:SPT2}).

\begin{figure}[h!]
        \centering
                        \includegraphics[scale=0.71]{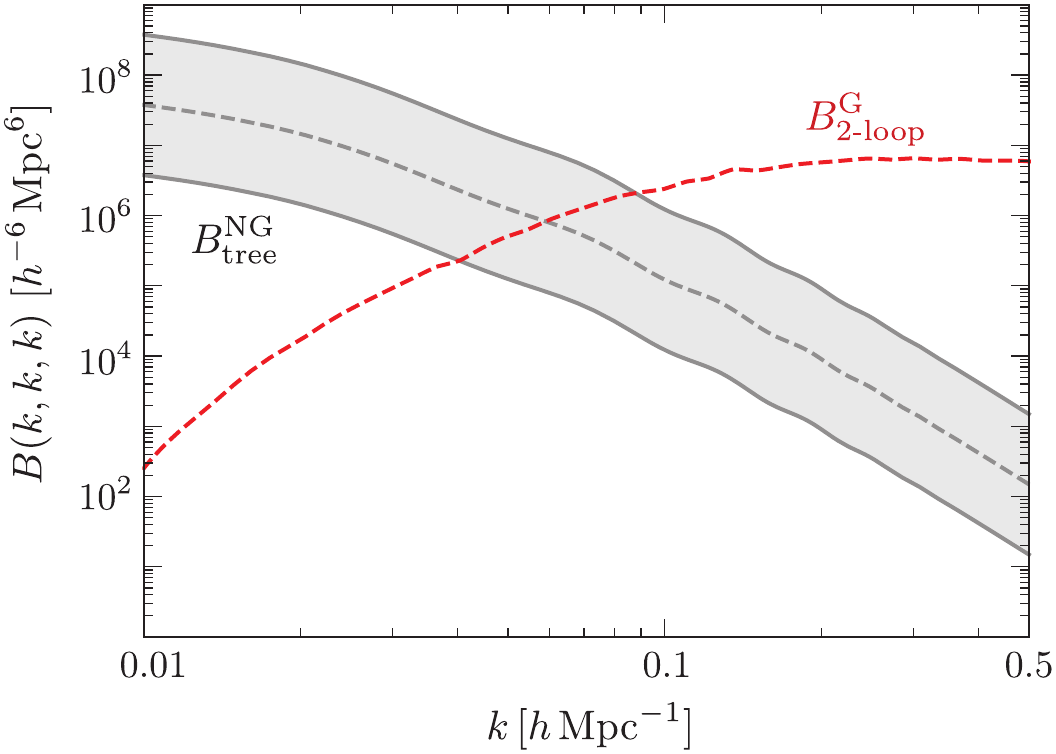} \hspace{0.5cm}  \includegraphics[scale=0.71]{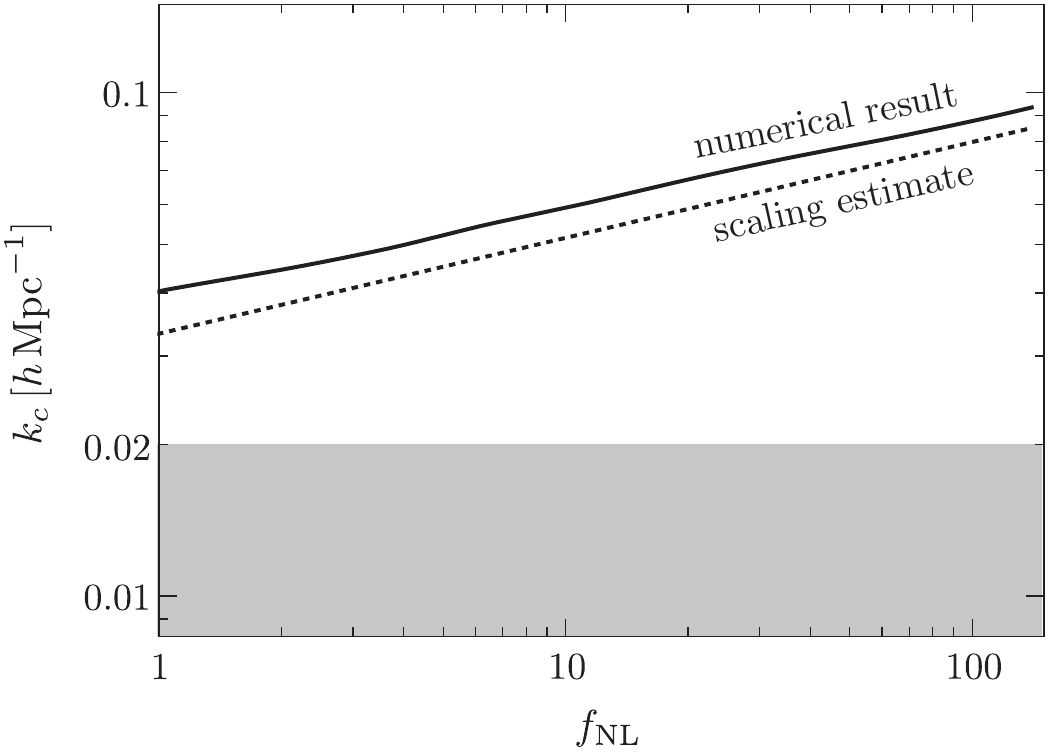}\\[14pt]
                          \includegraphics[scale=0.71]{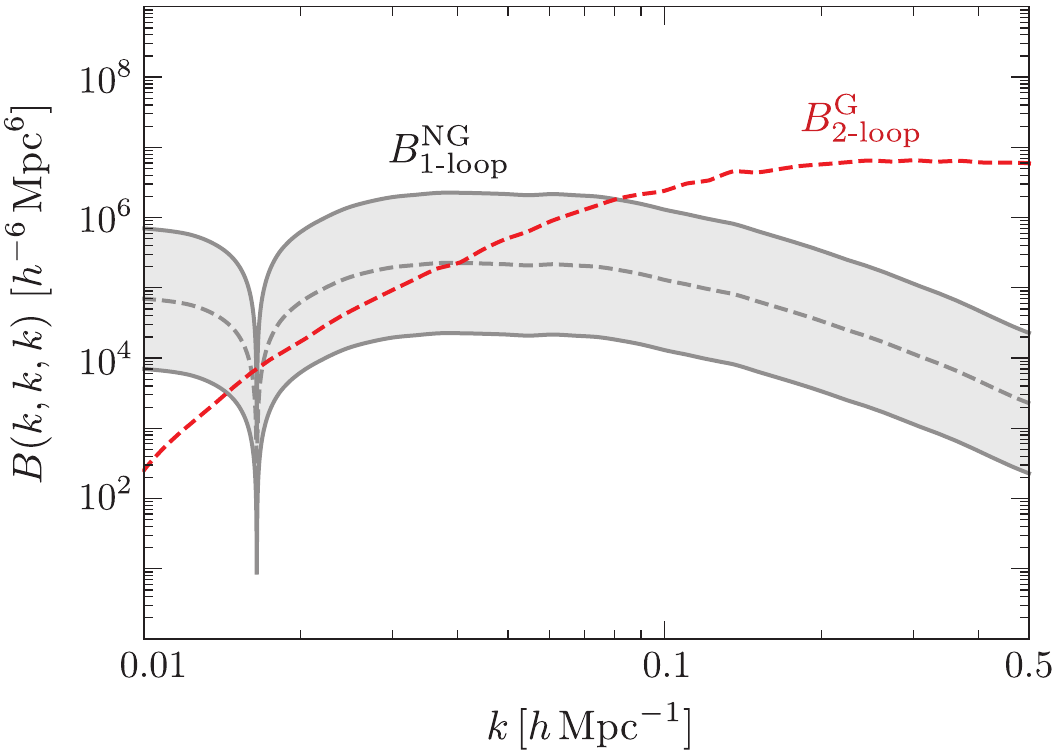} \hspace{0.5cm}  \includegraphics[scale=0.71]{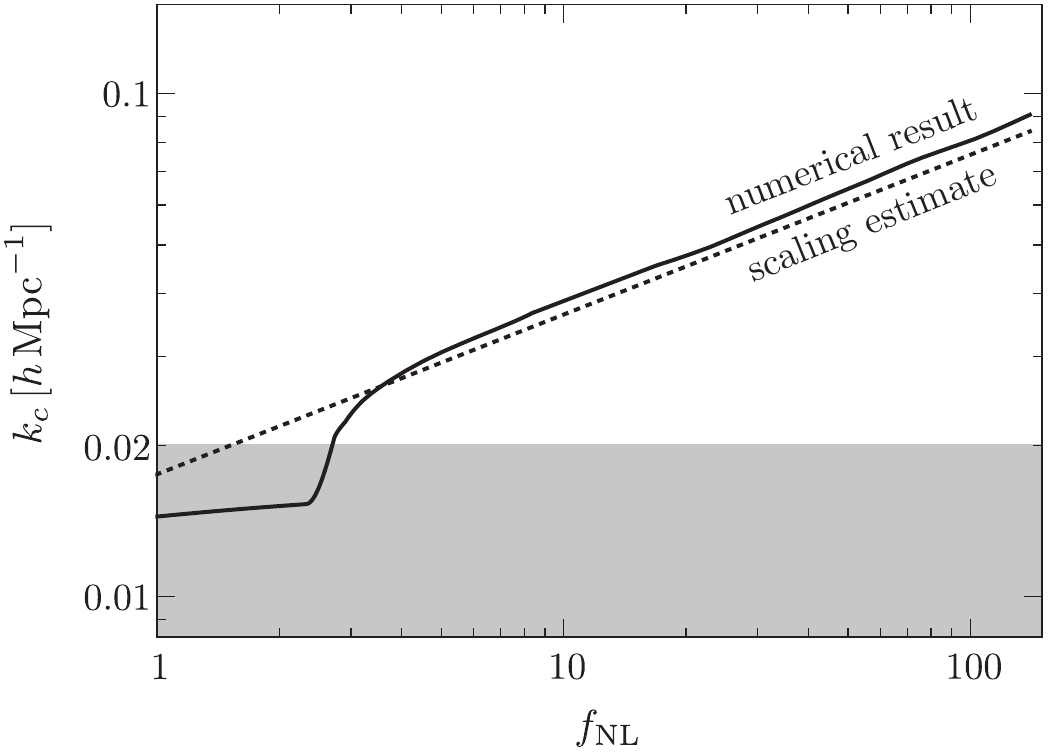}
  \caption{{\it Top}: Comparison between the tree-level contribution to the bispectrum for local PNG with  $\fnl \in [1,100]$ (grey band) and an estimate of the 2-loop Gaussian contribution (red dashed line). 
  {\it Bottom}: Comparison between the 1-loop non-Gaussian contribution (gray band) and the 2-loop Gaussian contribution (red dashed line). 
  The curves in the right plots have the same meaning as in fig.~\ref{fig:Scaling-PS}.  }\label{fig:Scaling-B1}
 \end{figure}

\vskip 4pt
Since the non-Gaussian contributions to the bispectrum already appear at tree level, they are less suppressed than in the power spectrum.
Again, we should ask up to which point we can neglect the Gaussian two-loop corrections. To estimate the two-loop corrections, we will consider one representative contribution, namely $B_{233}^{(\rm I)}$~\cite{Angulo:2014tfa}:

\beq
B_{\text{2-loop}}^{\rm G} \ \subset\ B_{233}^{\rm (I)} \ = \ \raisebox{-36.5 pt}{\includegraphics[scale=0.8]{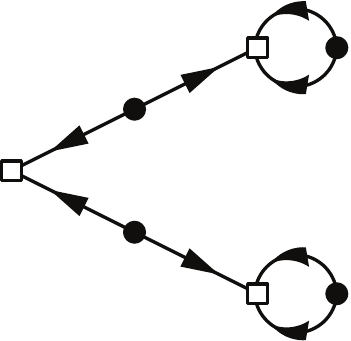}}
\eeq

\vskip 6pt
\noindent
We can then determine the scales at which the two-loop bispectrum starts to become important relative to tree-level bispectrum $B_{111}$ and the non-Gaussian one-loop contribution~$B_{\text{1-loop}}^{\rm NG}$:
\begin{align}
\frac{B_{\text{2-loop}}^{\rm G}}{B_{\text{tree}}^{\rm NG}} &\, \sim\, \frac{1}{\fnl\Delta_\varphi}\frac{s_{233}^{(\rm I)}}{s_{111}}\left(\frac{k_c}{\knl}\right)^{\frac{5}{2}(n+3)} = 1 \quad \ \ \, \longrightarrow  \quad k_c \approx 0.32\hskip 1pt\knl  \left(\frac{\fnl}{10}\right)^{1/5}\ , \\[6pt]
\frac{B_{\text{2-loop}}^{\rm G}}{B_{\text{1-loop}}^{\rm NG}} &\,\sim\, \frac{1}{\fnl\Delta_\varphi}\frac{s_{233}^{(\rm I)}}{s_{\text{1-loop}}^{\rm NG}}\left(\frac{k_c}{\knl}\right)^{\frac{3}{2}(n+3)} = 1 \quad \longrightarrow  \quad k_c \approx 0.30\hskip 1pt\knl \left(\frac{\fnl}{10}\right)^{1/3} \ . \label{eq:kc2}
\end{align}
As before, we have used $n\simeq -0.9 $ to estimate the numerical values of $k_c(\fnl)$. For the numerical prefactors we have used $s_{233}^{(\rm I)}/s_{111}\simeq 0.11$ and $s_{233}^{(\rm I)}/s_{\text{1-loop}}^{\rm NG} \simeq 0.01$. In fig.~\ref{fig:Scaling-B1} these estimates are compared to the exact result of a numerical computation in $\Lambda$CDM.  Again, we find that two-loop corrections becomes relevant at relatively large scales.

\begin{figure}[h!]
        \centering
                       \includegraphics[scale=0.9]{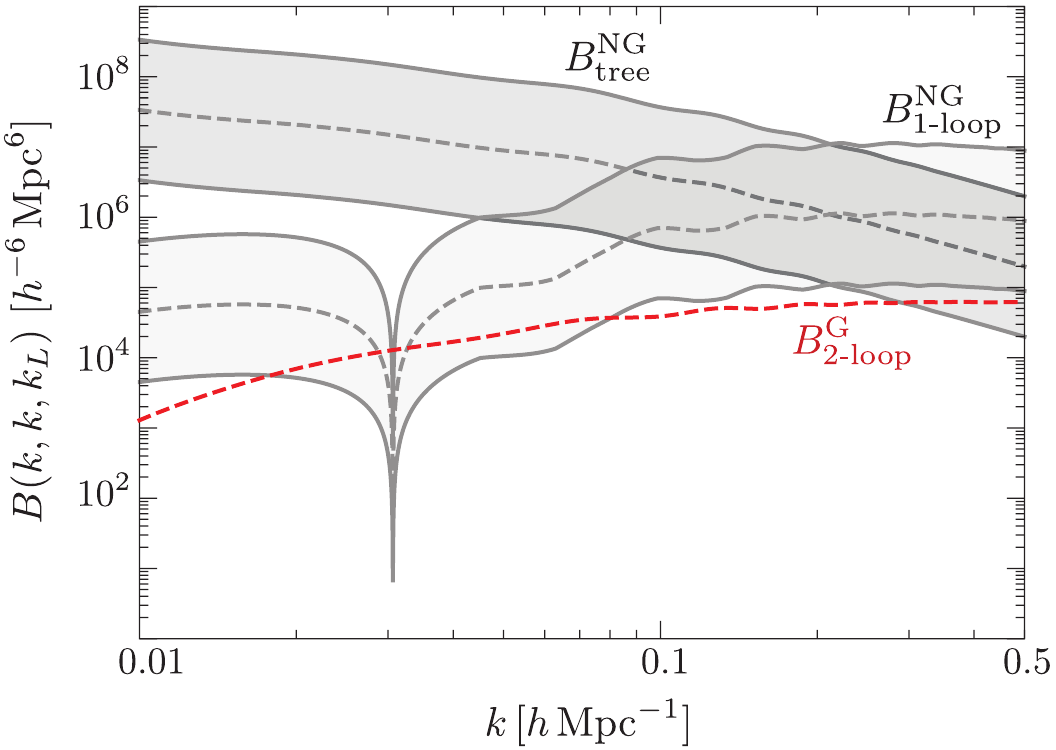}
  \caption{Comparison of the non-Gaussian tree-level and one-loop contributions with an estimate of the Gaussian two-loop contribution, evaluated for fixed $k_L \equiv 0.01\, h \hskip 1.5pt{\rm Mpc}^{-1}$.}\label{fig:Scaling-B3}
 \end{figure}

\vskip 4pt
We wish to emphasize that the importance of Gaussian two-loop corrections is maximal in the equilateral configuration.
Away from this limit the two-loop amplitude is significantly smaller. 
In fig.~\ref{fig:Scaling-B3}, we show tree-level and one-loop contributions associated with local non-Gaussianity for fixed $k_L \equiv 0.01\, h \hskip 1.5pt{\rm Mpc}^{-1}$. This time the two-loop correction is subdominant for all scales, even for relatively small values of $\fnl$.
 
\subsection{Comments on Primordial Shapes}
\label{sec:results}

The squeezed limit of the bispectrum contains interesting information about the spectrum of particles during inflation. In particular, it can reveal the masses~\cite{Chen:2009zp, Baumann:2011nk} and spins~\cite{Arkani-Hamed:2015bza} of particles to which the inflaton couples.  This requires an accurate measurement of the scaling dimension~$\Delta$ in the primordial bispectrum (\ref{eq:angSL}).  In~\cite{Sefusatti:2012ye}, the detectability of this signature was discussed for future galaxy surveys.  However, their analysis only used the tree-level form of the dark matter bispectrum, so we should ask when the nonlinear corrections discussed in this paper become important. (See~\cite{Sefusatti:2010ee} for related observations in the special case of local non-Gaussianity.)

\begin{figure}[h!]
        \centering
                 \hspace{-0.5cm}  \includegraphics[scale=0.75]{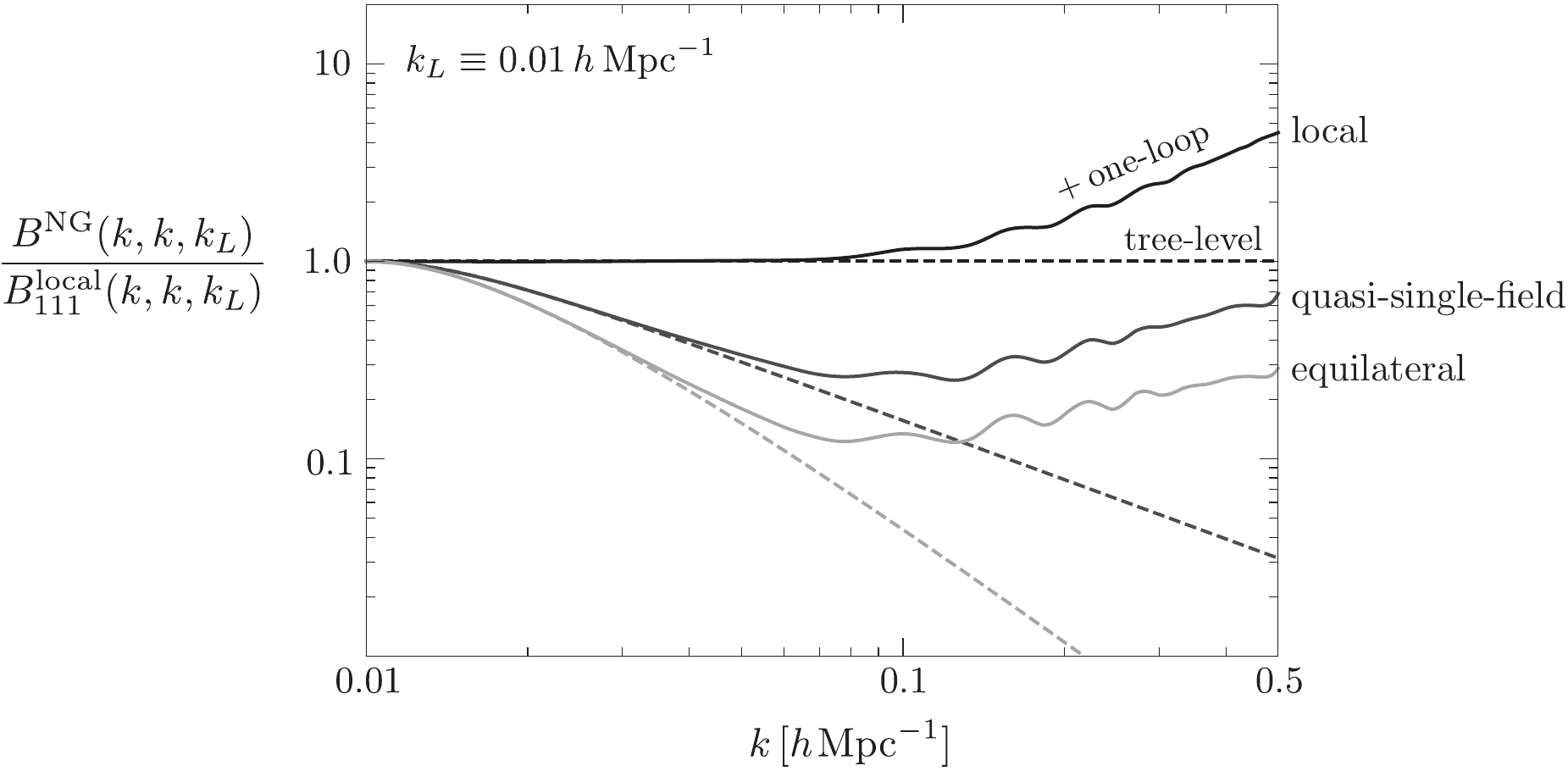}
  \caption{Comparison of the non-Gaussian SPT contribution $ B_{\mathrm{SPT}}^{\mathrm{NG}} $ for local, equilateral and QSF shape, normalized to the local tree-level bispectrum $ B_{111}^{\mathrm{local}} $. Dashed lines refer to non-Gaussian tree-level contributions, while solid lines include one-loop corrections. We see that the loop corrections become relevant on relatively large scales and tend to \textit{decrease} the difference among the three primordial signals.}
  \label{fig:Spectroscopy}
 \end{figure} 
 
 \vskip 4pt
In fig.~\ref{fig:Spectroscopy}, we show the non-Gaussian contributions to the bispectrum $ B_{\mathrm{SPT}}^{\mathrm{NG}} $, normalized with respect to the tree-level bispectrum of the local type $ B_{111}^{\mathrm{local}} $.  We have plotted three different primordial bispectra --- local, equilateral and quasi-single-field --- whose shapes were defined in~\S\ref{sec:summary}.
The dashed lines show the different scaling behavior of the tree-level contributions. We see clearly that the one-loop corrections become important at relatively large scales, especially as we go further away from the scaling of local PNG.
Moreover, the loop contributions are much less sensitive to the value of $\Delta$. In fact, the one-loop contributions for equilateral and quasi-single-field scale almost the same.
This suggests that adding the one-loop corrections is important and that it can make it harder to extract $\Delta$ (relative to the tree-level expectation).  We will study this question in more detail in~\cite{future}.

\section{Conclusions}
\label{sec:conclusions}

Measurements of primordial non-Gaussianity in the cosmic microwave background are nearly saturated --- correlations on large angular scales can't be measured more precisely because of cosmic variance, while Silk damping limits the information that can be extracted from small angular scales.  In contrast, constraints on PNG from LSS surveys are only starting to become available and have the potential to improve significantly in the future~\cite{Alvarez:2014vva}.
However, to extract this information requires pushing the theoretical understanding of gravitational clustering to smaller scales into the mildly nonlinear regime. 

\vskip 4pt
Nonlinear gravitational evolution produces two types of effects: First, it generates non-Gaussianity even if the initial conditions are perfectly Gaussian.  This inevitable ``background" needs to be characterized precisely before any non-Gaussian ``signal" can be discovered. 
Second, any primordial non-Gaussianity in the initial conditions gets distorted.  These distortions can affect how well we can distinguish between distinct types of non-Gaussianity (cf.~fig.~\ref{fig:Spectroscopy}).
Both of these effects are captured by the EFT-of-LSS.
The first was studied in~\cite{Baldauf:2014qfa, Angulo:2014tfa}, while, in this work, we have developed the tools to address the second.


\vskip 4pt
A number of open questions have been left for future work~\cite{future}:

\begin{itemize}

\item  For small levels of PNG, two-loop Gaussian corrections become comparable to the one-loop non-Gaussian corrections already on relatively large scales. (This is especially true for the power spectrum and the equilateral limit of the bispectrum.)  Extending the theoretical treatment to $k \sim 0.1\, h\hskip 1.5pt {\rm Mpc}^{-1}$ (and beyond), therefore requires a more complete treatment of the two-loop corrections for Gaussian initial conditions.  

\item The finite parts of the EFT contributions cannot be predicted, but have to be measured in N-body simulations or in observations.  For Gaussian initial conditions this one-loop matching has been performed in~\cite{Carrasco:2012cv, Baldauf:2014qfa, Angulo:2014tfa}.  The one-loop matching for non-Gaussian initial conditions is still outstanding.

\item Naively, the number of useful modes in LSS scales as the cube of the maximum wavenumber, $k_{\rm max}$, at which theory errors are still under control.  However, in practice, nonlinear evolution moves information from low-order correlations to higher-order correlations~\cite{Rimes:2005xs, Crocce:2005xz}.  It remains to be quantified how much information can actually be extracted from measurements of the dark matter bispectrum.

\item Most LSS observations only provide access to the statistics of biased tracers (galaxies, halos, etc.) of the underlying dark matter density field.  The biasing between these tracers and the dark matter introduces an additional source of nonlinearity.  On the other hand, the scale-dependence of the biasing also provides the opportunity of seeing specific imprints of the non-Gaussian initial conditions~\cite{Dalal:2007cu}.
A consistent model of nonlinear biasing requires renormalization~\cite{McDonald:2006mx, McDonald:2009dh, Schmidt:2012ys, Assassi:2014fva}.  
The first steps towards defining a self-consistent nonlinear biasing model in the presence of non-Gaussian initial conditions have appeared in~\cite{Angulo:2015eqa}.

\end{itemize}

\subsubsection*{Acknowledgements}

We thank Daniel Green, Hayden Lee and Fabian Schmidt  for discussions.
D.B.~thanks the theory groups at CERN (Geneva), IAP (Paris), IoP (Amsterdam), BCTP (Bonn) and ICTP (Trieste) for their hospitality while parts of this work were being performed. V.A.~thanks the ICTP (Trieste) for hospitality. D.B. and V.A.~acknowledge support from a Starting Grant of the European Research Council (ERC STG grant 279617). E.P.~and D.W.~are supported by the D-ITP consortium, a program of the Netherlands Organization for Scientific Research (NWO) that is funded by the Dutch Ministry of Education, Culture and Science~(OCW).
Y.W.~is supported by a de Sitter Fellowship of the Netherlands Organization for
Scientific Research (NWO).
\newpage
\appendix

\section{Odd Spin in the Squeezed Bispectrum}
\label{sec:oddspin}

In this appendix, we make a few comments about the allowed angular dependence in the squeezed limit of the bispectrum (see also~\cite{Shiraishi:2013vja,Lewis:2011au}). We will explain why, at leading order, we should expect only even-spin contributions to the squeezed limit.

\vskip 4pt
The following two types of squeezed limits are commonly used in the literature:
\begin{align}
&\lim_{k\to 0}B_\varphi(|\tfrac{1}{2}\k+\p|,|\tfrac{1}{2}\k-\p|,k)\ , \label{equ:Bi1}\\
&\lim_{k\to0}B_\varphi(p',|\p'-\k|,k)\ . \label{equ:Bi2}
\end{align} 
In this work, we have defined the Legendre expansion of (\ref{equ:Bi1}),
\beq
\lim_{k\to0}B_\varphi(|\tfrac{1}{2}\k+\p|,|\tfrac{1}{2}\k-\p|,k)= \Bigg[\sum_{L,i}a_{L,i} \left(\frac{k}{p}\right)^{\Delta_i} P_L(\hat \k\cdot\hat \p)\Bigg]\Bigg[1+{\cal O}\left(\frac{k^2}{p^2}\right)\Bigg]P_\varphi(p)P_\varphi(k)\ .\label{eq:symSL}
\eeq
Since the left-hand side is invariant under $\p\mapsto-\p$ (regardless of whether $p$ is large or not), we must have $a_{L,i}=0$ for $L$ odd.
Comparing (\ref{equ:Bi1}) and (\ref{equ:Bi2}), we find that $\p'=\p+\tfrac{1}{2}\k$, and hence
\beq
\hat\k\cdot\hat\p = \hat\k\cdot\hat\p' - \frac{k}{2p'}\left(1-(\hat\k \cdot\hat\p')^2\right)+{\cal O}\left(\frac{k^2}{(p')^2}\right)\ . \label{A4}
\eeq
Substituting (\ref{A4}) into~(\ref{eq:symSL}), we get
\beq
P_L(\hat\k\cdot\hat\p)  = P_L(\hat\k\cdot\hat\p')+\frac{k}{2p'}(L+1)\left[P_{L+1}(\hat\k \cdot\hat\p')- (\hat\k \cdot \hat \p')\hskip 1pt P_{L}(\hat\k\cdot\hat\p')\right]\ .
\eeq
This shows that we do generate Legendre polynomials of odd powers in the squeezed limit (\ref{equ:Bi2}). However, was also see that they are always subleading since they are forced to come together with a leading Legendre polynomial of even degree.

\newpage
\section{Coarse Graining in Perturbation Theory}
\label{sec:topdown}

In this appendix, we explicitly integrate out perturbative short scales in the Euler equation and show that all terms consistent with the symmetries are generated in the resulting effective theory.

\vskip 4pt
Suppose that the fluid equations are defined at a scale $\hat\Lambda<\knl$ and we wish to integrate out modes in the momentum shell $p\in[\Lambda,\hat\Lambda]$. The result is an additional stress tensor in the Euler equation~\cite{Baumann:2010tm}:
\beq
\tau^{ij} = \frac{1}{8\pi Ga^2}\big[2\partial^i\phi_s\partial^j\phi_s-\delta^{ij}(\partial_k\phi_s)^2\big]_\Lambda \,+\, \big[\rho\hskip 1pt v^i_sv^j_s\big]_\Lambda\ .\label{eq:ST}
\eeq
We wish to expand $\tau^{ij}$ in terms of the long-wavelength fields.
We will illustrate the computation for a single term 
\beq
\sigma^{ij}\equiv \big[\partial^i\phi_s \partial^j\phi_s\big]_\Lambda\ .\label{eq:sigmaij}
\eeq
For simplicity, we will work in an Einstein-de Sitter universe and compute  $\sigma^{ij}$ to first order in the long-wavelength fluctuations.

\subsection{Nonlinear Evolution}
\label{sec:evolution}

Since all modes are perturbative, $\sigma^{ij}$ can be computed in perturbation theory.
Below we show that 
\beq
\sigma^{ij}(\k,\tau)= \int_{\Lambda}^{\hat\Lambda} \frac{\d^3
\p}{(2\pi)^3}\left[p ^ip^j\sigma_\p(\k)+\int_0^\Lambda \frac{\d^3 \tilde \k}{(2\pi)^3}\, {\cal S}^{ij}_\p(\k,\tilde\k)\,\sigma_{\p}(\k-\tilde\k)\,\delta(\tilde\k,\tau) + \cdots\right]\ ,\label{eq:sigma1}
\eeq
where we have defined
\begin{align}
\sigma_\p(\k) &\equiv \big[\phi_{s}(\p+\tfrac{1}{2}\k,\tau_{in})\phi_{s}(-\p+\tfrac{1}{2}\k,\tau_{in})\big]_\Lambda\ , \\[4pt]
{\cal S}^{ij}_\p(\k,\tilde\k) &\equiv  p^ip ^j\left[\frac{20}{7}+\frac{(\k-\tilde\k)\cdot\tilde\k}{\tilde k^2}-\frac{20}{7}\frac{(\p\cdot \tilde\k)^2}{p^2\tilde k^2}\right] + \left[p^i\tilde k ^j+p^j\tilde k ^i\right]\frac{\p\cdot\tilde\k}{\tilde k^2}\ .\label{eq:Skernel}
\end{align}
The first term in~(\ref{eq:sigma1}) simply arises from the linear evolution. The second term comes from replacing one of the $\phi$ by its second-order solution $\phi_{(2)}$ and extracting the dependence on the long-wavelength fluctuations.  We give the details of the derivation in the following insert.

\vskip 10pt
\begin{framed}
\noindent
\small
{\it Derivation.}\ \  
The first-order contribution to~(\ref{eq:sigmaij}) is
\begin{align}
\frac{4}{9\H^{4}}\big[\sigma^{ij}\big]_{(1)}
&=\ a^3\int_{\Lambda}^{\hat\Lambda}  \frac{\d^3
\p}{(2\pi)^3}\, \Pi^{ij}(\p,\k)\Big[\delta_{1}(-\p+\tfrac{1}{2}\k)\delta_{2}(\p+\tfrac{1}{2}\k)+\delta_{2}(-\p+\tfrac{1}{2}\k)\delta_{1}(\p+\tfrac{1}{2}\k)\Big]\ ,
\end{align}
where we have used $- k^2 \phi(\k) = \frac{3}{2} \H^2 \delta(\k)$ 
and defined
\beq
\Pi^{ij}(\p,\k)\equiv\frac{(p^i-\tfrac{1}{2}k^i)(p^j+\tfrac{1}{2}k^j)}{(\p-\tfrac{1}{2}\k)^2(\p+\tfrac{1}{2}\k)^2}\ .
\eeq
Substituting~(\ref{dn}) for $\delta_{2}$, we get
\beq
\frac{4}{9\H^{4}}\big[\sigma^{ij}\big]_{(1)} = 2\hskip 1pta^3\int_{\Lambda}^{\hat\Lambda} \frac{\d^3
\p}{(2\pi)^3} \int_{\tilde\k}\ \Pi^{ij}(\p,\k)F_2(\tilde\k,\p+\tfrac{1}{2}\k-\tilde\k)\,\delta_{1}(-\p+\tfrac{1}{2}\k)\delta_{1}(\p+\tfrac{1}{2}\k-\tilde\k)\delta_{1}(\tilde\k)\ .
\eeq
\newpage
\noindent
There are two different regions of integration for $\tilde\k$\hskip 1pt:
\begin{enumerate}[(i)]
\item The UV region where all three fluctuations $\delta_1$ are short.
This contributes at cubic orders in an expansion of the short-scale fluctuations.
The dependence of these short scales on long-wavelength fluctuations is captured by a primordial trispectrum. Since we are working at first order in non-Gaussianity, we will not keep track of these contributions in the rest of this appendix.

\item The region where one of the modes is a long-wavelength fluctuation. This happens either for $|\tilde \k|<\Lambda$ or for $|\tilde\k-\p-\tfrac{1}{2}\k|<\Lambda$. We assume that these two regions do not overlap substantially and therefore consider them as disjoint. (The error made in using this assumption is captured by higher-derivative terms). Since both of these regions give exactly the same contribution to $\sigma^{ij}$, we focus on $\tilde k<\Lambda$ and multiply the result by two.
\end{enumerate}
Extracting the long-wavelength mode, the first-order contribution to $\sigma^{ij}$ becomes\begin{align}
\frac{4}{9\H^4}\big[\sigma^{ij}\big]_{(1)}&\,\simeq\, 4\hskip 1pta^3\int_0^\Lambda \frac{\d^3 \tilde \k}{(2\pi)^3} \bigg[\ \int_{\Lambda}^{\hat \Lambda} \frac{\d^3  \p}{(2\pi)^3}  \, \Pi^{ij}(\p,\k)F_2(\tilde\k,\p+\tfrac{1}{2}\k-\tilde\k)  \nonumber\\[-4pt]
&\hspace{4.5cm} \times \delta_{1}(-\p+\tfrac{1}{2}\k)\delta_{1}(\p+\tfrac{1}{2}\k-\tilde\k)\,\bigg]\,\delta_{1}(\tilde\k)\ .
\end{align}
Next, we shift the variable of integration, $\p\mapsto \p+\tfrac{1}{2}\tilde\k$, and express the two short fluctuations $\delta_1$ in terms of the initial potential $\phi_s(\tau_{in})$. Expanding the integrand in the limit of large $\p$, we get the answer in~(\ref{eq:sigma1}).  \hfill $\square$
\end{framed}

It is instructive to go back to position space. 
Eq.~(\ref{eq:sigma1}) then becomes
\beq
\sigma^{ij}(\x,\tau) = c_0^{ij}(\x,\tau) + c_{1}^{ij}(\x,\tau)\delta(\x,\tau) + c^{ij}{}_{kl}(\x,\tau)\,\partial^k\partial^l\phi(\x,\tau) + \cdots\ ,\label{eq:sigmaexp}
\eeq
which is to be compared to the structure of the stress tensor in~(\ref{eq:effstresstensor}).
Up to corrections of order $\phi_s^3$, we have 
\begin{align}
c_{a}^{ij}(\x,\tau)&= c_{a}\hskip 1pt[\partial^i\phi_s(\q)\partial^i\phi_s(\q)]_\Lambda \ , \label{eq:cij}\\
\hskip  1ptc^{ij}{}_{kl}(\x,\tau)&=-\frac{40}{21\H^2}[\partial^i\partial^j\phi_s(\q)\partial_{k}\partial_l\partial^{-2}\phi_s(\q)]_\Lambda+\frac{1}{3\H^2}\left([\partial^i\phi_s(\q)\partial_k\phi_s(\q)]_\Lambda \hskip 1pt \delta^{j}_{l}+ \text{perms}\right)\, ,\label{eq:cijkl}
\end{align}
where $\q(\x,\tau)$ is the Lagrangian position of the Eulerian coordinate $\x$, $a\in\{0,1\}$ and $c_a$ are unimportant numerical factors. Let us make a few comments:
\begin{itemize}
\item We see that the coefficients are in general non-local functions of $\phi_s$ --- in this example, this is the case only for the coefficient $c^{ij}{}_{kl}$. Since the inverse Laplacian in~(\ref{eq:cijkl}) acts only on short-scale fluctuations, $c^{ij}{}_{kl}$ is a non-local function of $\phi_s$ within a region of size $\Lambda^{-1}$ centered around the Lagrangian coordinate $\q$.
\item Some terms such as $\theta$ and $\partial^iv^j$ seem to be missing in~(\ref{eq:sigmaexp}). This is just an artefact of working to lowest order in perturbation theory, where these terms are indistinguishable from $\delta$ and $\partial^i\partial^j\phi$.
\end{itemize}

\normalsize
Eq.~(\ref{eq:sigma1}) shows how the late-time $\sigma^{ij}(\x,\tau)$ depends on the long-wavelength fluctuations and the initial short-scale fluctuations. So far, this is very general, in the sense that we haven't specified the initial conditions for the short scales. Next, we will determine how these initial fluctuations are modulated by long-wavelength fluctuations in the presence of PNG.

\subsection{Non-Gaussian Initial Conditions}
\label{sec:IC}

To determine the dependence of $\sigma_{\p}(\k)$ on the long-wavelength fluctuations, we average~(\ref{eq:sigma1}) over the short scales. This boils down to replacing $\sigma_\p(\k)$ by $\vev{\sigma_\p(\k)}_s$.  We then substitute~(\ref{eq:1ptshort}) into the expression of $\vev{\sigma_\p(\k)}_s$. The result receives contributions from different spins
\beq
\vev{\sigma_\p(\k)}_s =\bar\sigma_\p(\k)+ \fnl\big[\vev{\sigma_\p(\k)}_s^{[0]} +\vev{\sigma_\p(\k)}_s^{[2]} +\vev{\sigma_\p(\k)}_s^{[4]}\big] \ ,
\eeq
where $\bar\sigma_\p(\k)\propto \delta(\k)$ is the Gaussian contribution (which is independent of the long-wavelength fluctuations) and $\vev{\sigma_\p(\k)}_s^{[L]}$ represent the spin-$L$ contribution
\begin{align}
\vev{\sigma_\p(\k)}_s^{[0]} &= \left({\mu/p}\right)^{\Delta}P_\phi(p)\times \, \psi(\k)\ ,\\[4pt]
\vev{\sigma_\p(\k)}_s^{[2]} &= \left({\mu/p}\right)^{\Delta}P_\phi(p)\hskip 2pt(\hat p_i \hat p_j)\times\, \psi^{ij}(\k)\ ,\\[4pt]
\vev{\sigma_\p(\k)}_s^{[4]} &= \left({\mu/p}\right)^{\Delta}P_\phi(p)\hskip 2pt(\hat p_i \hat p_j\hat p_k\hat p_l)\times\, \psi^{ijkl}(\k)\ ,
\end{align}
where the fields $\psi$, $\psi^{ij}$ and $\psi^{ijkl}$ were defined in eqs.~(\ref{equ:psi}), (\ref{psiij}) and (\ref{eq:psiijkl}), respectively.
As explained in the main text, higher-spin contributions are captured by higher-order tensors which can only contribute at higher order in the fluctuations and/or derivatives. For this reason, we did not consider them in this work.

\subsection{EFT Operators}
\label{sec:operators}

We split $\sigma^{ij}$ into Gaussian and non-Gaussian contributions
\beq
\vev{\sigma^{ij}}_s=\vev{\sigma^{ij}}_s^{\rm G}+\vev{\sigma^{ij}}_s^{\rm NG}\ .
\eeq
To obtain the Gaussian contribution we replace $\vev{\sigma_\p(\k)}_s$ by its Gaussian contribution $\bar\sigma_\p(\k)$. The non-Gaussian part gets contributions from the different spins:
\beq
\vev{\sigma^{ij}}_s^{\rm NG} =\fnl\big[\vev{\sigma^{ij}}_s^{[0]} +\vev{\sigma^{ij}}_s^{[2]} +\vev{\sigma^{ij}}_s^{[4]}\big]\ .
\eeq
To compute the spin-$L$ contribution to $\vev{\sigma^{ij}}_s$, we replace $\sigma_\p(\k)$ in (\ref{eq:sigma1}) by $\vev{\sigma_\p(\k)}_s^{[L]}$.  Going to real space, we find 
\begin{align}
\vev{\sigma^{ij}}_{s}^{[0]} &=\frac{1}{3}\beta \bigg[\Big(\Psi + \frac{16}{7}\Psi\delta\Big)\delta^{ij} + \frac{6}{7}\Psi\partial^i\partial^j\Phi\bigg]\ , \\[5pt]
\vev{\sigma^{ij}}_{s}^{[2]}  &=\frac{2}{15}\beta \bigg[\Psi^{ij}+\frac{120}{49}\Psi^{ij}\delta -\frac{20}{49}\Psi^{kl}\partial_k\partial_l\Phi\,\delta^{ij} + \frac{9}{49}\Psi^{k(i}\partial^{j)}\partial_k\Phi\bigg]\ ,\\[5pt]
\vev{\sigma^{ij}}_{s}^{[2]}  &=- \frac{32}{441}\beta\, \Psi^{ijkl}\partial_l\partial_k\Phi\ ,
\end{align}
where we have defined the coefficient
\beq
\beta(\Lambda,\hat\Lambda) \equiv \int_\Lambda^{\hat\Lambda}\frac{\d p}{2\pi^2}\,p^4 \frac{P_\phi(p)}{(p/\mu)^\Delta}\ .
\eeq
We see that all terms consistent with the symmetries are generated in the effective theory --- see eqs.~(\ref{eq:tauNG0V2}), (\ref{eq:tauNG2}) and (\ref{eq:tauNG4}).

\newpage
\section{Perturbation Theory and Counterterms}
\label{sec:SPT-Appendix}

In this appendix, we collect a few results from standard perturbation theory (SPT) and derive explicit expressions for the one-loop counterterms in the EFT-of-LSS.

\subsection{Equations of Motion}
The equations of motion for the density contrast $\delta$ and the velocity divergence $\theta$ are
\begin{align}
\partial_\tau \delta+ \theta &\,=\, {\cal S}_\alpha \ ,\\
 (\partial_\tau + \H) \theta + \frac{3}{2}\Omega_m \H^2 \delta &\,=\, {\cal S}_\beta + \tau_{\theta} \ , \label{equ:Eqb}
\end{align}
where $\tau_\theta\equiv-\partial_i\big[\rho^{-1}\partial_j\tau^{ij}\big]$, and ${\cal S}_{\alpha,\beta}$
are nonlinear source terms
\begin{align}
{\cal S}_\alpha(\k,\tau) &\equiv - \int_\p \alpha(\p,\k-\p) \theta(\p,\tau) \delta(\k-\p,\tau)\ , \qquad \alpha(\k_1,\k_2) \equiv \frac{\k_1\cdot(\k_1+\k_2)}{k_1^2} \ , \label{Sa}\\
{\cal S}_\beta(\k,\tau) &\equiv - \int_\p \beta(\p,\k-\p) \theta(\p,\tau) \theta(\k-\p,\tau) \ , \qquad \beta(\k_1,\k_2) \equiv \frac{(\k_1+\k_2)^2}{2}\frac{\k_1\cdot \k_2}{k_1^2 k_2^2} \ . \label{Sb}
\end{align}
Using the scale factor $a(\tau)$ as the evolution variable, the equations of motion become
\begin{align}
\overbrace{\H^2 \left[-a^2 \partial_a^2 + \left(\frac{3}{2}\Omega_m -3\right) a \partial_a + \frac{3}{2}\Omega_m \right]}^{\displaystyle {\cal D}_\delta} \delta  &\,=\, {\cal S}_\beta  + \tau_{\theta} - \H \partial_a(a{\cal S}_\alpha)\ , \label{equ:E1}\\[4pt]
\underbrace{\H^2 \left[+a^2 \partial_a^2 + \left(4- \frac{3}{2}\Omega_m \right) a \partial_a + (2-3\Omega_m) \right]}_{\displaystyle {\cal D}_\theta} \theta  &\,=\, \partial_a(a {\cal S}_\beta  + a\tau_{\theta}) - \frac{3}{2}\Omega_m\H \hskip 1pt{\cal S}_\alpha\ . \label{equ:E2}
\end{align}
The linearized equation of motion for $\delta$ (obtained by setting the right-hand side of~(\ref{equ:E1}) to zero) has the following growing mode solution 
\beq
\delta_{(1)}(\k,a) = D_{1}(a) \delta_{1}(\k)\ , \label{dSPT1}
\eeq
where $ \delta_{1}(\k)$ 
describes the initial condition and $D_1(a)$ is the linear growth factor
\beq
D_1(a) = \frac{5}{2}\H_0^2 \Omega_m^0  \frac{\H}{a} \int_{a_{in}}^a \frac{\d a'}{\H^3(a')}\ . \label{D1}
\eeq
In Einstein-de Sitter, the result in (\ref{D1}) reduces to $D_1(a) = a/a_{in}$.

\vskip 4pt
Solving (\ref{equ:E1}) with a delta-function source, $\delta_D(a-a')$, gives the Green's function for the evolution of $\delta$:
\begin{align}
G_\delta(a,a') &\,=\, \Theta(a-a')\, \frac{2}{5} \frac{1}{\H_0^2 \Omega_m^0} \frac{D_1(a')}{a'} \left[\frac{D_-(a)}{D_-(a')} - \frac{D_1(a)}{D_1(a')} \right] \ , \label{eq:Gdelta}
\end{align}
and $D_- \equiv \H/(a\H_0) \approx D_1^{-3/2}$. 
A similar Green's function $G_\theta(a,a')$ exists for $\theta$, but it won't be needed in this work.
To a remarkably good approximation, we have
\begin{align}
\int_{a_{in}}^a\d a'\ G_\delta(a,a')\H^2(a')f^2(a')[D_1(a')]^n&\approx -\frac{2}{(n-1)(2n+3)}[D_1(a)]^n\label{eq:approx1}\ ,
\end{align}
where $f \equiv d\ln D_1/d\ln a$. The result in (\ref{eq:approx1}) will be useful below. We have checked that  this approximation is valid to better than 2\% accuracy, for the values of $n$ considered in this paper.

\subsection{Perturbative Solution}
\label{sec:perturb}

For $\delta_{(1)} < 1$, the solution to the nonlinear equations can be written as a series in powers of the initial dark matter contrast $\delta_1$ (and integrals over the Green's function $G_\delta$):
\beq
\delta(\k,a) = \sum_{n=1}^\infty \delta_{(n)}(\k,a) \ , \qquad \theta(\k,a) = - \H f(a) \sum_{n=1}^\infty \theta_{(n)}(\k,a)\ .
\eeq
For a non-vanishing stress tensor in (\ref{equ:Eqb}), the $n$-th order solution can be written as
\beq
\delta_{(n)}(\k,a)=\delta^{{\rm SPT}}_{(n)}(\k,a)+\delta^c_{(n)}(\k,a)+\delta^{J}_{(n)}(\k,a)\ ,\label{eq:deltan}
\eeq
where $\delta^{{\rm SPT}}$ is the SPT result obtained with $\tau_\theta \equiv \tau_v + \tau_n =0$, $\delta^c$ is the solution sourced by the viscosity part of the stress tensor $\tau_v$, while $\delta^J$ is sourced by the noise component of the stress tensor $\tau_n$. 
We first review the SPT solution (\S\ref{sec:SPT}) and then derive the expression for the one-loop counterterms (\S\ref{sec:ctr1} and \S\ref{sec:ctr2}).

\subsubsection{SPT Solution}
\label{sec:SPT}

The $n$-th order SPT solution can be written in terms of a convolution of the Green function~(\ref{eq:Gdelta}) and lower-order SPT solutions. 
Using (\ref{eq:approx1}), one finds that the $n$-th order solution is proportional to the $n$-th power of the linear growth factor $D_1$: 
\beq
\delta_{(n)}^{{\rm SPT}}(\k,a) \approx D_1^n(a) \delta_n(\k) \ .\label{equ:dSPT1}
\eeq
A similar result holds for the velocity divergence
\beq
\theta_{(n)}^{{\rm SPT}}(\k,a) \approx D_1^n(a) \theta_n(\k)\ .
\eeq
The initial conditions $\delta_n$ and $\theta_n$ can be written as a convolution of powers of $\delta_1$:
\begin{align}
\delta_n^{{\rm SPT}}(\k) &\,=\, \int_{\k_1} \ldots \int_{\k_n} (2\pi)^3 \delta_D\big(\k - \k_{1  \ldots n}\big) \, F_n(\k_1,\ldots, \k_n) \, \delta_1(\k_1) \ldots \delta_1(\k_n)\ ,  \label{dn} \\
\theta_n^{{\rm SPT}}(\k) &\,=\, \int_{\k_1} \ldots \int_{\k_n} (2\pi)^3 \delta_D\big(\k - \k_{1  \ldots n}\big) \, G_n(\k_1,\ldots, \k_n)\, \delta_1(\k_1) \ldots \delta_1(\k_n)\ , \label{tn}
\end{align}
where $\k_{1\ldots n} \equiv \k_1 + \cdots + \k_n$.
Explicit expressions for the kernel functions $F_n$ and $G_n$  can be found in~\cite{Bernardeau:2001qr}. For instance, $F_1=G_1=1$, while the (symmetrized) second-order kernel functions are
\begin{align}
F_2(\k_1,\k_2) &= \frac{5}{7}+\frac{1}{2}\left(\frac{\k_1\cdot\k_2}{k_1^2}+\frac{\k_1\cdot\k_2}{k_2^2}\right)+\frac{2}{7}\frac{(\k_1\cdot\k_2)^2}{k_1^2k_2^2}\ , \label{F2} \\
G_2(\k_1,\k_2) &= \frac{3}{7}+\frac{1}{2}\left(\frac{\k_1\cdot\k_2}{k_1^2}+\frac{\k_1\cdot\k_2}{k_2^2}\right)+\frac{4}{7}\frac{(\k_1\cdot\k_2)^2}{k_1^2k_2^2}\ .
\end{align}

\subsubsection{Viscosity Counterterms}
\label{sec:ctr1}

In order to renormalize the one-loop bispectrum, we need to compute $\delta^{c}$ and $\delta^{J}$ up to second order. This requires knowing the stress tensor up to second order.
First, let us focus on the viscosity contribution to the stress tensor 
\begin{align}
\, \tau_{v}^{(1)}&\equiv -d^2(a) \hskip -1pt \bigtriangleup \hskip -1pt \delta_{(1)} - g(a)\fnl\hskip -1pt \bigtriangleup \hskip -2pt\Psi_{(1)} \ ,\\[4pt]
\, \tau_{v}^{(2)}&\equiv -d^2(a) \hskip -1pt \bigtriangleup \hskip -1pt\delta_{(2)} -e_1(a) \hskip -1pt \bigtriangleup \hskip -1pt (\delta_{(1)})^2 -e_2(a) \hskip -1pt \bigtriangleup \hskip -1pt(s_{(1)})^2 - e_3(a)\partial_i(s^{ij}_{(1)}\partial_j\delta_{(1)} )\nonumber\\
&\ \ \ \, -\fnl\Big[g(a)\big[ \hskip -1pt \bigtriangleup \hskip -2pt\Psi_{(2)} - \partial_i(\delta_{(1)}\partial^i\Psi_{(1)})\big]+g_1(a) \hskip -1pt \bigtriangleup \hskip -1pt(\Psi_{(1)}\,\delta_{(1)})+ g_2(a)\partial_{i}\partial_{j}(s^{ij}_{(1)}\Psi_{(1)})\Big]\ , \label{A21}
\end{align}
where $\delta_{(1,2)}$ and $s^{ij}_{(1)}\equiv \partial^i\partial^j\Phi_{(1)}-\frac{1}{3}\delta^{ij}\delta_{(1)}$ refer to the SPT solutions and {\it not} the full solution~(\ref{eq:deltan}). We have defined the parameter
$d^2\equiv c_s^2+f(c_{vis}^2+\hat c_{vis}^2)$ as the sum of the sound speed and the viscosity parameter (for more details, see \cite{Baldauf:2014qfa, Angulo:2014tfa}). Note that the velocity divergence~$\theta$ does not appear in these equations because at second order it is completely degenerate with $\delta$.  The field $\Psi(\x,\tau)\equiv\psi(\q(\x,\tau))$ admits an expansion in powers of the fluctuations, with $\Psi_{(1)}\equiv\psi$ and $\Psi_{(2)}\equiv\D\psi\cdot\D\Phi$; cf.~eq.~(\ref{eq:Psiexp}). 
\vskip 4pt
The $n$-th order counterterms $\delta^c_{(n)}$ can be written as
\begin{align}
\delta^c_{(n)}(\k,a) &= \int_{\k_1}\ldots\int_{\k_n}(2\pi)^3\delta_D\big(\k - \k_{1  \ldots n}\big)\,F_n^{c}(\k_1,\ldots,\k_n|a)\,\delta_{(1)}(\k_1,a)\ldots\delta_{(1)}(\k_n,a)
\nonumber\\
 			&\  + \fnl\int_{\k_1}\ldots\int_{\k_n}(2\pi)^3\delta_D\big(\k - \k_{1  \ldots n}\big)\,H_n^{c}(\k_1,\ldots,\k_n|a)\,\psi(\k_1)
			\ldots\delta_{(1)}(\k_n,a)  \ .
\end{align}
where $F_n^{c}$ and $H_n^{c}$ are kernel functions that are to be determined. In what follows, we will compute these kernels up to second order.

\subsubsection*{First order}

The first-order counterterm $\delta^c_{(1)}$ is the solution to ${\cal D}_\delta\, \delta^c_{(1)}=\tau_v^{(1)}$,
and can therefore be written as 
\begin{align}
\delta^c_{(1)}(\k,a)&= -\xi(a)\hskip 1pt k^2\delta_{(1)}(\k,a) -\gamma(a)\fnl\hskip 1pt k^2\Psi_{(1)}(\k) \ , \label{dc1}
\end{align}
where we have defined
\begin{align}
\xi(a)& \equiv-\frac{1}{D_1(a)}\int_{a_{in}}^a\d a'\ G_\delta(a,a') \hskip 2pt d^2( a') D_1( a') \label{xi}\ , 
 \\[4pt]
\gamma(a) &\equiv-\int_{a_{in}}^a\d a'\ G_\delta(a,a') \hskip 2pt g( a')\ . \label{gamma}
\end{align}
We see that the parameters $\xi$ and $\gamma$ will depend on the time dependence of $d^2(a)$ and $g(a)$. For the one-loop power spectrum, this is not very important, since in the end we just need to fit the values of $\xi$ and $\gamma$ at a given redshift. However, as we shall see in the next section, the second-order counterterm $\delta^c_{(2)}$ depends on the time dependence of these parameters.
A convenient ansatz for the time dependence of the parameters in (\ref{xi}) and (\ref{gamma}) is
\begin{align}
d^2(a) &= [\H(a)f(a)]^2[D_1(a)]^{m_d+1}\,\bar d^{\hskip 1.5pt 2}\ , \\[4pt]
g(a) &= [\H(a)f(a)]^2[D_1(a)]^{m_g+1}\,\bar g\ , \label{eq:timeansatz}
\end{align}
where $\bar d$ and $\bar g$ are constants.
Using (\ref{eq:approx1}), we then have
\begin{align}
\xi(a) &= \frac{2}{(m_{d}+1)(2m_{d}+7)}[D_1(a)]^{m_{d}+1}\,\bar d^{\hskip 1.5pt 2}\ ,\label{eq:gamma}\\
\gamma(a) &= \frac{2}{m_g(2m_g+5)}[D_1(a)]^{m_g+1}\,\bar g\ .\label{eq:beta1}
\end{align}

\subsubsection*{Second order}

At second order, things are a bit more complicated, since the second-order solution will depend on the precise time dependence of the first-order solution $\delta_{(1)}^c$. Consider the equation of motion for $\delta_{(2)}^c$:
\beq
{\cal D}_\delta\,\delta^c_{(2)}={\cal S}_\beta^{(2)} + \tau_v^{(2)} - \H \partial_a(a{\cal S}_\alpha^{(2)})\ ,
\eeq
where ${\cal S}_{\alpha}^{(2)}$ and ${\cal S}_\beta^{(2)}$ are obtained by replacing one of the $\delta$ or $\theta$ in the convolution by their linear SPT solution and the other by the corresponding linear counterterm. 
Using (\ref{equ:dSPT1}) and (\ref{dc1}) for $\delta^{\rm SPT}_{(1)}$ and $\delta^c_{(1)}$, respectively, and replacing by $\theta^{\rm SPT}_{(1)}$ and $\theta^{c}_{(1)}$ by $\theta_{(1)} =-\dot\delta_{(1)}$, the solution for $\delta^c_{(2)}$ can be written as
\begin{align}
\delta^{c,{\rm G}}_{(2)}(\k,a) &=\int\limits_{\p}F_2^{c}(\p,\k-\p|a)\,\delta_{(1)}(\p,a)\delta_{(1)}(\k-\p,a) \ , \\
\delta^{c,{\rm NG}}_{(2)}(\k,a) &=\fnl \int\limits_{\p}H_2^{c}(\p,\k-\p|a)\,\psi(\p)\delta_{(1)}(\k-\p,a)\ .
\end{align}
We have separated the solution into a Gaussian part $\delta^{c,{\rm G}}_{(2)}$ and a non-Gaussian term $\delta^{c,{\rm NG}}_{(2)}$. We look at each of these terms in turn.

\begin{itemize}
\item {\bf Gaussian contributions}

The kernel functions $F_2^{c}$ were computed in \cite{Carrasco:2013mua,Angulo:2014tfa,Baldauf:2014qfa}:
\beq
F_2^{c}(\k_1,\k_2|a)=F_2^{\tau}(\k_1,\k_2|a)+F_{2}^{\alpha\beta}(\k_1,\k_2|a)+F_2^{\delta}(\k_1,\k_2|a)\ ,\label{eq:F2c}
\eeq
where $F_2^{\tau}$ is sourced by the nonlinear terms in $\tau_v^{(2)}$, $F_2^{\alpha\beta}$ is sourced by ${\cal S}_{\alpha,\beta}$ and $F_2^{\delta}$ is sourced by the second-order SPT solution $\delta_{(2)}^{{\rm SPT}}$ which appears in the stress tensor (\ref{A21}) as $\tau_v^{(2)}\supset - d^2 \hskip -1pt \bigtriangleup \hskip -1pt\delta_{(2)}^{\rm SPT}$.

\begin{itemize}\renewcommand{\labelitemii}{$\circ$}
\item The kernel $F_2^{\tau}$ can be written as
\beq
F_2^{\tau}(\k_1,\k_2|a)= -\sum_{i=1}^3\epsilon_i(a) E_{i}(\k_1,\k_2)\ ,
\eeq
where the time-dependent coefficients are
\beq
\epsilon_i(a)\equiv-\frac{1}{[D_1(a)]^2}\int_{a_{in}}^a\d a'\ G_\delta(a,a')\, [D_1(a')]^2\,e_i(a')\ . \label{equ:ei}
\eeq
and the momentum kernels are
\begin{align}
E_1(\k_1,\k_2)&\equiv \k_{12}^2\ ,\\
E_2(\k_1,\k_2)&\equiv \k_{12}^2\left[\frac{(\k_1\cdot\k_2)^2}{k_1^2k_2^2}-\frac{1}{3}\right]\ ,\\
E_3(\k_1,\k_2)&\equiv \left[-\frac{1}{6}\k_{12}^2+\frac{1}{2}\k_1\cdot\k_2\left[\frac{\k_{12}\cdot\k_2}{k_2^2}+\frac{\k_{12}\cdot\k_1}{k_1^2}\right]\right]\ .
\end{align}
\item The kernel $F_{2}^{\alpha\beta}$ can be written as
\beq
F_2^{\alpha\beta}(\k_1,\k_2|a)=-\xi(a)E_{\alpha\beta}(\k_1,\k_2)\ ,
\eeq
where $\xi(a)$ was defined in (\ref{xi}), and
\begin{equation}
\begin{aligned}
E_{\alpha\beta}(\k_1,\k_2)&\equiv\frac{1}{2m_{d}+9}\bigg[2\beta(\k_1,\k_2)(k_1^2+k_2^2) \\
&\hskip 25pt+\frac{2m_{d}+7}{2(m_{d}+2)}\Big(\alpha(\k_1,\k_2)\big(k_2^2+(m_{d}+2)k_1^2\big)
+\{1\leftrightarrow 2\}\Big)\bigg]\ .
\end{aligned}
\end{equation}
\item Finally, the kernel $F_2^{\delta}$ reads
\beq
F_2^{\delta}(\k_1,\k_2|a) = -\xi(a)E_{\delta}(\k_1,\k_2)\ ,
\eeq
where we have defined
\beq
E_{\delta}(\k_1,\k_2)=\frac{(m_{d}+1)(2m_{d}+7)}{(m_{d}+2)(2m_{d}+9)}\,\k_{12}^2\,F_2(\k_1,\k_2)\ ,
\eeq
and $F_2$ is the SPT kernel (\ref{F2}).
\end{itemize}

\item {\bf Non-Gaussian contributions}

The kernel function $H_2^{c}$ can also be written as a sum of terms
\beq
H_2^{c}(\k_1,\k_2|a)=H_2^{\tau}(\k_1,\k_2|a)+H_2^{\alpha\beta}(\k_1,\k_2|a)+H_2^{\Psi}(\k_1,\k_2|a)\ ,\label{eq:H2c}
\eeq
where $H_2^{\tau}$ and $H_2^{\alpha\beta}$ have the same meanings as before, and $H_2^{\Psi}$ is sourced by the term proportional to $g(a)$ in the stress tensor.

\begin{itemize}\renewcommand{\labelitemii}{$\circ$}
\item The kernel $H_2^{\tau}$ can be written as
\beq
H_2^{\tau}(\k_1,\k_2|a)=-\sum_{i=1}^2\gamma_i(a) G_{i}(\k_1,\k_2)\ ,
\eeq
where the time-dependent coefficients are
\beq
\gamma_i(a)\equiv-\frac{1}{D_1(a)}\int_{a_{in}}^a\d a'\ G_\delta(a,a')\,D_1(a')g_i(a')\ ,
\eeq
and the momentum kernels are
\begin{align}
G_1(\k_1,\k_2)&=\k_{12}^2\ ,\\
G_2(\k_1,\k_2)&=\frac{(\k_{12}\cdot\k_{2})^2}{k_2^2}-\frac{1}{3}\k_{12}^2\ .
\end{align}
\item The kernel $H_2^{\alpha\beta}$ can be written as
\begin{align}
H_2^{\alpha\beta}(\k_1,\k_2|a)	&=-\gamma(a) G_{\alpha\beta}(\k_1,\k_2)\ ,
\end{align}
where
\begin{equation}
\begin{aligned}
G_{\alpha\beta}(\k_1,\k_2) \equiv \,\, &\frac{4}{2m_g+7}\,\beta(\k_1,\k_2)k_1^2 \\
&+\frac{2m_g+5}{(m_g+1)(2m_g+7)}\big[(m_g+1)\alpha(\k_1,\k_2)+\alpha(\k_2,\k_1)\big]k_1^2\ .
\end{aligned}
\end{equation}
\item Finally, the kernel $H^{\Psi}_2$ reads
\beq
H_2^{\Psi}(\k_1,\k_2|a)	=-\gamma(a) G_{\Psi}(\k_1,\k_2)\ ,
\eeq
where
\beq
G_{\Psi}(\k_1,\k_2) = \frac{m_g(2m_g+5)}{(m_g+1)(2m_g+7)}\,\left[\k_{12}^2\frac{\k_1\cdot\k_2}{k_2^2}-\k_{12}\cdot\k_1\right]\ .
\eeq
\end{itemize}
\end{itemize}

\subsubsection{Noise Counterterms}
\label{sec:ctr2}

Up to second order, the noise contributions to the stress tensor are 
\begin{align}
\, \tau_{n}^{(1)}&\equiv -\partial_{i}\partial_{j}J^{ij}_0\ , \label{t1}\\[4pt]
\, \tau_{n}^{(2)}&\equiv \partial_{i}(\delta\partial_{j}J^{ij}_0) -\partial_{i}\partial_{j}(J^{ij}_1\delta_{(1)}) -\partial_{i}\partial_{j}(J^{ij}_2{}_{kl} s^{kl}_{(1)})-\fnl\partial_{i}\partial_{j}(J^{ij}_\psi\Psi_{(1)})\ . \label{t2}
\end{align}
The  $n$-th order noise counterterm $\delta^J_{(n)}$ can be written as the sum of Gaussian and non-Gaussian contributions
\beq
\delta^{J}_{(n)}(\k,a) = \delta^{J,\hskip 1pt {\rm G}}_{(n)}(\k,a) + \delta^{J,\hskip 1pt {\rm NG}}_{(n)}(\k,a)\ .
\eeq
The non-Gaussian part receives contributions from both noise terms in the initial conditions~(\ref{eq:philong}) and the noise terms in the stress tensor (\ref{eq:noise0}).
We will derive explicit expressions for the counterterms up to second order.

\subsubsection*{First order}

 Solving ${\cal D}_\delta\hskip 2pt \delta^{J,\hskip 1pt \rm G}_{(1)} = \tau_{n}^{(1)}$, gives the first-order Gaussian solution
\begin{align}
\delta_{(1)}^{J,\hskip 1pt {\rm G}}(\k,a) &=  k_i k_j\int_{a_{in}}^a\d a'\ G_{\delta}(a,a') \hskip 1pt J_0^{ij}(\k,a')\ \equiv\ k_ik_jN^{ij}_0(\k,a)\ .
\end{align}
The non-Gaussian contribution to $\delta^{J}_{(1)}$ only comes from the initial conditions. In fact, the solution is obtained by replacing $\delta_{(1)}(\k,a)\mapsto \fnl\hskip 1ptM(k)\psi_J(\k)$ in the SPT expansion. At first order, we get
\beq
\delta_{(1)}^{J,\hskip 1pt {\rm NG}}(\k,a) = \fnl\hskip 1ptM(k) \hskip 2pt \psi_J(\k)\ .
\eeq

\subsubsection*{Second order}

The second-order Gaussian contribution is obtained by solving~(\ref{equ:E1}) with $\tau_{\theta}$ replaced by the Gaussian contribution to $\tau^{(2)}_n$ and by replacing one of the $\delta$ (or $\theta$) in ${\cal S}_{\alpha,\beta}$ by their linear SPT solution $\delta^{\rm SPT}_{(1)}$ and the other by the first-order Gaussian noise contribution~$\delta^{J,\hskip 1pt \rm G}_{(1)}$ (or $\theta^{J,\hskip 1pt \rm G}_{(1)}$). Since we will not require the second-order Gaussian noise counterterm for this work, we will not explicitly compute it here and refer the reader to \cite{Baldauf:2014qfa} for an explicit expression.

The second-order non-Gaussian contribution is obtained in a similar way. The contribution coming from the initial noise term $\psi_J$ is obtained by replacing $\delta_{(1)}\mapsto \fnl\hskip 1ptM(k)\psi_J(\k)$ in the SPT expansion. 
Adding the term sourced by the non-Gaussian contribution in (\ref{t2}), we get
\beq
\delta^{J,\hskip 1pt {\rm NG}}_{(2)}(\k,a) =\fnl \int_{\p}\left[k_ik_jN^{ij}_\psi(\k-\p,a)\psi(\p)+2F_{2}(\k-\p,\p)\delta_{(1)}(\k-\p,a)M(p)\psi_{J}(\p)\right]\, ,
\eeq
where 
\beq
N^{ij}_\psi(\k,a) \equiv \int_{a_{in}}^a\d a'\ G_{\delta}(a,a')J^{ij}_\psi(\k,a')\ .
\eeq

\newpage
\section{IR-Safe Integrands}
\label{sec:IR}

The one-loop integrals can have divergences as the loop momentum $\p$ or one of the external momenta $\pm \k_i$ approach~$\0$. Although individual diagrams can be IR divergent, the equivalence principle guarantees that the sum will be finite.
To avoid delicate cancellations of large integrals (which might affect the precision of the numerical computation) it is useful to define integrands which are well-behaved in the IR. For Gaussian initial conditions, this was done for the power spectrum in~\cite{Carrasco:2013sva} and for the bispectrum in~\cite{Angulo:2014tfa, Baldauf:2014qfa}. In this appendix, we extend these results to non-Gaussian diagrams.

\subsection{Power Spectrum}

Let us first consider the non-Gaussian contribution to the one-loop power spectrum~(\ref{equ:P12})
\beq
P_{12}(k)=2\int_{\p}F_{2}(\k-\p,\p) B_{111}(k,|\k-\p|,p)\equiv\int_\p p_{12}(\p,\k)\ .
\eeq
The kernel function $F_2$ is divergent when either of its argument vanishes~\cite{Bernardeau:2001qr}, cf.~eq.~(\ref{F2}).
This means that the integrand is divergent in the limits $\p\to \0$ and $\p\to\k$. However, these IR divergences are unphysical and cancel in the integral. To make this cancelation manifest at the level of the integrand, we will first map the divergence at $\p=\k$ into a divergence at $\p=\0$.  To do so, let us split the region of integration as follows
\begin{align}
P_{12}(k)&=\int\limits_{p<|\k-\p|}p_{12}(\p,\k) \ \, +\int\limits_{p>|\k-\p|}p_{12}(\p,\k)\ .
\end{align}
Changing the integration variable in the second integral, $\p\mapsto \k-\p$, we get
\begin{align}
P_{12}(k)&=2\int_{\p} p_{12}(\p,\k)\,\Theta(|\k-\p|-p)\ ,
\end{align}
where $\Theta$ is the Heaviside function. Although this integrand no longer has a divergence at $\p=\k$, there is still a divergence at $\p=\0$. To remove this divergence, we first notice that
\beq
\lim_{\p\to\0}F_{2}(\k-\p,\p) = \frac{\k\cdot\p}{2p^{2}} + {\cal O}(p^0)\ .
\eeq
This means that the problematic divergence disappears if we make the integrand invariant under the exchange $\p\mapsto-\p$. We therefore write
\beq
 \tilde p_{12}(\p,\k) \equiv  p_{12}(\p,\k)\,\Theta(|\k-\p|-p)+p_{12}(-\p,\k)\,\Theta(|\k+\p|-p)\ .
\eeq
This integrand is IR-safe: it is well behaved in the limits $\p\to\0$ and $\p\to\pm\k$.

\subsection{Bispectrum}

In the main text, we have written the non-Gaussian one-loop bispectrum as
\beq
B^{\rm NG}_{\rm loop} = \int_\p \big[b_{311}^{(\rm I)}(\p,\k_i)+b_{311}^{(\rm II)}(\p,\k_i)+b_{122}^{(\rm I)}(\p,\k_i)+b_{122}^{(\rm II)}(\p,\k_i)+{\rm perms}\big]\ , \label{equ:BNGloop}
\eeq
where
\begin{align}
b_{311}^{(\rm I)}(\p,\k_i)&\equiv3\hskip 1pt F_{3}(\k_1+\p,-\p,\k_2)B_{111}(k_1,p,|\k_1+\p|)P_{11}(k_2)\ ,\\
b_{311}^{(\rm II)}(\p,\k_i)&\equiv3\hskip 1pt F_{3}(\k_1,\p,-\p)B_{111}(k_1,k_2,k_3)P_{11}(p)\ ,\\
b_{122}^{(\rm I)}(\p,\k_i)&\equiv4\hskip 1ptF_2(\k_3+\p,-\p)F_{2}(\p,\k_2-\p)B_{111}(k_1,|\k_3+\p|,|\k_2-\p|)P_{11}(p)\ ,\\
b_{122}^{(\rm II)}(\p,\k_i)&\equiv2\hskip 1ptF_{2}(\k_1,\k_2)F_2(\k_1-\p,\p)B_{111}(k_1,|\k_1-\p|,p)P_{11}(k_2)\ .
\end{align}
These integrands contain divergences when the loop momentum $\p$ approaches zero or $\pm\k_i$.
Following the same logic as for the power spectrum, we can write the bispectrum~(\ref{equ:BNGloop}) as an integral of two IR-safe integrands
\beq
B^{\rm NG}_{\rm loop} = \int_\p \big[b^{(A)}(\p,\k_i)+b^{(B)}(\p,\k_i)\big]\ .
\eeq
The integrands $b^{(A)}(\p,\k_i)$ and $b^{(B)}(\p,\k_i)$  can be written as 
\begin{align}
b^{(A)}(\p,\k_i)&=b^{(A)}_1(\p,\k_i)+b^{(A)}_2(\p,\k_i)+b^{(A)}_3(\p,\k_i)\ ,\\
b^{(B)}(\p,\k_i)&=b^{(B)}_1(\p,\k_i)+b^{(B)}_2(\p,\k_i)\ ,
\end{align}
where we have defined 
\begin{align}
b^{(A)}_1(\p,\k_i)&\equiv b_{122}^{\rm (I)}(\k_2-\p,\k_i)\Big[\Theta(|\k_{2}-\p|-p)\Theta(|\k_{3}+\p|-|\k_{2}-\p|)\nonumber\\
&\hskip 90pt+\Theta(|\k_{3}+\p|-p) \Theta(|\k_{2}-\p|-|\k_{3}+\p|)\Big]  + \text{5 perms}\ , \\[5pt]
b^{(A)}_2(\p,\k_i)&\equiv  2\hskip 1ptb_{122}^{\rm (II)}(\p,\k_i)\,\Theta(|\k_{1}-\p|-p)+\text{5 perms}\ ,\\[5pt]
b^{(A)}_3(\p,\k_i)&\equiv  2\hskip 1pt b_{311}^{\rm (I)}(\p,\k_i)\,\Theta(|\k_{1}+\p|-p) +\text{5 perms}\ ,\\[5pt]
b^{(B)}_1(\p,\k_i)&\equiv \frac{1}{2}\Big[b_{122}^{(\rm I)}(\p,\k_i)\Theta(|\k_3+\p|-p)\Theta(|\k_2-\p|-p) + \{\p\to-\p\}\Big]+\text{2 perms}\ ,\\[5pt]
b^{(B)}_2(\p,\k_i)&\equiv b_{311}^{\rm (II)}(\p,\k_i)+\text{2 perms}\ .
\end{align}
It is easy to check that the integrands $b^{(A,B)}_i$ do not have any divergences as $\p$ approaches one of the external momenta $\pm \k_i$. Moreover, while each of the individual integrands $b^{(A,B)}_i$ is divergent in the limit $\p\to\0$, the sums $b^{(A)}$ and $b^{(B)}$ are finite.

\newpage
\section{Notation and Conventions}
\label{sec:notation}

\begin{table}[h!]
\begin{tabular}{|>{\centering\arraybackslash}p{\dimexpr 0.09\textwidth}  | p{\dimexpr 0.26\textwidth-4\tabcolsep} | p{\dimexpr 0.535\textwidth-4\tabcolsep} | >{\centering\arraybackslash}p{\dimexpr 0.16\textwidth -4\tabcolsep} |}
\hline
Symbol    & Relation &	Meaning 	&  Equation\\ \hline\hline
$a$	  & 	& scale factor 	&		\\ \hline
$\tau$ & $a \, \d\tau=\d t$ & conformal time & \\ \hline
$\H$ &$\equiv d \ln (a)/ d\tau$ & conformal Hubble parameter & \\ \hline
$\H_0$ && present value of $\H$ & \\ \hline
$\x$ &  & comoving coordinate & \\ \hline
$\q$ &  & Lagrangian coordinate & \\ \hline
$\k$ &  & long momentum & \\ \hline
$\p$ &  & short momentum & \\ \hline
$\Omega_m$ &  & matter density in units of the critical density& \\ \hline
$\Omega_\Lambda$ &  & dark energy density& \\ \hline
$h$ &  & dimensionless Hubble constant & \\ \hline
$ \rho $ & & dark matter density&  \\ \hline
$ \delta $ & $ \equiv \delta \rho/\rho$ & dark matter density contrast &  \\ \hline
$ \theta $ & $ \equiv \partial_i v^i$ & velocity divergence &  \\ \hline
$ \delta_{(n)}  $ & & density contrast in SPT at order $ n $ & \eqref{dSPT}\\ \hline
$ \theta_{(n)} $ & & velocity divergence in SPT at order $ n $ & \eqref{dSPT} \\ \hline
$F_{n}$ & & kernel function in $\delta_{(n)}$ & \eqref{dn}\\ \hline
$G_{n}$ & & kernel function in $\theta_{(n)}$ & \eqref{tn} \\ \hline
$ P_{mn}$ &$\equiv \vev{\delta_{(m)} \delta_{(n)}}'$ & power spectrum in SPT & \eqref{equ:PSPT} \\ \hline
$ B_{lmn}$ &$\equiv \vev{\delta_{(l)} \delta_{(m)} \delta_{(n)}}'$ & bispectrum in SPT  &  \eqref{equ:BSPT} \\ \hline
$ P_{\delta}$ &$\equiv \vev{\delta \delta}'$  & nonlinear dark matter power spectrum & \eqref{equ:PSPT} \\ \hline
$ B_{\delta}$ & $\equiv \vev{\delta \delta \delta}'$ & nonlinear dark matter bispectrum & \eqref{equ:BSPT} \\ \hline
$ \Delta_{\delta}^2$ &  & dimensionless power spectrum & \eqref{equ:Deltad} \\ \hline
$ {\cal B}_{\delta}$ &  & dimensionless bispectrum &  \eqref{equ:calBd} \\ \hline
$ \phi $ & & Newtonian potential & \eqref{equ:poisson} \\ \hline
$ \Phi $ & $\bigtriangleup \Phi=\delta  $  & rescaled Newtonian potential & \eqref{equ:poisson}  \\ \hline
$ \varphi $ & $\phi = T(k)\varphi$  & primordial potential & \eqref{equ:T}  \\ \hline
$ \varphi_g $ &  & Gaussian primordial potential &  \eqref{eq:phiNG} \\ \hline
$\Delta_\varphi(k_0)$ & $\equiv 3.0\times 10^{-5}$ & amplitude of the primordial potential & \eqref{eq:Pvar} \\ \hline
$k_0$ & & pivot scale & \eqref{eq:Pvar} \\ \hline
$ T(k)$ &   & transfer function & \eqref{equ:T} \\ \hline
$ M(k)$ &   & transfer function in the Poisson equation & \eqref{equ:M} \\ \hline
$D_1$ &  & linear growth factor & \eqref{D1} \\ \hline
$f$ &  $\equiv d \ln D_1/\ln a$ & growth rate & \eqref{eq:tauNG0V2} \\ \hline
$ P_{\varphi}$ & & primordial power spectrum & \eqref{eq:Pvar} \\ \hline
$ B_{\varphi}$ & & primordial bispectrum & \eqref{eq:Bvar} \\ \hline 
$n_s$ & & scalar spectral index & \eqref{eq:Pvar} \\ \hline
$ s_{ij} $ & $\equiv \partial_i \partial_j \Phi - \frac{1}{3} \delta_{ij} \hskip -1pt  \bigtriangleup \hskip -2pt \Phi   $& tidal tensor & \eqref{eq:tauNG0V2}  \\ \hline

\end{tabular}
\end{table}

\newpage
\begin{table}[h!]
\begin{tabular}{|>{\centering\arraybackslash}p{\dimexpr 0.09\textwidth}  | p{\dimexpr 0.26\textwidth-4\tabcolsep} | p{\dimexpr 0.535\textwidth-4\tabcolsep} | >{\centering\arraybackslash}p{\dimexpr 0.16\textwidth -4\tabcolsep} |}
\hline
Symbol    & Relation &	Meaning 	&  Equation\\ \hline\hline
$ \psi $ & & correlation in the initial conditions & \eqref{equ:psi} \\ \hline
$ \Psi $ & $ \Psi(\x)\equiv \psi(\q(\x)) $ & Eulerian definition of $\psi$ & \eqref{PSI} \\ \hline
$ \psi_J $ & & noise term in the initial conditions & \eqref{eq:philong} \\ \hline
$ \psi^{ij} $ & & spin-2 correlation in the initial conditions & \eqref{psiij} \\ \hline
$ \psi^{ijkl} $ & & spin-4 correlation in the initial conditions & \eqref{eq:psiijkl} \\ \hline
$ K_{\mathsmaller{\rm NL}} $ & & kernel function of the primordial bispectrum & \eqref{eq:phiNG} \\ \hline
$a_L $ & & coefficient in the Legendre expansion of $ K_{\mathsmaller{\rm NL}} $ &  \eqref{eq:angSL}  \\ \hline
$\Delta$ & & scaling dimension in $ K_{\mathsmaller{\rm NL}} $  & \eqref{eq:angSL}   \\ \hline
$ f_{\mathsmaller{\rm NL}} $ & & amplitude of the primordial bispectrum & \eqref{eq:angSL}  \\ \hline
$P_L$ & & Legendre polynomial of order $L$ & \eqref{eq:angSL}  \\ \hline
$W_\Lambda$ &  & window function & \eqref{equ:W}  \\ \hline
$F_\Lambda$ & & Fourier transform of $W_\Lambda$  & \eqref{equ:W}  \\ \hline
$\tau^{ij}$ & & stress tensor of the EFT  & \eqref{eq:taueffX}  \\ \hline
$\vev{\tau^{ij}}_s$ & & viscosity part of $\tau^{ij}$  & \eqref{eq:tau2}   \\ \hline
$\Delta \tau^{ij}$ & & noise part of $\tau^{ij}$  & \eqref{eq:tau2}   \\ \hline
$J^{ij}_{\psi}$ & & parameter in $\Delta \tau^{ij}$ & \eqref{eq:noise0} \\ \hline
$ \tau_{\theta} $ & & EFT source in the Euler equation & \eqref{tauT} \\ \hline
$ \tau_{v} $ & & viscosity part of $\tau_\theta$ & \eqref{eq:tauv} \\ \hline
$ \tau_{n} $ & & noise part of $\tau_\theta$ & \eqref{eq:taun} \\ \hline
  $c_s$ &  & sound speed & \eqref{equ:tau}\\ \hline
    $c_{vis}, \hat c_{vis}$ &  & viscosity parameters & \eqref{equ:tau}\\ \hline
  $d^2$ & $\equiv c_s^2 + f (c_{vis}^2+\hat c_{vis}^2)$  & parameter in $\tau_v$  & \eqref{eq:tauv}\\ \hline
    $e_i$, $g$, $g_i$ &  & parameters in $\tau_v$ & \eqref{eq:tauv}\\ \hline
  $\xi$ &  & parameter in $\delta^c_{(1)}$  & \eqref{equ:xi}\\ \hline
      $\gamma$ &  & parameter in $\delta^c_{(1)}$ & \eqref{equ:gamma}\\ \hline
    $\epsilon_i$ &  &  parameter in $\delta^c_{(2)}$ & \eqref{equ:ei}\\ \hline
     $\gamma_i$ &  & parameter in $\delta^c_{(2)}$ & \eqref{equ:gammai}\\ \hline
$ \mathcal{S}_{\alpha,\beta} $ & & SPT quadratic source terms & \eqref{Sa} \\ \hline
$\delta^c_{(n)}$ & & viscosity counterterm at order $n$ & \eqref{equ:deltac(n)}\\ \hline
$\delta^J_{(n)}$ & & noise counterterm at order $n$ & \eqref{eq:deltaJ}\\ \hline
$F^{c}_{n}$ & & kernel function in $\delta^c_{(n)}$ &\eqref{eq:F2c}\\ \hline
$H^{c}_{n}$ & & kernel function in $\delta^c_{(n)}$ & \eqref{eq:H2c}\\ \hline
$G_\delta$ &  & Green's function for $\delta$ & \eqref{eq:Gdelta} \\ \hline
${\cal D}_\delta$ &  & evolution operator in the fluid equation & \eqref{equ:E1} \\ \hline
$P_{1\psi}$ & $\equiv \vev{\delta_{(1)} \psi}'$  & correlation of $\delta_{(1)}$ and $\psi$ & \eqref{eq:psidelta} \\ \hline
$P_{1c}$ & $\equiv \vev{\delta_{(1)} \delta_{(1)}^c}'$  & correlation of $\delta_{(1)}$ and $\delta_{(1)}^c$ & \eqref{eq:P12renorm} \\ \hline
 $\sigma^2(\Lambda)$ &  & divergence in $P_{12}$ & \eqref{eq:Div1} \\ \hline
  $\sigma_\psi^2(\Lambda)$ &  & divergence in $P_{12}$ & \eqref{eq:Div2} \\ \hline
\end{tabular}
\end{table}

\newpage
\begin{table}[h!]
\begin{tabular}{|>{\centering\arraybackslash}p{\dimexpr 0.09\textwidth}  | p{\dimexpr 0.26\textwidth-4\tabcolsep} | p{\dimexpr 0.535\textwidth-4\tabcolsep} | >{\centering\arraybackslash}p{\dimexpr 0.16\textwidth -4\tabcolsep} |}
\hline
Symbol    & Relation &	Meaning 	&  Equation\\ \hline\hline
 $\hat \sigma^2(\Lambda)$ &  & divergence in $B_{122}$ & \eqref{equ:Div3} \\ \hline
 $\hat \sigma_\psi^2(\Lambda)$ &  & divergence in $B_{122}$ & \eqref{equ:Div4} \\ \hline
$\sigma_{v}^2$ &  & velocity dispersion & \eqref{equ:sigmav}\\ \hline
$P_{\rm SPT}^{\rm G}$ & $\equiv P_{11} + P_{13}+P_{22}$  & Gaussian SPT contributions to $P_\delta$ &  \eqref{equ:PGEFT} \\ \hline
$P_{\rm SPT}^{\rm NG}$ & $\equiv P_{12} $ & non-Gaussian SPT contribution to $P_\delta$ &  \eqref{equ:PNGEFT} \\ \hline
$P_{\rm EFT}^{\rm G}$ &$\equiv -2 \xi\hskip 1pt k^2 P_{11}$ & Gaussian EFT counterterm &  \eqref{equ:PGEFT} \\ \hline
$P_{\rm EFT}^{\rm NG}$ & $\equiv -2 \gamma\hskip 1pt k^2 P_{1\psi}$ & non-Gaussian EFT counterterm &  \eqref{equ:PNGEFT} \\ \hline
$B_{\rm SPT}^{\rm G}$ & & Gaussian SPT contributions to $B_\delta$ & \eqref{eq:BGloop} \\ \hline
$B_{\rm SPT}^{\rm NG}$ & & non-Gaussian SPT contributions to $B_\delta$  & \eqref{eq:BNGloop} \\ \hline
$B_{\rm EFT}^{\rm G}$ & & sum of Gaussian EFT counterterms  &  \eqref{BG}\\ \hline
$B_{\rm EFT}^{\rm NG}$ & & sum of non-Gaussian EFT counterterms     &  \eqref{BNG} \\ \hline
$B_0^{\rm G}$ & $\equiv B_{\rm SPT}^{\rm G}+\xi B_{\xi}^{\rm G}$  & Gaussian SPT contributions plus $B_\xi^{\rm G}$ &\eqref{equ:B0} \\ \hline
$B_0^{\rm NG}$ & $\equiv B_{\rm SPT}^{\rm NG}+\xi B_{\xi}^{\rm NG}$  & non-Gaussian SPT contributions plus $B_\xi^{\rm NG}$ &\eqref{equ:B0} \\ \hline
$B_c^{\rm G}$ & $\equiv B_{\rm EFT}^{\rm G} - \xi B_\xi^{\rm G}$ & sum of Gaussian counterterms ($- B_\xi^{\rm G}$)  &\eqref{equ:BcG} \\ \hline
$B_c^{\rm NG}$ & $\equiv B_{\rm EFT}^{\rm NG} - \xi B_\xi^{\rm NG}$ & sum of non-Gaussian counterterms ($- B_\xi^{\rm NG}$) &\eqref{equ:BcNG} \\ \hline
\end{tabular}
\end{table}

\newpage

\addcontentsline{toc}{section}{References}
\bibliographystyle{utphys}
\bibliography{NG_EFT}

\end{document}